\documentclass{elsarticle}

\usepackage{jcomp}
\usepackage[utf8]{inputenc}
\usepackage{amsmath}
\usepackage{amsfonts}
\usepackage[table]{xcolor}
\usepackage{hyperref}

\let\vec\mathbf
\let\eps\varepsilon
\newcommand{\R}{\mathbb{R}}

\newcommand{\tol}{_{tol}}
\newcommand{\fltmin}{\texttt{FLT\_MIN}}
\newcommand{\stepn}{^{(n)}}
\newcommand{\stepns}{^{(n\star)}}
\newcommand{\stepnp}{^{(n+1)}}
\newcommand{\stepnnp}{^{(n,n+1)}}
\newcommand{\stepnns}{^{(n,n\star)}}

\newcommand{\abs}[1]{\left|#1\right|}
\newcommand{\norm}[1]{\lVert#1\rVert}
\newcommand{\invariant}[1]{\norm{#1}_{II}}
\newcommand{\shrate}{\dot{\vec\gamma}}

\newcommand{\fluidparts}{{\mathcal F}}
\newcommand{\wallparts}{{\mathcal W}}

\providecommand{\norm}[1]{\lVert#1\rVert}
\DeclareMathOperator{\fma}{fma}
\DeclareMathOperator{\erf}{erf}

\providecommand{\path}[1]{\url{#1}}

\title{A numerically robust, parallel-friendly variant of BiCGSTAB
for the semi-implicit integration of the viscous term in Smoothed
Particle Hydrodynamics}

\author[1]{Giuseppe \snm{Bilotta}\corref{ingv}}
\cortext[ingv]{Corresponding author: giuseppe.bilotta@ingv.it}
\author[2]{Vito \snm{Zago}}
\author[1]{Veronica \snm{Centorrino}}
\author[2]{Robert A. \snm{Dalrymple}}
\author[1,3]{Alexis \snm{H\'erault}}
\author[1]{Ciro \snm{Del Negro}}
\author[4]{Elie \snm{Saikali}}

\address[1]{Istituto Nazionale di Geofisica e Vulcanologia, Sezione di
Catania, 2 Piazza Roma, Catania 95125, Italy}
\address[2]{Department of Civil and Environmental Engineering,
Northwestern University,
2145 Sheridan Road
Evanston, IL 60208-3109, USA}
\address[3]{Laboratoire Mod\'elisation math\'ematique et numerique,
Conservatoire National des Arts et M\'etiers, 292 Rue Saint-Martin,
Paris 75003, France}
\address[4]{Université Paris-Saclay, CEA,
Service de Thermo-hydraulique et de Mécanique des Fluides,
91191, Gif-sur-Yvette, France}

\accepted{21 Jun 2022}
\availableonline{}

\renewenvironment{abstract}{%
  \global\setbox\absbox=\vtop\bgroup%
  \hsize=\textwidth%
  \leftskip=.335\textwidth%%
  \noindent\unskip\ignorespaces%
  }
 {\mbox{}\newline%
   \mbox{}\newline%
   \copyright~\the\year.
   This manuscript version is made available under the CC-BY-NC-ND 4.0 license\\
   \url{https://creativecommons.org/licenses/by-nc-nd/4.0/}
   \egroup%
 }%

\begin{document}

\begin{abstract}
 Implicit integration of the viscous term can significantly improve
 performance in computational fluid dynamics for highly viscous fluids
 such as lava. We show improvements over our previous proposal for
 semi-implicit viscous integration in Smoothed Particle Hydrodynamics,
 extending it to support a wider range of boundary models. Due to the
 resulting loss of matrix symmetry, a key advancement is a more robust
 version of the biconjugate gradient stabilized method to solve the
 linear systems, that is also better suited for parallelization in both
 shared-memory and distributed-memory systems. The advantages of the new
 solver are demostrated in applications with both Newtonian and
 non-Newtonian fluids, covering both the numerical aspect (improved
 convergence thanks to the possibility to use more accurate boundary
 model) and the computing aspect (with excellent strong scaling and
 satisfactory weak scaling).
\end{abstract}


\maketitle

\section{Introduction}

Smoothed Particle Hydrodynamics (SPH) is a Lagrangian, mesh-free
numerical method for computational fluid dynamics originally designed
for astrophysics~\cite{monaghan1977, monaghan1992} that has since been
adopted in contexts spanning from oceanography to geophysics and
industrial applications~\cite{monaghan2005, violeau2012, shadloo2016}.

For liquids, the SPH numerical method is usually associated with a
weakly compressible formulation (WCSPH), so that the system of
Navier--Stokes and mass conservation equations is closed by introducing
an equation of state that allows direct computation of the pressure from
the other state variables (usually just the density), but density
variations are kept small enough (in the order of $1\%$) to be
considered negligible~\cite{monaghan1994}.

A significant computational benefit of the standard WCSPH formulations,
when paired with an explicit integration scheme, is that the method
becomes trivially parallelizable, and can easily take advantage of
high-performance parallel computing hardware~\cite{harada2007}.

The GPUSPH particle engine~\cite{herault_sph_2010} was the first
implementation of WCSPH to run entirely on Graphics Processing Units
(GPUs) using CUDA, leveraging the ease of parallelization to achieve
significant speed-ups over equivalent serial code. Originally limited to
single-GPU simulations of Newtonian fluids, GPUSPH has since been
extended with the aim of becoming a high-performance, numerically robust
and physically accurate simulation engine for complex fluids: it now
allows the distribution of computation across multiple GPUs on the same
machine~\cite{rustico_advances_2014} as well as across GPU-equipped
nodes in a cluster~\cite{rustico_multi-gpu_2014}, and it covers a wider
range of physical models, including support for non-Newtonian rheologies
and thermo-dynamics, with the specific intent of becoming an essential
tool for the study of lava
flows~\cite{bilotta_2016,zago_aog_2017,zago_aog_2019}.

The downside of adopting an explicit integration scheme is the
potentially small time-step, constrained by CFL-like conditions for the
sound speed, maximum forces magnitude, and diffusive coefficients
(kinematic viscosity, thermal diffusitivity, etc).
In particular, for simulations with a very low Reynolds number, the
dominant factor in the time-step for explicitly integrated WCSPH is
given by the viscous contribution. To work around this issue, several
alternatives have been proposed, centered around an implicit approach to
the integration of the viscous term,
e.g.~\cite{fan2010,litvinov2010,pvl2013,peer2015,weiler2018,monaghan2019visc}.

For GPUSPH, the importance of this limitation has emerged especially in
the simulation of the cooling phase of lava flows~\cite{ganci_egu_2017},
which led us to the development of a semi-implicit integration scheme
that leverages the symmetry of the matrix associated with the implicit
viscous term under appropriate conditions to solve for the velocity
using the Conjugate Gradient (CG) method~\cite{zago_jcp_2018}.

The limitations of the CG are that it can only be reliably applied in
case of numerically symmetric matrix. In the context of the
semi-implicit scheme presented by~\cite{zago_jcp_2018}, this is only
possible if the following requirements are satisified: (1) reciprocal
particle contributions must be symmetric, i.e. the viscous contribution
of particle $\alpha$ to particle $\beta$ must be equal to the
contribution of particle $\beta$ to particle $\alpha$, for all
pairs $\alpha,\beta$ of interacting particles; (2) all particles must
have the same mass and (3) analytical symmetry must be preserved
numerically.

\iffalse
Condition (1) is ensured with specific SPH boundary formulations (e.g.
dynamic~\cite{dalrymple2001sph} or Lennard--Jones
boundaries~\cite{monaghan1994}) where the boundary particles' velocity is
prescribed. This allows us to effectively eliminate the boundary
particles from the linear system and move their contribution to the
right-hand side.

Condition (2) is ensured for single-fluid test-cases when geometries are
all perfectly regular, all lengths are exact multiples of the
inter-particle spacing $\Delta p$ and particles are initialized with a
constant mass (as opposed to e.g. hydrostatic initialization with a
regular spacing, that leads to relative variations in mass of the order
of $10^{-2}$ or less).

Finally, condition (3) is ensured if all operations are done in the same
sequence regardless of the central particle. Some hints about ways to
ensure this, and surprising ways in how it can fail, are discussed in
appendix~\ref{sec:numsym}.
\fi

If any of these conditions is not satisfied, the matrix will be (at
least numerically) non-symmetric, which may prevent the CG from
converging. In the case of (2) and (3), however, the deviation from an
exactly symmetric matrix is generally small enough to allow CG to
converge anyway, or at least to provide sufficiently accurate results even
when stalling. Because of this, in the original implementation presented
in~\cite{zago_jcp_2018}, CG was adopted as a solver, and the limitation
was circumvented by keeping track of the ``best'' (lowest residual
norm) solution found during iteration, to be selected if the CG
stopped converging.

We present here several improvements over our previous
proposal. These include in particular the adoption of a non-standard
formulation for the CG that is more amenable to distribution of the
computation across multiple computational devices
(section~\ref{sec:new-cg}), and most importantly a similar redesign
of the biconjugate gradient stabilized method (BiCGSTAB)
(section~\ref{sec:new-bicgstab}) that is not only better for
parallelization, but also more stable numerically, even when using
single precision. Finally, some caveats on how to avoid the
computational pitfalls that can break numerically the symmetry of
analytically symmetric matrices are presented in ~\ref{sec:numsym}.

The adoption of the BiCGSTAB as a possible solver for the linear system
associated to the implicit integration of the viscous term allows us to
extend the usage of the semi-implicit formulation to a larger number of
cases and SPH formulations. We illustrate this by applying it to boundary
conditions that cannot be eliminated or symmetrized, the ``dummy
boundary'' formulation introduced in~\cite{adami2012}
(sections~\ref{sec:bc}, \ref{sec:semi-implicit}), and show applications
to the Newtonian and a non-Newtonian plane Poiseuille flow
(section~\ref{sec:results})
and to the flow in a periodic lattice of cylinders (section~\ref{sec:lattice}).

\section{Physical-mathematical equations and their discretization}

We focus on the simulation of a weakly compressible,
isothermal fluid. The evolution of the fluid over time is described by
the mass continuity equation
\[
 \frac{D\rho}{Dt} = -\rho \nabla\cdot \vec u
\]
and the Navier--Stokes equations for momentum conservation,
\[
 \rho \frac{D\vec u}{Dt} = - \nabla P + \nabla( \mu \nabla \vec u) +
 \rho \vec g,
\]
where $\rho$ is the fluid density, $\vec u$ the velocity, $\mu$ the
(possibly non-homogeneous) dynamic viscosity, $P$ the pressure, $\vec g$
the external body forces (e.g. gravity), and $D/Dt$ represents the total
(i.e. Lagrangian) derivative with respect to time.

These equations are closed under the assumption of weak compressibility
by adding an equation of state. The most commonly used in WCSPH
is Cole's equation of state~\cite{cole1948}:
\begin{equation}\label{eq:cole-pressure}
P(\rho)=c_0^2 \frac{\rho_0}{\zeta} \left(\left(\frac{\rho}{\rho_0}\right)^\zeta-1\right),
\end{equation}
where $\rho_0$ is the at-rest fluid density, $\zeta$ the polytropic
constant and $c_0$ the speed of sound. Weak compressibility requires
$c_0$ to be at least an order of magnitude higher than the maximum
velocity experienced during the~flow, which should include
the maximum theoretical free-fall velocity in the domain,
to avoid column collapse issues. Other equations of state are also possible,
as discussed in section~\ref{sec:lattice}.

Finally, the dynamic viscosity $\mu$ may be constant (in which case it
can be taken out of the $\nabla$ operator), or may be non-homogeneous,
or even shear-rate dependent, as in the case of a non-Newtonian
fluid~\cite{xu2013,bilotta_2016,xu2017,xu2019}.

\subsection{Rheological models}

In the test-cases presented for validation of our new
semi-implicit formulation we will make use of two different rheological
models: the standard Newtonian model (constant viscosity) and the
Bingham plastic rheological model with Papanastasiou
regularization~\cite{papanastasiou1987}.

Let $\vec\tau$ be the shear stress tensor and $\shrate$ the
shear strain rate. Denote by $\invariant{\tau}$ the square root of the second
invariant of the tensor $\vec\tau$:
\[
\invariant{\tau} = \sqrt{ \frac12
 \left(\tau_{ii} \tau_{jj} - \tau_{ij}\tau_{ji}\right) }.
\]
The Bingham plastic rheological model is defined by two parameters: a
\emph{yield strength} (or \emph{yield stress}) $\tau_0$ and a
\emph{consistency index} $\mu_0$.
For shear stress $\invariant{\tau} < \tau_0$, the fluid behaves like a
rigid body, $\shrate = 0$, i.e. effectively as a fluid with infinite
apparent viscosity. When $\invariant{\tau} \ge \vec \tau_0$,
the apparent viscosity of the Bingham fluid is
\begin{equation}\label{eq:bingham}
\mu(\shrate) = \frac{\tau_0}{\invariant{\shrate}} + \mu_0.
\end{equation}

Due to the discontinuity for $\shrate \to 0$, the Bingham rheology is
difficult to handle numerically.
Papanastasiou~\cite{papanastasiou1987} proposes a regularization of
the yield strength contribution to the apparent viscosity in the form
\begin{equation}\label{eq:papa}
\mu(\shrate) = \tau_0 \frac{1 - \exp(-m\invariant{\shrate})}{\invariant{\shrate}} + \mu_0,
\end{equation}
for some large exponent $m$. With \eqref{eq:papa}, the limiting
viscosity for vanishing shear rates is $\lim_{\invariant{\shrate}\to0}
\mu(\shrate) = m \tau_0 + \mu_0$, which is large but finite.

\subsection{Spatial discretization}

With SPH, the fluid domain $\Omega$ is discretized by a set of nodes
(particles) at a given average inter-particle distance $\Delta p$;
each particle carries information about a virtual portion of fluid, such
as its mass, density, velocity, etc, evolving according to the discretized
version of the continuum equations.

The key to the SPH discretization is the approximation of Dirac's
$\delta$ with a family of \emph{smoothing kernel}s $W(\cdot,
h):\Omega\to\R$ parametrized by a \emph{smoothing length} $h > 0$. The
kernels are positive, with unit integral and $\lim_{h\to0} W(\cdot, h) =
\delta$ in the sense of distributions, which in particular gives us
\[
 \lim_{h\to0} \int_\Omega \phi(\vec r) W(\vec r' - \vec r, h) d\vec r =
 \int_\Omega \phi(\vec r) \delta(\vec r' - \vec r) d\vec r =
 \phi(\vec r')
\]
for any scalar field $\phi : \Omega \to \R$ (and with the usual abuse of
notation for the convolution with $\delta$). The integral is further
discretized as a summation over the particles, leading to the standard
SPH discretization of scalar fields
\[
<\phi(\vec r')> = \sum_\alpha \phi(\vec r_\alpha) W(\vec r' - \vec
r_\alpha, h) V_\alpha,
\]
where the summation is extended to all particles, $\vec r_\alpha$
represents their position and $V_\alpha$ their volume.

The kernel is usually taken to be radially symmetric,
non-increasing in $r = \lVert \vec r \rVert$ and with compact support.
Radial symmetry among other things ensures
first-order consistency~\cite{liu-liu}
and since $W(\vec r, h) = \tilde W(r, h)$ we also have
$\nabla_{\vec r} W (\vec r, h) = \vec r F(r, h)$ where
\[
 F(r, h) = \frac{1}{r} \frac{\partial}{\partial r} \tilde W(r, h).
\]
To improve numerical stability, it is convenient to choose a kernel such
that $F(r, h)$ can be computed without an actual division by $r$, to
avoid degenerate cases for $r \to 0$.

SPH discretization is affected by two sources of error: the
approximation of Dirac's delta, controlled by the smoothing length $h$,
and the approximation of the integral, controlled by the particle
spacing $\Delta p$. Consistency is achieved for $h\to0$ and $\Delta
p\to0$, but $\Delta p$ must go to zero faster than $h$
(i.e. $h/\Delta p$ must also go to zero),
see e.g.~\cite{raviart1985,lanson2008}.

In practical applications, $h$ is often computed from $\Delta p$
multiplying it by some constant \emph{smoothing factor} $\alpha$,
with $1.3$ being a common choice, and the influence radius (half the
diameter of the compact support) of the kernel is $k \alpha \Delta p$,
for some kernel radius $k$.

Further details about our choices of kernel will be discussed in
section~\ref{sec:kernel-choice}, with the corresponding implications in
terms of computational performance and numerical accuracy.

\subsection{Discretized equations}

The reader is referred to~\cite{morris1997,monaghan2005,violeau2012,zago_jcp_2018}
for further details about the derivation of
the discretized equations. The ones we will use in our work are
\begin{equation}\label{eq:part_continuity}
\frac{D\rho_\beta}{Dt} = \sum_{\alpha} m_\alpha \vec u_{\alpha\beta}
\cdot \nabla_\beta W_{\alpha\beta}
\end{equation}
for the mass continuity equation, and
\begin{equation}\label{eq:part_NS}
\frac{D\vec u_\beta}{Dt} = - \sum_{\alpha}
\left( \frac{P_\alpha}{\rho_\alpha^2} +
\frac{P_\beta} {\rho_\beta^2}\right)
F_{\alpha\beta} m_\alpha \vec r_{\alpha\beta} + \sum_{\alpha}
\frac{2\bar\mu_{\alpha\beta}}{\rho_\alpha \rho_\beta}
F_{\alpha\beta} m_\alpha \textbf{u}_{\alpha\beta} + \vec  g
\end{equation}
for the momentum equation, using the standard SPH notation where for any
particle $\alpha$ we denote by $\vec r_\alpha$ its position, $V_\alpha =
m_\alpha/\rho_\alpha$ its volume as the ratio of its (constant) mass to its
density, $P_\alpha$ its pressure and $\vec u_\alpha$ its velocity;
additionally, given two particles $\alpha, \beta$ we have $\vec
r_{\alpha\beta} = \vec r_\alpha - \vec r_\beta$, $\vec u_{\alpha\beta} =
\vec u_\alpha - \vec u_\beta$, $W_{\alpha\beta} = W(\vec
r_{\alpha\beta}, h)$, $F_{\alpha\beta} = F(\vec r_{\alpha\beta}, h)$,
whereas by $\bar\mu_{\alpha\beta}$ we denote the average of the dynamic
viscosity of the two particles, $\mu_\alpha$ and $\mu_\beta$.

To compute the averaged viscosity $\bar\mu_{\alpha\beta}$, several
options are possible. The arithmetic average may be more efficient to
compute, but the harmonic average gives more accurate results for large
differences between $\mu_\alpha$ and $\mu_\beta$~\cite{morris1997,huAdams2006,violeau2009,ghaitanellisphd}.

\paragraph{Non-Newtonian viscosity}

In the non-Newtonian case, the viscosity is computed for each particle
$\beta$ from the shear rate tensor. We first compute the velocity gradient as
\[
\nabla \vec u_\beta = \sum_\alpha \frac{m_\alpha}{\rho_\alpha} F_{\alpha\beta} \vec r_{\alpha\beta} \vec u_{\alpha\beta},
\]
which is used to compute the shear rate $\shrate_\beta = (\nabla \vec u_\beta + (\nabla \vec
u_\beta)^\top)/2$ (where $(\cdot)^\top$ denotes the transpose of a
tensor), and finally its second invariant $D_\beta = \invariant{\shrate_\beta}$.

For the Papanastasiou regularization, $D$ is then used to compute yield
strength contribution to the apparent viscosity, i.e. the first term of
the right-hand side in \eqref{eq:papa}:
\[
Y(D) = \frac{1 - \exp(-m D)}{D}.
\]

Commonly adopted strategies to avoid numerical instabilities as $D\to0$
are to replace $Y(D)$ with $Y_\eps(D) = (1 - \exp(-m D))/(D + \eps)$ for
some small value of $\eps$, or thresholding on $D$ with
\[
\bar Y_\eps(D) = \begin{cases}
(1 - \exp(-m D))/D & D \ge \eps,\\
m & D < \eps.
\end{cases}
\]
Our choice instead is to approximate $Y(D)$ with
\[
\bar Y(D) = \begin{cases}
(1 - \exp(-m D))/D & D \ge m,\\
m \,T(m D) & D < m,
\end{cases}
\]
where $T(x)$ is the MacLaurin series for $(1 - \exp(-x))/x$, truncated
after~8 terms, computed in Horner form~\cite{horner1819}:
\[
T(x) = 1 - \frac x2 \left( 1 - \frac x3 \left( \dots
\left(1 - \frac x7\left(1 - \frac x8\right)\right)\dots\right)\right).
\]
The threshold on $D$ and order of expansion are chosen to ensure that the
maximum error of the truncated series is in the order of
single-precision machine epsilon.

\subsection{Boundary conditions}
\label{sec:bc}

Equations~\eqref{eq:part_continuity} and \eqref{eq:part_NS} are only
correct for particles for which the kernel support does not intersect
the domain boundary. Otherwise, an integral term representing the flux
through the domain boundary is needed.

For particles near the free surface, this can usually be ignored
assuming null fields outside of the fluid body. However, near a solid
wall the additional term must be taken into consideration, in such a way
that the proper boundary conditions are applied on those sections of the
interface.

Several boundary models have been proposed for SPH (see
e.g.~\cite{violeau2016} and references within). In this work we will
focus on two specific particle-based models: the so-called \emph{dynamic
boundary} model~\cite{dalrymple2001sph}, and the so-called \emph{dummy
boundary} model~\cite{adami2012}.

For the dynamic boundary, solid walls are discretized with particles
in the same manner as the fluid. These particles have a prescribed
velocity (the velocity of the corresponding section of boundary), and
their density evolves according to the (discretized) mass continuity
equation~\eqref{eq:part_continuity}. To ensure that each fluid particle
has a complete kernel support, the solid wall is discretized with
several layers of particles, sufficient to cover a full influence
radius, i.e. half of the diameter of the kernel compact support.

For the dummy boundary, a similar discretization is used, but the
properties of the boundary particle do \emph{not} evolve according to
some differential equations; instead, they are computed from the
properties of the neighboring fluid particles in a way that improves the
imposition of the correct boundary conditions.

Following~\cite{adami2012}, denoting by $\fluidparts$ the set of fluid particles,
the pressure of the dummy boundary particles is prescribed as
\begin{equation}\label{eq:dummyP}
 P_\beta = \frac{
  \sum_{\alpha\in\fluidparts} P_\alpha W_{\beta\alpha} +
  (\vec g - \vec a_\beta) \cdot \sum_{\alpha\in\fluidparts} \rho_\alpha
  \vec r_{\beta\alpha} W_{\beta\alpha}
 }{
  \sum_{\alpha\in\fluidparts} W_{\beta\alpha}
 },
\end{equation}
where $\vec a_\beta$ represents the acceleration of the boundary at the
position of boundary particle $\beta$. Since summations are extended only
to neighboring \emph{fluid} particles, the pressure value for each
dummy boundary particle can be computed independently (no need to solve linear systems).

To impose no-slip boundary conditions (as required for low Reynolds
number simulations), \emph{two} velocities are considered for dummy boundary
particles in~\cite{adami2012}: the  \emph{wall velocity} $\vec v_{\beta,w}$, i.e. the
velocity of the wall at the position of the boundary particle $\beta$,
and the \emph{viscous velocity}, obtained as
\begin{equation}\label{eq:dummyvel}
 \vec u_{\beta,v} = 2 \vec u_{\beta,w} - \frac{
  \sum_{\alpha\in\fluidparts} \vec u_\alpha W_{\beta\alpha}
 }{
  \sum_{\alpha\in\fluidparts} W_{\beta\alpha}
 }.
\end{equation}
This velocity is \emph{only} used in the computation of the viscous term
of adjacent fluid particles, i.e. fluid particles with at least a
boundary neighbor; the discretized momentum equation of these particles
can then be written as:
\begin{equation}\label{eq:part_NS_dummy}
\frac{D\vec u_\beta}{Dt} = - \sum_{\alpha}
\left( \frac{P_\alpha}{\rho_\alpha^2} +
\frac{P_\beta} {\rho_\beta^2}\right)
F_{\alpha\beta} m_\alpha \vec r_{\alpha\beta} +
\sum_{\alpha \in
\fluidparts}
\frac{2\bar\mu_{\alpha\beta}}{\rho_\alpha \rho_\beta}
F_{\alpha\beta} m_\alpha \textbf{u}_{\alpha\beta} +
\sum_{\alpha \in
\wallparts}
\frac{2\bar\mu_{\alpha\beta}}{\rho_\alpha \rho_\beta}
F_{\alpha\beta} m_\alpha (\vec u_{\alpha,v} - \vec u_\beta) +
\vec  g.
\end{equation}

\section{Integration scheme}

\subsection{Explicit integration scheme}\label{sec:explicit-scheme}

In GPUSPH we use a predictor\slash corrector integration scheme, that
can be described in the following way:
\begin{enumerate}[1.]
\item compute acceleration and density derivative at instant $n$:
\begin{enumerate}[a)]
\item $\vec a\stepn = \vec a(\vec r\stepn, \vec u\stepn, \rho\stepn, \mu\stepn)$,
\item $\dot \rho\stepn = \dot \rho(\vec r\stepn, \vec u\stepn, \rho\stepn)$,
\end{enumerate}
\item compute half-step intermediate positions, velocities, density, viscosity:
\begin{enumerate}[a)]
\item $\vec r\stepns = \vec r\stepn + \vec u\stepn \frac{\Delta t}2$,
\item $\vec u\stepns = \vec u\stepn + \vec a\stepn \frac{\Delta t}2$,\label{predictor-newvel}
\item $\rho\stepns = \rho\stepn + \dot \rho\stepn \frac{\Delta t}2$,
\item $\mu\stepns = \mu(\vec r\stepns, \vec u\stepns)$,
\end{enumerate}
\item compute corrected acceleration and density derivative:
\begin{enumerate}[a)]
	\item $\vec a\stepns = \vec a(\vec r\stepns, \vec u\stepns, \rho\stepns, \mu\stepns)$,
	\item $\dot \rho\stepns = \dot \rho(\vec r\stepns, \vec u\stepns, \rho\stepns)$,
\end{enumerate}
\item compute new positions, velocities, density, viscosity:
\begin{enumerate}[a)]
\item $\vec r\stepnp = \vec r\stepn + (\vec u\stepn + \vec a\stepns \frac{\Delta t}2) \Delta t$, \label{trapezoidal_explicit}
\item $\vec u\stepnp = \vec u\stepn + \vec a\stepns \Delta t$\label{corrector-newvel},
\item $\rho\stepnp = \rho\stepn + \dot \rho\stepns \Delta t$,
\item $\mu\stepnp = \mu(\vec r\stepnp, \vec u\stepnp)$.
\end{enumerate}
\end{enumerate}

The integration scheme is fully explicit, since both
equations~\eqref{eq:part_continuity} and~\eqref{eq:part_NS}
(or~\eqref{eq:part_NS_dummy}) are computed using the data from the
particle system at step $n$ to produce the results for step $n\star$,
and at step $n\star$ to produce the results for step $n+1$. If dummy
boundaries are being used, the properties of the boundary particles are
computed before the computation of the new accelerations (steps~1.a
and~3.a).

\subsection{From explicit to semi-implicit}
\label{sec:semi-implicit}

The semi-implicit scheme first presented in~\cite{zago_jcp_2018} can be
introduced by looking at the forces computation in either the predictor
or corrector as if the individual sub-step was an application of the
forward Euler method, from $n$ to $n+1$. In the explicit approach
we have, for each fluid particle,
\[
 \frac{\vec u_\beta\stepnp - \vec u_\beta\stepn}{\Delta t} =
  \vec f_{\beta,P}\stepn + \vec f_{\beta,\nu}\stepn + \vec g
\]
where $\vec f_{\beta,P}\stepn$ and $\vec f_{\beta,\nu}\stepn$ represent
respectively the pressure and viscous contribution to the acceleration
of particle $\beta$, computed with data at time-step $n$. The new value
for the velocity is then computed as
\[
 \vec u_\beta\stepnp = \vec u_\beta\stepn + \Delta t \left(
  \vec f_{\beta,P}\stepn + \vec f_{\beta,\nu}\stepn + \vec g
  \right).
\]

\paragraph*{Dynamic boundary case}

Assume for the time being that dynamic boundary conditions are in effect
(we shall see momentarily what changes in the dummy boundary case).
We isolate the viscous term and write it out in full,
from~\eqref{eq:part_NS}:
\[
 \vec u_\beta\stepnp = \vec u_\beta\stepn +
 \Delta t \left(\vec f_{\beta,P}\stepn + \vec g\right) +
 \Delta t \sum_\alpha
\frac{2\bar\mu_{\alpha\beta}\stepn}{\rho_\alpha\stepn \rho_\beta\stepn}
F_{\alpha\beta}\stepn m_\alpha \vec u_{\alpha\beta}\stepn
\]
and transition to the semi-implicit approach by using the velocity at
step $n+1$ to compute the viscous term:
\begin{equation}\label{eq:implicit-verbose}
 \vec u_\beta\stepnp = \vec u_\beta\stepn +
 \Delta t \left(\vec f_{\beta,P}\stepn + \vec g\right) +
 \Delta t \sum_\alpha
\frac{2\bar\mu_{\alpha\beta}\stepn}{\rho_\alpha\stepn \rho_\beta\stepn}
F\stepn_{\alpha\beta} m_\alpha \vec u_{\alpha\beta}\stepnp.
\end{equation}

We now introduce
\begin{equation}\label{eq:kab}
 K_{\alpha\beta}\stepn =
-\frac{2\bar\mu_{\alpha\beta}\stepn}{\rho_\alpha\stepn \rho_\beta\stepn}
F_{\alpha\beta}\stepn m_\alpha
\end{equation}
to simplify notation, and remark that the coefficient is computed from
values at step $n$, while the velocities it multiplies are taken at step
$n+1$. If we move all $n+1$ terms to the left-hand side of~\eqref{eq:implicit-verbose}, 
writing out the expression for $\vec u_{\alpha\beta}\stepnp = \vec u_\alpha\stepnp -
\vec u_\beta\stepnp$ and collecting similar terms, we finally get
\begin{equation}\label{eq:implicit-system-base}
 \left(1 - \Delta t \sum_\alpha K_{\alpha\beta}\stepn \right)\vec u_\beta\stepnp
 + \Delta t \sum_\alpha K_{\alpha\beta}\stepn \vec u_\alpha\stepnp =
 \vec u_\beta\stepn +
 \Delta t \left(\vec f_{\beta,P}\stepn + \vec g\right) \quad
 \forall\beta\in\fluidparts.
\end{equation}

With our assumption of dynamic boundary conditions, the new velocity $\vec
u_\alpha\stepnp = \vec u_{\alpha,w}\stepnp$ is known (prescribed) for
all boundary particles. A possible approach is then to rewrite
\eqref{eq:implicit-system-base} isolating the unknowns to get
\begin{equation}\label{eq:implicit-system-dyn-reduced}
 \left(1 - \Delta t \sum_{\alpha\in\fluidparts\cup\wallparts} K_{\alpha\beta}\stepn \right)\vec u_\beta\stepnp
 + \Delta t \sum_{\alpha\in\fluidparts} K_{\alpha\beta}\stepn \vec u_\alpha\stepnp =
 \vec u_\beta\stepn +
 \Delta t \left(\vec f_{\beta,P}\stepn + \vec g\right)
 - \Delta t \sum_{\alpha\in\wallparts} K_{\alpha\beta}\stepn
 \vec u_\alpha\stepnp
\end{equation}
obtaining a linear system where the only unknowns are the fluid particle
velocities, and there is one (vector) equation per fluid particle.

However, both for consistency with~\cite{zago_jcp_2018} and to simplify
the extension to the dummy boundary case, we consider the boundary
particle velocities as unknowns as well, resulting in the following
system of equations:
\begin{equation}\label{eq:implicit-system-dyn-full}
\left\{
\begin{aligned}
 \left(1 - \Delta t \sum_\alpha K_{\alpha\beta}\stepn \right)
  & \vec u_\beta\stepnp
 + \Delta t \sum_\alpha K_{\alpha\beta}\stepn \vec u_\alpha\stepnp && {}=
 \vec u_\beta\stepn +
 \Delta t \left(\vec f_{\beta,P}\stepn + \vec g\right) &
 \forall\beta\in\fluidparts,\\
& \vec u_\beta\stepnp && {}= \vec u_{\beta,w}\stepnp &
\forall\beta\in\wallparts.
\end{aligned}\right.
\end{equation}

The velocity at step $n+1$ is therefore obtained solving one linear
system per spatial component. These systems all have the same
coefficient matrix $\vec A\stepn(\Delta t)$ (\emph{viscous matrix}
to advance from step $n$ with time-step $\Delta t$), with diagonal entries
\begin{equation}\label{eq:diag}
a_{\beta\beta}\stepn(\Delta t) = \begin{cases}
1 - \Delta t \displaystyle\sum_{\alpha\in\fluidparts\cup\wallparts} K_{\alpha\beta}\stepn
&\forall\beta\in\fluidparts,\\
1 & \forall\beta\in\wallparts,
\end{cases}
\end{equation}
and off-diagonal entries
\begin{equation}\label{eq:offdiag}
a_{\beta\alpha}\stepn(\Delta t) = \begin{cases}
\Delta t  K_{\alpha\beta}\stepn &
	\text{$\beta \in \fluidparts$ and $\alpha$ is its neighbor,}\\
0 & \text{otherwise.}
\end{cases}
\end{equation}
The matrix is strictly diagonally dominant, as shown
in~\cite{zago_jcp_2018}, and therefore non-singular.

\paragraph*{Dummy boundary case}

In the dummy boundary case, \eqref{eq:implicit-system-base} is still
valid if the boundary particle contribution is taken to be the
\emph{viscous velocity} defined in~\eqref{eq:dummyvel}, i.e. if we
rewrite it as
\[
 \left(1 - \Delta t \sum_\alpha K_{\alpha\beta}\stepn \right)\vec v_\beta\stepnp
 + \Delta t \sum_\alpha K_{\alpha\beta}\stepn \vec v_\alpha\stepnp =
 \vec v_\beta\stepn +
 \Delta t \left(\vec f_{\beta,P}\stepn + \vec g\right) \quad
 \forall\beta\in\fluidparts,
\]
where
\[
\vec v_\alpha = \begin{cases}
\vec u_\alpha & \forall\alpha\in\fluidparts,\\
\vec u_{\alpha,v} & \forall\alpha\in\wallparts.
\end{cases}
\]
This must be paired with \eqref{eq:dummyvel}, giving us the system
\begin{equation}\label{eq:implicit-system-dummy-full}
\left\{
\begin{aligned}
 \left(1 - \Delta t \sum_\alpha K_{\alpha\beta}\stepn \right)
  & \vec v_\beta\stepnp
 + \Delta t \sum_\alpha K_{\alpha\beta}\stepn \vec v_\alpha\stepnp && {}=
 \vec v_\beta\stepn +
 \Delta t \left(\vec f_{\beta,P}\stepn + \vec g\right) &
 \forall\beta\in\fluidparts,\\
 & \vec v_\beta\stepnp +
 \frac{
  1
 }{
  \sum_{\alpha'\in\fluidparts} W_{\beta\alpha'}
 } \sum_{\alpha\in\fluidparts}
 W_{\beta\alpha} \vec v_\alpha\stepnp
 && {} = 2 \vec u_{\beta,w}\stepnp &
\forall\beta\in\wallparts.
\end{aligned}\right.
\end{equation}

Observe that in contrast to \eqref{eq:implicit-system-dyn-full}, this
time the boundary particle velocities are actual unknowns, and their
value must be solved for together with the velocity of the fluid
particles. We still need to solve one linear system per spatial
component, and all systems have the same coefficient matrix
$\vec A\stepn(\Delta t)$. The diagonal entries are still given by
equation~\eqref{eq:diag},
but the off-diagonal entries are now
\[
a_{\beta\alpha}\stepn(\Delta t) = \begin{cases}
\Delta t  K_{\alpha\beta}\stepn &
	\text{if $\beta\in\fluidparts$ and $\alpha \in\fluidparts\cup\wallparts$ is its neighbor,}\\
  W_{\beta\alpha} / \sum_{\alpha'\in\fluidparts} W_{\beta\alpha'}
 & \text{if $\beta\in\wallparts$ and $\alpha\in\fluidparts$ is its neighbor,}\\
0 & \text{otherwise.}
\end{cases}
\]

In this case the matrix is only \emph{weakly} diagonally dominant: the
rows corresponding to fluid particles are still strictly diagonally
dominant (SDD), but for the boundary particles we have
\[
 \sum_\alpha \abs{a_{\beta\alpha}\stepn(\Delta t)} =
 \sum_\alpha a_{\beta\alpha}\stepn(\Delta t) =
 \sum_{\alpha\in\fluidparts}
 \frac{
  W_{\beta\alpha}
 }{
  \sum_{\alpha'\in\fluidparts} W_{\beta\alpha'}
 } =
 \frac{
 \sum_{\alpha\in\fluidparts}
  W_{\beta\alpha}
 }{
  \sum_{\alpha'\in\fluidparts} W_{\beta\alpha'}
 } = 1 = a_{\beta\beta}\stepn(\Delta t) =
 \abs{ a_{\beta\beta}\stepn(\Delta t) }
\]
so only the equality holds.
However, all non-zero off-diagonal contributions to these rows
come from fluid particles, and thus correspond to indices of SDD rows:
therefore, for each non-SDD row there is a directed graph walk (of
length exactly 1) to SDD rows, ensuring that the matrix is \emph{weakly
chained diagonally dominant} (WCDD), and hence non-singular~\cite{wcdd}.

\paragraph*{Shared formulation}
We can describe the approach for both the dynamic and dummy boundary
models as the resolution of $d$ linear systems ($d$ being the
dimensionality of the problem, in our case $d=3$) described by the
equation
\[
\vec A\stepn(\Delta t) \vec V\stepnp = \vec B\stepnnp(\Delta t)
\]
where $\vec A\stepn(\Delta t)$ is an $N\times N$ matrix ($N$ being the
number of particles) and $\vec V\stepnp, \vec B\stepnnp(\Delta t)$ are
$N\times d$ matrices.

The rows of $\vec V\stepnp$ are
\[
\vec V_\beta\stepnp = \begin{cases}
\vec u_\beta\stepnp & \forall\beta\in\fluidparts,\\
\vec u_\beta\stepnp
	& \text{$\forall\beta\in\wallparts$ for the dynamic model,}\\
\vec u_{\beta,v}\stepnp
	& \text{$\forall\beta\in\wallparts$ for the dummy model,}\\
\end{cases}
\]
whereas the rows of $\vec B\stepnnp(\Delta t)$ are
\[
\vec B_\beta\stepnnp(\Delta t) = \begin{cases}
\vec u_\beta\stepn +
\Delta t \left(\vec f_{\beta,P}\stepn + \vec g\right)
	& \forall\beta\in\fluidparts,\\
\vec u_{\beta,w}\stepnp
	& \text{$\forall\beta\in\wallparts$ for the dynamic model,}\\
2 \vec u_{\beta,w}\stepnp
	& \text{$\forall\beta\in\wallparts$ for the dummy model.}\\
\end{cases}
\]
We observe that for fluid particles the right-hand side is always the
explicit integration of the \emph{inviscid} acceleration (i.e. the
acceleration computed without considering the viscous term).

\subsection{Semi-implicit integration scheme}
\label{sec:semi-implicit-scheme}

Given the non-singularity of the coefficient matrix for the viscous
systems \eqref{eq:implicit-system-dyn-full} and
\eqref{eq:implicit-system-dummy-full}, we can adapt the explicit
integration scheme into a semi-implicit predictor\slash corrector
scheme, for either the dummy or dynamic boundary model, with the
following structure:
\begin{enumerate}[1.]
\item compute \emph{inviscid} acceleration and density derivative at instant $n$:
\begin{enumerate}[a)]
\item $\bar{\vec a}\stepn = \bar{\vec a}(\vec r\stepn, \vec u\stepn, \rho\stepn, \mu\stepn)$,
\item $\dot \rho\stepn = \dot \rho(\vec r\stepn, \vec u\stepn, \rho\stepn)$,
\end{enumerate}
\item compute half-step intermediate velocities, positions, density, viscosity:
\begin{enumerate}[a)]
\item compute right-hand side $\vec B\stepnns(\Delta t/2)$,
\item solve
 $\vec A\stepn(\Delta t/2) \vec V\stepns = \vec B\stepnns(\Delta t/2)$
and get $\vec u\stepns$ from $\vec V\stepns$,
\item $\vec r\stepns = \vec r\stepn + \vec u\stepn \frac{\Delta t}2$,
\item $\rho\stepns = \rho\stepn + \dot \rho\stepn \frac{\Delta t}2$,
\item $\mu\stepns = \mu(\vec r\stepns, \vec u\stepns)$,
\end{enumerate}
\item compute corrected \emph{inviscid} acceleration and density derivative
\begin{enumerate}[a)]
\item $\bar{\vec a}\stepns = \bar{\vec a}(\vec r\stepns, \vec u\stepns, \rho\stepns, \mu\stepns)$,
\item $\dot \rho\stepns = \dot \rho(\vec r\stepns, \vec u\stepns, \rho\stepns)$,
\end{enumerate}
\item compute new positions, velocities, density, viscosity:
\begin{enumerate}[a)]
\item compute right-hand side $\vec B\stepnnp(\Delta t)$,
\item solve
$\vec A\stepns(\Delta t) \vec V\stepnp = \vec B\stepnnp(\Delta t)$
and get $\vec u\stepnp$ from $\vec V\stepnp$,
\item $\vec r\stepnp = \vec r\stepn + (\vec u\stepn + \vec u\stepnp)\frac{\Delta t}2$,
\item $\rho\stepnp = \rho\stepn + \dot \rho\stepns \Delta t$,
\item $\mu\stepnp = \mu(\vec r\stepnp, \vec u\stepnp)$.
\end{enumerate}
\end{enumerate}

An important thing to remark is that for non-Newtonian fluids, the
apparent dynamic viscosity is computed from the strain rate at the
\emph{previous} time-step. Using the \emph{current} values of $\vec u$
would make the systems non-linear, since the $K_{\alpha\beta}$
coefficients would then depend on the new velocity itself.

\section{Linear system properties and resolution}

In~\cite{zago_jcp_2018} it was observed that under appropriate
conditions, the coefficient matrix was not only strictly diagonally
dominant, but also symmetric. This is the case for the dynamic boundary
model under the assumption that the particles all have the same mass
(single fluid simulation with uniform initialization). In
fact, even under these assumptions the matrix is only symmetric if the
\emph{reduced} system \eqref{eq:implicit-system-dyn-reduced} that only
involves fluids particles is considered, \emph{or} with the full system
in the case of null boundary velocity.
Indeed, with the full system \eqref{eq:implicit-system-dyn-full}, if the
wall boundary has non-zero speed, the symmetry of the matrix is broken
by the fact that the contribution of boundary neighbors to fluid
particles is non-zero, while the contribution of fluid neighbors to
boundary particles is null.

The symmetry of the matrix was leveraged in~\cite{zago_jcp_2018} to
solve the linear system using the Conjugate Gradient (CG) method. To take
into account the possibility that the matrix may not be perfectly
symmetric in practical applications (e.g. due to hydrostatic particle
initialization leading to small fluctuations in the particle mass), the
CG implementation in~\cite{zago_jcp_2018} kept track of the best
solution found at each solver iteration, to be restored in case the CG
stopped converging or even diverged.

Since the symmetry of the matrix cannot be leveraged anymore in the
dummy boundary case, we have extended the work from~\cite{zago_jcp_2018}
to offer the stabilized biconjugate gradient (BiCGSTAB) as an
alternative solver. Additionally, both the original CG implementation
and the new BiCGSTAB have been restructured to improve numerical
stability and allow the implementation to take advantage of the distributed
(multi-GPU and multi-node) computing features of GPUSPH.

\subsection{Restructuring the Conjugate Gradient}\label{sec:new-cg}

The standard implementation of the Conjugate Gradient to solve a generic
linear system
\[
\vec A \vec x = \vec b
\]
can be described as follows.

Initialize $\vec x_0$ initial guess, $\vec r_0 = \vec b - \vec A\vec x_0$
residual and $\vec p_0 = \vec r_0$ (first basis vector). Also compute
the convergence threshold $\tau = \norm{\vec b}\eps\tol$ (in practice,
we avoid computing square roots and use its square $\tau^2 = \eps\tol^2
\vec b\cdot \vec b$; we use $\eps\tol = 2^{-23} \simeq
1.1920929\cdot10^{-7}$, i.e. the single-precision machine epsilon).

Then, for each iteration $i \ge 0$ compute:
\begin{eqnarray*}
 \alpha_i &=& \frac{\vec r_i \cdot \vec r_i}{\vec p_i\cdot \vec A\vec p_i},\\
 \vec r_{i+1} &=& \vec r_i - \alpha_i \vec A\vec p_i,\\
 \vec x_{i+1} &=& \vec x_i + \alpha_i \vec p_i,\\
 \beta_i &=& \frac{\vec r_{i+1} \cdot \vec r_{i+1}}{\vec r_i \cdot \vec r_i},\\
 \vec p_{i+1} &=& \vec r_{i+1} + \beta_i \vec p_i.
\end{eqnarray*}

For an easier distribution across multiple computational devices, the
logic can be improved by “rotating” the algorithm so that the $\beta_i$
update and consequent check on the residual norm are the last
steps of the algorithm, and the $\vec p$ update is the first.
The new implementation of CG in GPUSPH can therefore be described as
follows.

Initialize $\vec x_0$ initial guess, $\vec r_0 = \vec b - \vec A\vec x_0$
residual and $\beta_0 = 0$. Also compute the convergence
threshold $\tau = \norm{\vec b}\eps\tol$ (in practice, we avoid
computing square roots and use its square $\tau^2 = \eps\tol^2 \vec
b\cdot \vec b$).

Then, for each iteration $i \ge 0$ compute:
\begin{eqnarray*}
 \vec p_i &=& \vec r_i + \beta_i \vec p_{i-1}\\
 \alpha_i &=& \frac{\vec r_i \cdot \vec r_i}{\vec p_i\cdot \vec A\vec p_i}\\
 \vec r_{i+1} &=& \vec r_i - \alpha_i \vec A\vec p_i\\
 \vec x_{i+1} &=& \vec x_i + \alpha_i \vec p_i\\
 \beta_{i+1} &=& \frac{\vec r_{i+1} \cdot \vec r_{i+1}}{\vec r_i \cdot \vec r_i}
\end{eqnarray*}

Note that since $\beta_0 = 0$, it does not matter how $\vec p_{-1}$ is
initialized. In fact, our implementation of the BLAS \texttt{axpby}
routine optimizes for this case by not loading the vector if the
coefficient is null, avoiding potential IEEE not-a-number values that
may pop up when reading uninitialized memory, without paying the cost of
e.g. setting it to zero. Effectively, this turns the $\vec p$ update on
the first step into a straight copy, with the same computational cost as
the $\vec p_0 = \vec r_0$ initialization used in the standard
version of the algorithm.

\paragraph{Stopping criterion}

In the “rotated” CG, convergence can be checked concurrently to the
computation of $\beta_i$ using $\vec r_{i+1} \cdot \vec r_{i+1} <
\tau^2$ as stopping condition.

(In fact, $\beta_i$ can be computed even before computing $\vec x_{i+1}$,
since $\beta_i > 1$ indicates that the CG might have stopped converging,
which is the condition under which the “last good value” of $\vec x_i$
used in~\cite{zago_jcp_2018} was updated. In our new implementation, we
have decided not to adopt this solution, since the cases where the
strategy was necessary are now covered by the adoption of the BiCGSTAB
method instead.)

In the reordered sequence, therefore, convergence is checked at the end
of the algorithm. In addition, we also check if the algorithm is
stalling when computing $\alpha_i$, i.e. we check if $\alpha_i = 0$ in
finite precision even though $\vec r_i$ does not have a sufficiently
small norm. In this case, we complete the iteration and then exit.

Avoiding an early bailout, we pay a computational cost (due to the
vector updates that are effectively no-ops), in exchange for a more
streamlined implementation of the scheme (single exit point at the end),
which simplifies our multi-GPU implementation.

In practice we have never experienced the stalling condition when the
matrix is numerically symmetric, except when $\tau^2$ is exceptionally
low, which can be avoided by setting a lower bound to $\tau^2$. We use
the smallest representable \emph{normal} number:
\[
 \tau^2 = \min\left\{\eps\tol^2 \vec b\cdot\vec b, \fltmin\right\}
\]
where $\fltmin = 2^{-126} \simeq 1.17549435\cdot10^{-38}$ in single
precision. Note that \fltmin\ is \emph{not} squared. Further notes on
the implementation can be found in~\ref{sec:impl-notes}

\subsection{Restructuring the stabilized biconjugate gradient method}
\label{sec:new-bicgstab}

When the matrix is not symmetric (be it numerically, or analytically),
the conjugate gradient is not a feasible solver. Our choice
has been to rely on the stabilized biconjugate gradient (BiCGSTAB) in
this case. The main driving forces behind this choice have been the
relative simplicity of the algorithm compared to other general linear
system solvers, especially in the context of parallel and distributed
implementation, as well as its similarity with our implementation of~CG,
requiring only a small number of changes to reach full implementation.

Like we did for the CG, we have rearranged the traditional BiCGSTAB
algorithm, this time with a double purpose: streamline the
implementation to simplify the distributed (multi-GPU\slash multi-node)
version, but also to improve its numerical robustness. This was
particularly important after observing that a direct implementation of
the standard algorithm would frequently stall with our default choice of
convergence tolerance, leading to much noisier numerical results. Our
reordering of the algorithm significantly reduced the instances of
stalling convergence, leading also to very clean results.

The standard implementation of BiCGSTAB can be described in the
following way.

Initialize $\vec x_0$ initial guess, $\vec r_0 = \vec b - \vec A\vec x_0$
residual, $\hat{\vec r}_0 = \vec r_0$ (in general, any vector such that
$\hat{\vec r}_0 \cdot \vec r_0 \ne 0$), $\vec p_0 = \vec 0$.
Also compute the convergence threshold $\tau = \norm{\vec b}\eps\tol$
and set $\gamma_0 = \alpha_0 = \omega_0 = 1$.

Then, for each iteration $i \ge 1$ compute:
\begin{eqnarray*}
 \gamma_i &=& \hat{\vec r}_0 \cdot \vec r_{i-1}\\
 \beta_i &=& \frac{\gamma_i}{\gamma_{i-1}} \frac{\alpha_{i-1}}{\omega_{i-1}}\\
 \vec p_i &=& \vec r_{i-1} + \beta_i\left( \vec p_{i-1} - \omega_{i-1}
 \vec A \vec p_{i-1} \right)\\
 \delta_i &=& \hat{\vec r}_0 \cdot \vec A \vec p_i\\
 \alpha_i &=& \frac{\gamma_i}{\delta_i}\\
 \vec r_\star &=& \vec r_{i-1} - \alpha_i \vec A \vec p_i \\
 \vec x_\star &=& \vec x_{i-1} + \alpha_i \vec p_i \\
 \omega_i &=& \frac{\vec r_\star \cdot \vec A \vec r_\star}{(\vec A \vec
 r_\star)\cdot(\vec A \vec r_\star)}\\
 \vec r_i &=& \vec r_\star - \omega_i \vec A \vec r_\star\\
 \vec x_i &=& \vec x_\star + \omega_i \vec r_\star
\end{eqnarray*}
until convergence. Convergence is generally checked each time a $\vec x$
update is computed, so both after computing $\vec x_\star$ and when
computing $\vec x_i$.

To introduce our redesign of the BiCGSTAB algorithm, observe first of all
that
\[
 \beta_i =
 \frac{\gamma_i}{\gamma_{i-1}} \frac{\alpha_{i-1}}{\omega_{i-1}} =
 \frac{\gamma_i}{\gamma_{i-1}} \frac{\gamma_{i-1}}{\delta_{i-1}\omega_{i-1}} =
 \frac{\gamma_i}{\delta_{i-1}\omega_{i-1}} =
 \frac{\alpha'_{i-1}}{\omega_{i-1}}
\]
where
\[
 \alpha'_{i-1} = \frac{\gamma_i}{\delta_{i-1}}.
\]
We can take advantage of this by computing
$\gamma, \alpha', \beta$ at the end of the loop, leading to the
following version.

Initialize $\vec x_0$ initial guess, $\vec r_0 = \vec b - \vec A\vec x_0$
residual, $\hat{\vec r}_0 = \vec r_0$ (in general, any vector such that
$\hat{\vec r}_0 \cdot \vec r_0 \ne 0$). Also compute the convergence
threshold $\tau = \norm{\vec b}\eps\tol$ and set $\gamma_0 = \hat{\vec
r}_0 \cdot \vec r_0 = \vec r_0\cdot\vec r_0, \alpha'_0 = \beta_0 = 0$.

Then for each iteration $i \ge 1$ compute:
\begin{eqnarray*}
 \vec p_i &=& \vec r_{i-1} + \beta_{i-1} \vec p_{i-1} - \alpha'_{i-1}
 \vec A \vec p_{i-1}\\
 \delta_i &=& \hat{\vec r}_0 \cdot \vec A \vec p_i\\
 \alpha_i &=& \frac{\gamma_i}{\delta_i} \\
 \vec r_\star &=& \vec r_{i-1} - \alpha_i \vec A \vec p_i \\
 \vec x_\star &=& \vec x_{i-1} + \alpha_i \vec p_i \\
 \omega_i &=& \frac{\vec r_\star \cdot \vec A \vec r_\star}{(\vec A \vec
 r_\star)\cdot(\vec A \vec r_\star)}\\
 \vec r_i &=& \vec r_\star - \omega_i \vec A \vec r_\star\\
 \vec x_i &=& \vec x_\star + \omega_i \vec r_\star\\
 \gamma_{i+1} &=& \hat{\vec r}_0 \cdot \vec r_i\\
 \alpha'_i &=& \frac{\gamma_{i+1}}{\delta_i} \\
 \beta_i  &=& \frac{\alpha'_i}{\omega_i}
\end{eqnarray*}

We have observed that, with the very stringent thresholds we have on the
residuals, the standard BiCGSTAB has a tendency to stall frequently in
single precision, whereas the formulation we propose allows us to solve
the system without stalls except at the highest resolutions (see also
section~\ref{sec:results}). The simplified, and more robust, definition of
$\beta_i$ is a key element of the improved stability.

\paragraph{Stopping criterion}

Compared to CG, BiCGSTAB updates $\vec x$ twice, so the convergence
threshold must be checked twice. Additionally, the squared norm of the
residual is \emph{not} computed as part of the algorithm, and it has to
be computed specifically for the check.

In our implementation, we only check if the \emph{final} $\vec x$ update
is good enough, but will also stop if any of the coefficients zeroes out,
indicating that the algorithm is stalling.
A particularly pernicious case is $\delta_i = 0$, since this leads to
degenerate $\alpha_i, \alpha'_i$ and consequently $\beta_i$. In these
situations, we set $\alpha_i = \alpha'_i = \beta_i = 0$ and, since we
are concurrently solving three systems, make a stopping decision based
on the condition for the other systems.

Each system is considered in one of three states: converged (if at the
last evaluation $\vec r_i\cdot\vec r_i < \tau^2$), stalled ($\delta_i =
0$) or progressing (if neither of these conditions is satisfied). If all
systems are in \emph{either} converged or stalled state, we complete the
iteration and stop. If \emph{any} of the systems is still progressing,
we complete the iteration and \emph{continue to the next}, again for all
systems. In both cases, the user is informed about which of the systems
has stalled.

\section{Test case: the plane Poiseuille flow}
\label{sec:results}

We validate our implementation using the single-fluid plane Poiseuille
flow test case. Four variants are considered, changing the fluid
rheology (Newton and Papanastasiou) and the boundary model (dynamic and
dummy).

\begin{figure}
\centerline{\includegraphics{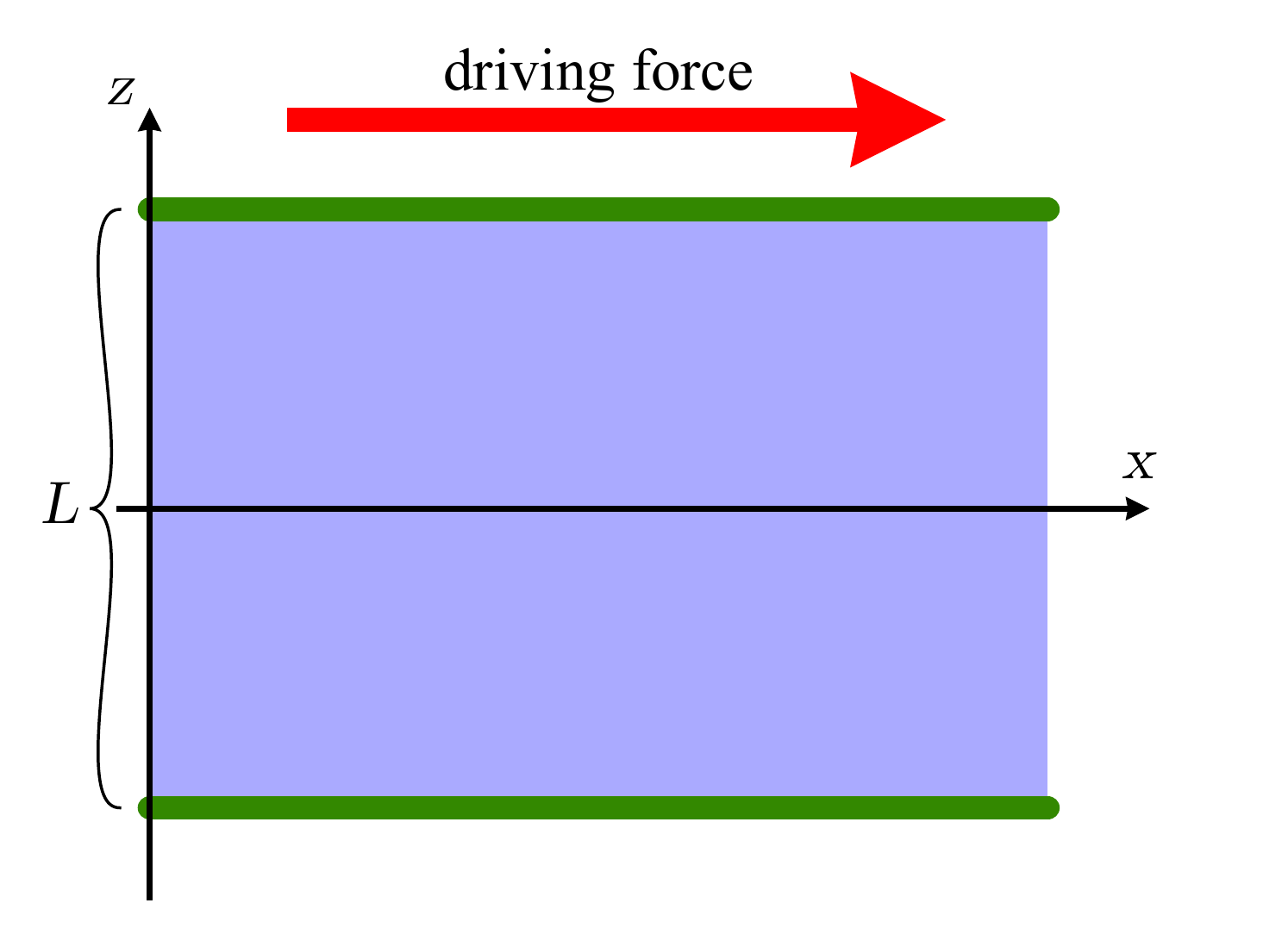}}
\caption{Cross-section along the $y=0$ plane of the domain for the
analytical test case.}
\label{fig:plane-poiseuille}
\end{figure}

The domain is defined by two parallel planes separated by a distance
$L = 1\,\text{m}$
in the vertical ($z$) direction. The origin is placed halfway between
the planes, so that the two planes have equations $z = -L/2$ and $z =
L/2$.
A driving force per unit mass $\vec g = (g, 0, 0)$
with $g = 0.05 \,\text{m}\,\text{s}^{-2}$
is applied to the contained fluid.
Motion is thus expected to be uni-directional in the $x$ direction
($\vec u = (u(z), 0, 0)$), and the domain is periodic both in the $x$
and $y$ directions (figure~\ref{fig:plane-poiseuille}).

In the Newtonian case, the fluid rheology is defined by the dynamic
viscosity $\mu = 0.1 \,\text{Pa}\,\text{s}$. In the Papanastasiou case
we have consistency index $\mu_0 = 0.1 \,\text{Pa}\,\text{s}$ and yield
strength $\tau_0 = 0.0125 \,\text{Pa}$, with regularization
constant $m=1000\,\text{s}$. In both cases the
at-rest density is $\rho_0 = 1 \,\text{kg}\,\text{m}^{-3}$.

With the resulting Reynolds number $\text{Re} = 6.25\cdot10^{-1}$,
the maximum flow velocity is
$u_{\text{max}} = 6.25\cdot10^{-2}\,\text{m}\,\text{s}^{-1}$
(Newton) or $u_{\text{max}} = 1.56\cdot10^{-2}\,\text{m}\,\text{s}^{-1}$
(Papanastasiou), which is in both cases lower than the maximum
free-fall velocity across the domain
$u_h = 3.16\cdot10^{-1}\,\text{m}\,\text{s}^{-1}$.
The sound speed is thus taken 20 times higher than $u_h$
($c_0 = 6.32\,\text{m}\,\text{s}^{-1}$).

We validate our results by comparing the velocity profile obtained in
GPUSPH against the velocity profile expected by the theory for a
stationary flow. The simulations start with a still fluid ($\vec u =
0$); stationary flow is expected to be reached after the viscous
relaxation time $t^\star = \rho_0 L^2/\mu = 10\,\text{s}$. (In the
Bingham case a lower relaxation time would be sufficient, due to the
higher effective viscosity, but for simplicity we adopt the same
$t^\star$ used for the Newtonian fluid.)

\subsection{Domain discretization}

The whole domain is discretized with an initially regular
distribution of particles, with constant spacing $\Delta p$, but
with slightly different positions depending on the choice of boundary
conditions.

\begin{figure}
\centerline{\includegraphics[width=\textwidth]{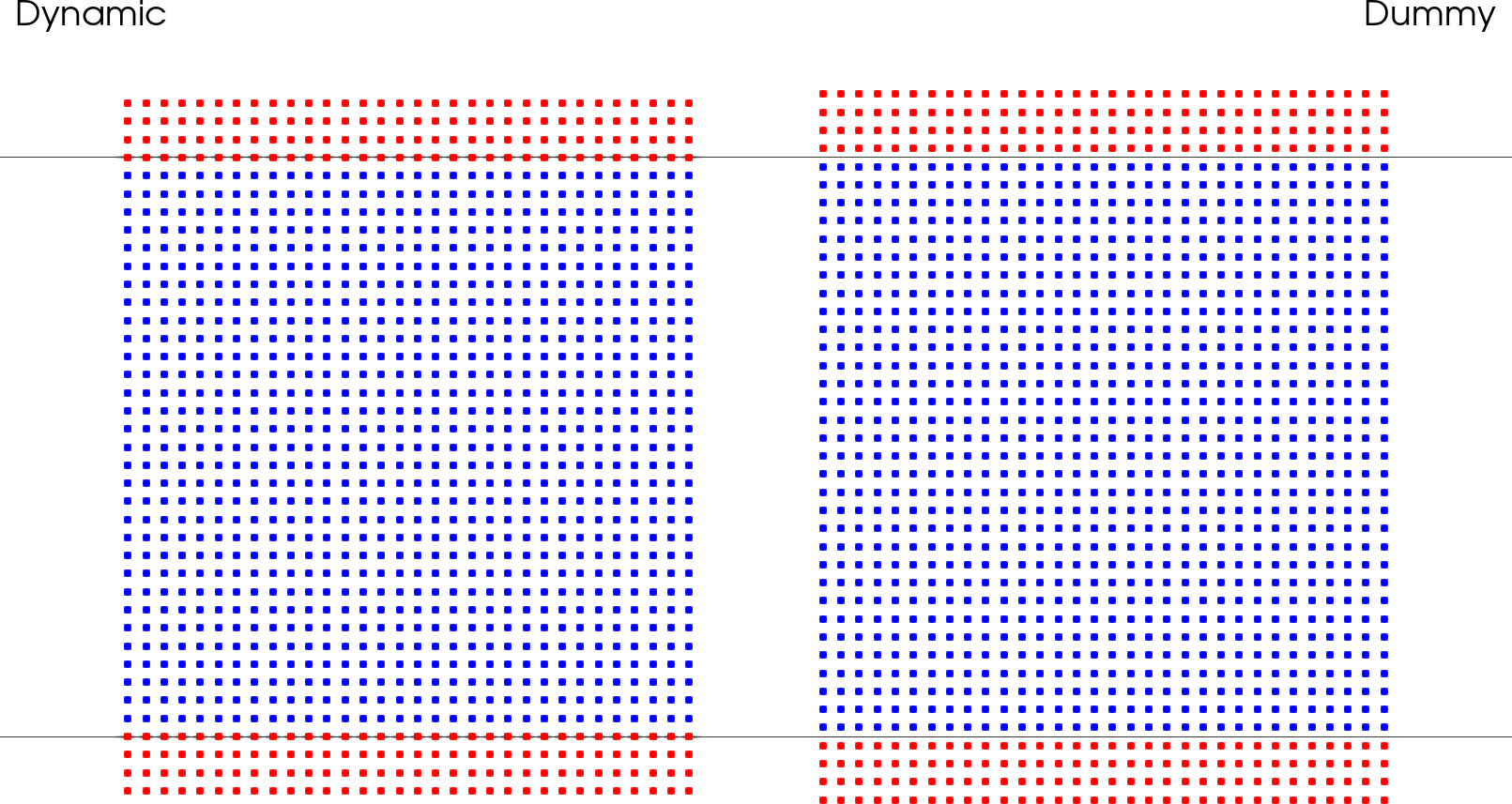}}
\caption{Cross-section of the domain discretized with the dynamic (left)
and dummy (right) boundary models when using the Wendland kernel with
smoothing length $1.3 \Delta p$. Particles are colored by type (blue
for fluid, red for boundary). The number of boundary particle layers is
determined by the kernel's influence radius, and is higher for the other
kernel choices discussed in section~\ref{sec:kernel-choice}.}
\label{fig:poiseuille-sph}
\end{figure}

When using the dynamic boundary condition, the no-slip condition $\vec u
= \vec u_w = 0$ is obtained by placing the first layer of boundary
particles at the analytical coordinates of the wall. In
contrast, with dummy boundary conditions the analytical position of the
wall is halfway between the boundary and fluid particles: the
first layer of boundary particles is thus placed $\Delta p/2$
\emph{inside} the wall (figure~\ref{fig:poiseuille-sph}).

We test with three different resolutions: $\Delta p = 1/16, 1/32, 1/64$,
with the measurement unit being meters.
Due to the different discretization, in the dynamic boundary case there
is one layer of particles at $z=0$, but not when using dummy boundaries.
A central layer of particles can be achieved with dummy boundary
conditions by using $1/17, 1/33, 1/65$ spacings, but this choice leads
to somewhat noisier simulations, and the results are not reported here.

\subsection{Smoothing kernel choice}
\label{sec:kernel-choice}

The choice of smoothing kernel in SPH has implications in terms of
numerical accuracy, stability and performance.
According to~\cite{morris1997}, the commonly used spline of
order~3 with radius~2 produces noisy results at low Reynold numbers,
and the authors of that paper prefer a computationally more intensive
spline of order~5 with radius~3.

The default kernel used in GPUSPH is based on the $\psi_{3,1}$ function
defined by Wendland~\cite{Wendland1995}, a fifth-order spline of
radius~2. We have $W(r, h) = w(r/h)/h^3$ with
\[
w(q) = \begin{cases}
\frac{21}{16\pi} \left(1 - \frac q2\right)^4(2q+1) & q \le 2\\
0 & q > 2
\end{cases},
\]
and $F(r, h) = f(r/h)/h^5$ with
\[
f(q) = \begin{cases}
\frac{105}{128\pi} (q-2)^3 & q \le 2\\
0 & q > 2
\end{cases}.
\]
With a smoothing factor of $1.3$, and thus an influence radius of
$2.6\Delta p$, each particle will have around 80 neighbors.

Another option is the normalized truncated Gaussian of radius $k=3$. We
have then
\[
w(q) = \begin{cases}
(\exp(-q^2) - \exp(-k^2))/w_G & q \le k\\
0 & q > k
\end{cases}
\]
where $w_G = \pi^{3/2} \erf(k) - 2\pi/3 \exp(-k^2) k (3 + 2 k^2)$ is the
normalization factor, and
\[
f(q) = \begin{cases}
2 \exp(-q^2)/w_G & q \le k\\
0 & q > k
\end{cases}.
\]
With the same smoothing factor used for the Wendland kernel, and thus an
influence radius of $3.9\Delta p$, each particle will have around 250
neighbors.

Due to the larger influence radius and the use of transcendental
functions, the Gaussian kernel is around 4 times more computationally
expensive than the Wendland kernel. The larger influence radius also
requires about 2 times more memory to store the neighbors list of each
particle (in the default configuration, GPUSPH will reserve memory for
up to 128 neighbors per particle). However, as we shall see, the
Gaussian kernel can give more accurate results, and there are
additional benefits to a larger influence radius in the semi-implicit
formulation.

\subsection{Newtonian results}

The expected velocity profile for a Newtonian fluid in the plane
Poiseuille flow is a parabola with maximum velocity in the middle, and
null velocity at the planes (due to the no-slip boundary condition):
\[
u(z) = \frac12 \frac{\rho_0 g}{\mu} \left(\frac{L^2}{4} - z^2\right).
\]

\paragraph{Convergence}
\label{sec:convergence}

\begin{figure}
\centering{%
\includegraphics[width=.49\textwidth]{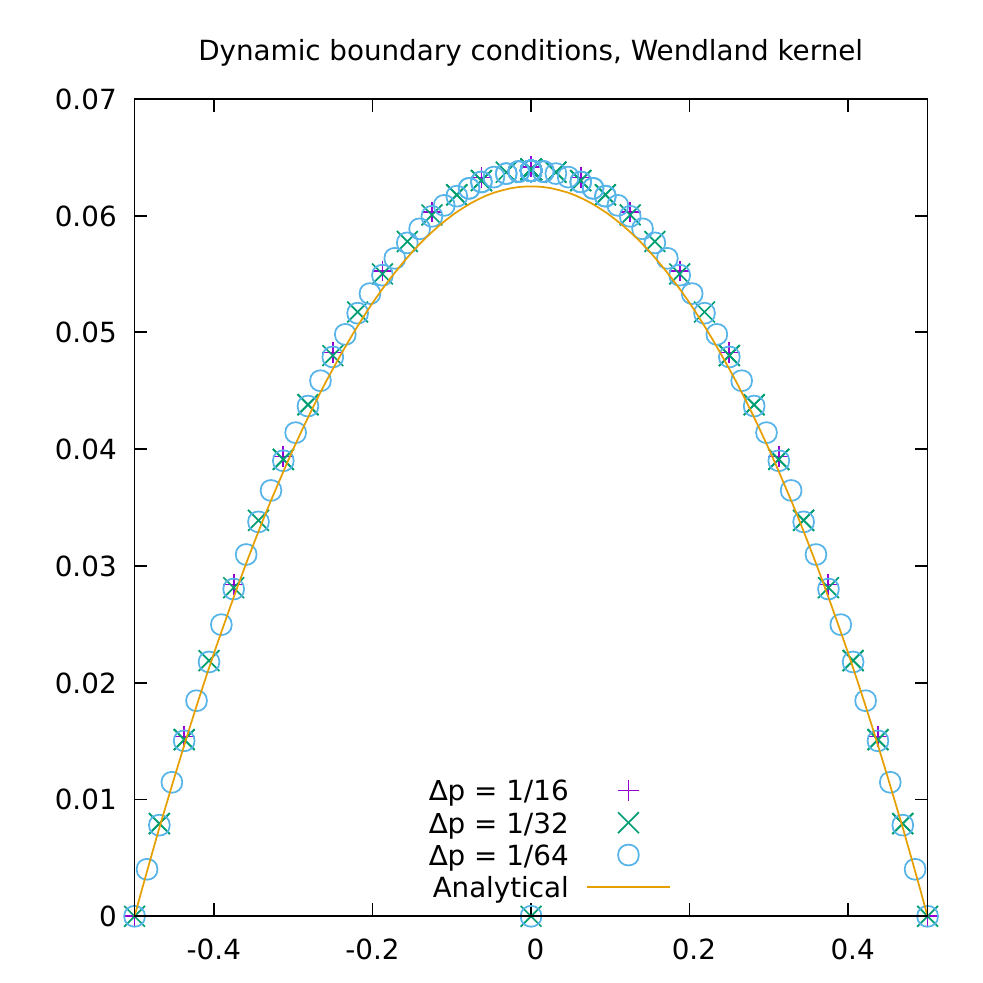}%
\hfill
\includegraphics[width=.49\textwidth]{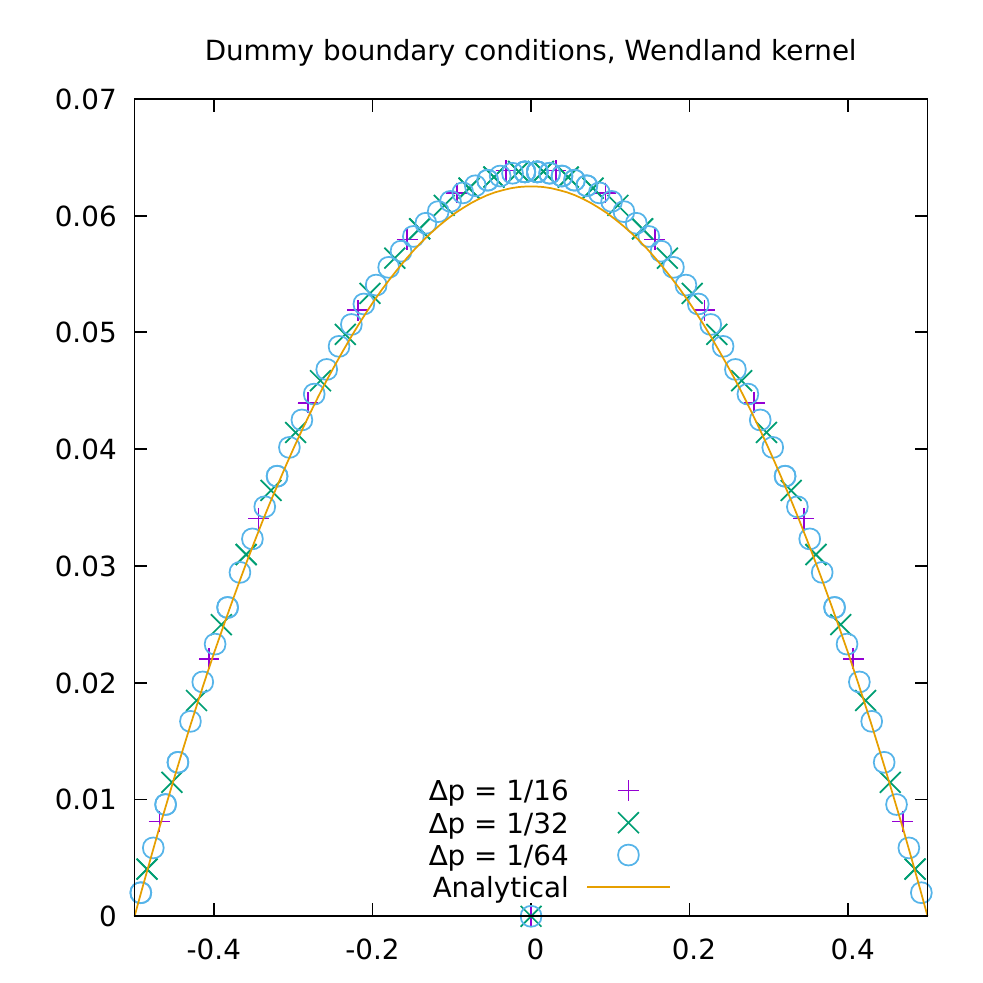}%
}
\caption{Analytical and numerical velocity profiles of the plane
Poiseuille flow for a Newtonian fluid, with dynamic (left) and dummy
boundary (right) conditions, using the Wendland smoothing kernel.}
\label{fig:newton-wendland}
\end{figure}

\begin{figure}
\centering{%
\includegraphics[width=.49\textwidth]{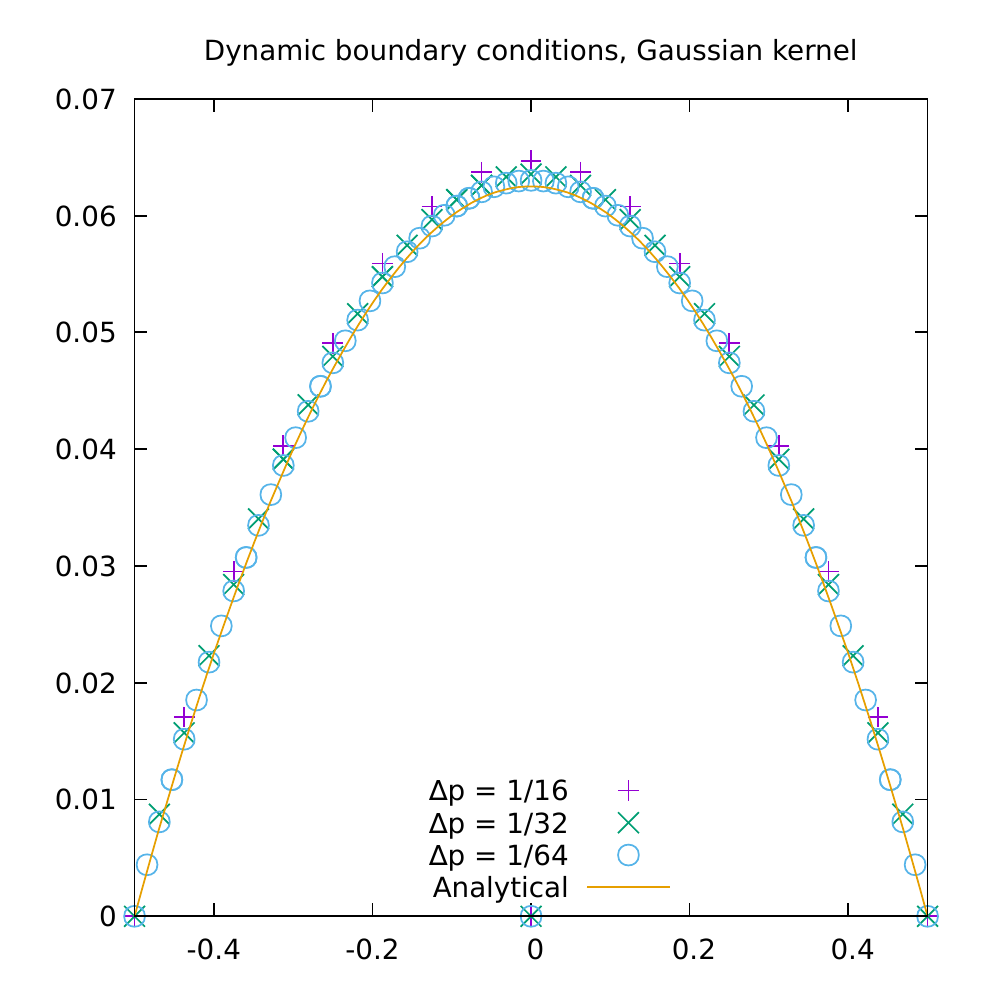}%
\hfill
\includegraphics[width=.49\textwidth]{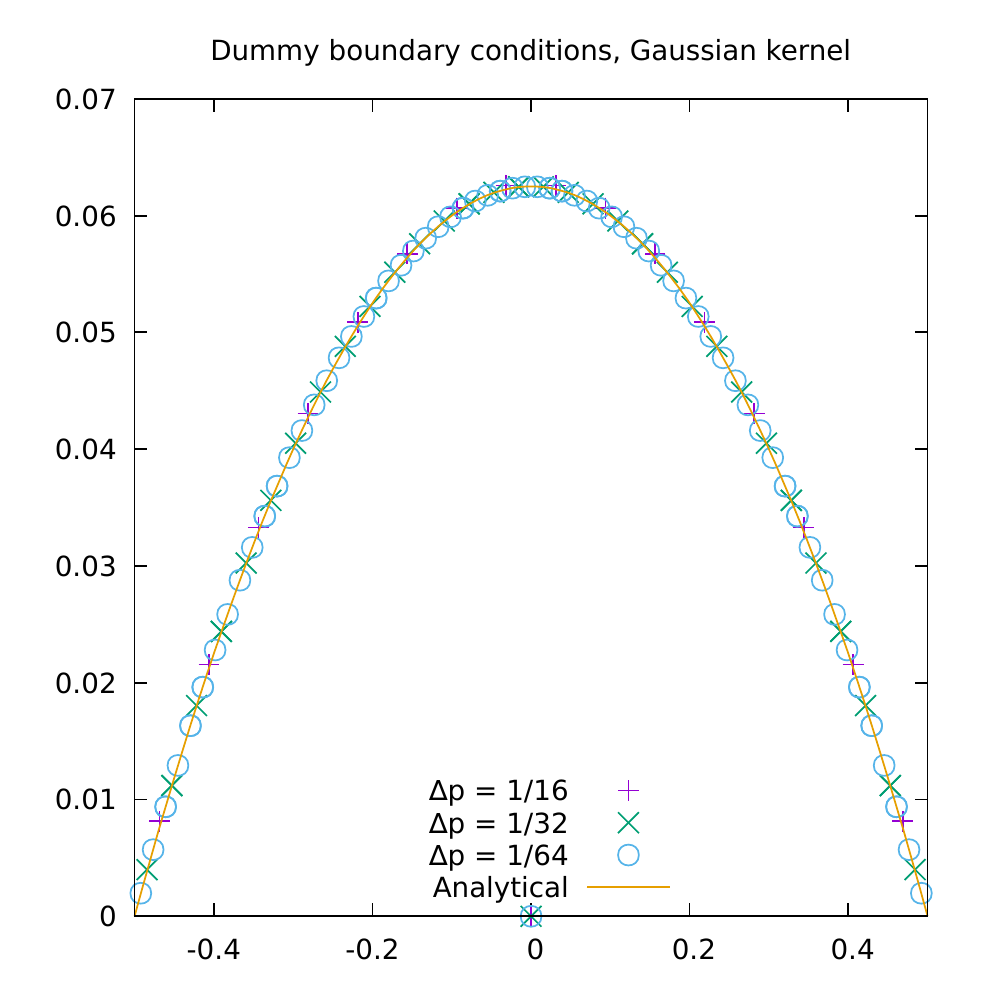}%
}
\caption{Analytical and numerical velocity profiles of the plane
Poiseuille flow for a Newtonian fluid, with dynamic (left) and dummy
(right) boundary conditions, using the Gaussian smoothing kernel.}
\label{fig:newton-gauss}
\end{figure}

In its simplicity, this test case presents some peculiarities,
that are illustrated in figures~\ref{fig:newton-wendland}
and~\ref{fig:newton-gauss}. We observe that using the Wendland kernel
(figure~\ref{fig:newton-wendland}) with a smoothing factor for $1.3$
produces results that are largely independent from the resolution;
additionally, the velocity profile, while qualitatively correct, is
consistently overestimated. The Gaussian kernel, on the other hand,
(figure~\ref{fig:newton-gauss}), shows convergence towards the correct
solution.

In both cases, we remark the lack of noise in the results: we are
plotting the results for all of the fluid particles (and, with dynamic
boundaries, the first layer of boundary particles), and there is no
spread in any direction, indicating that all particles with the same $z$
coordinate have numerically the same velocity, regardless of their $x$
or $y$ position.

The actual numerical performance with the two smoothing kernels is shown
in table~\ref{tab:newton-error}. We can remark that the dummy
boundary conditions consistently provide lower error and higher
convergence rates compared to the dynamic boundary conditions, as
expected. The Gaussian kernel also provides better results in all
respects, and in particular it shows convergence towards the analytical
solution, whereas the Wendland kernel seems to converge towards a
solution which is $2\%$ higher than the analytical solution.

\begin{table}
\centerline{%
\renewcommand{\arraystretch}{1.2}%
\setlength{\tabcolsep}{1ex}%
\begin{tabular}{r r | l l | l l | l l | l l }
Kernel & $\dfrac{1}{\Delta p}$ & \multicolumn{4}{c|}{Dynamic} & \multicolumn{4}{c}{Dummy} \\[1ex]
&&
\multicolumn{1}{c}{$l_1$} &
\multicolumn{1}{c}{rate $l_1$} &
\multicolumn{1}{c}{$l_2$} &
\multicolumn{1}{c|}{rate $l_2$} &
\multicolumn{1}{c}{$l_1$} &
\multicolumn{1}{c|}{rate $l_1$} &
\multicolumn{1}{c}{$l_2$} &
\multicolumn{1}{c}{rate $l_2$} \\
\hline
Gaussian & $16$
    & $2.083\cdot10^{-3}$ &  n.a.  & $2.153\cdot10^{-3}$ &  n.a. & $3.615\cdot10^{-4}$ &  n.a.  & $3.713\cdot10^{-4}$ &  n.a.  \\
& $32$
    & $1.033\cdot10^{-3}$ & $1.01$ & $1.050\cdot10^{-3}$ & $1.04$ & $4.081\cdot10^{-5}$ & $3.15$ & $5.087\cdot10^{-5}$ & $2.87$ \\
& $64$
    & $5.097\cdot10^{-4}$ & $1.02$ & $5.139\cdot10^{-4}$ & $1.03$ & $2.010\cdot10^{-5}$ & $1.02$ & $2.096\cdot10^{-5}$ & $1.28$ \\
\hline
Wendland & $16$
    & $1.210\cdot10^{-3}$ &  n.a.  & $1.325\cdot10^{-3}$ &  n.a. & $1.227\cdot10^{-3}$ &  n.a.  & $1.282\cdot10^{-3}$ &  n.a.  \\
& $32$
    & $1.061\cdot10^{-3}$ & $0.19$ & $1.133\cdot10^{-3}$ & $0.23$ & $9.598\cdot10^{-4}$ & $0.35$ & $1.032\cdot10^{-3}$ & $0.31$ \\
& $64$
    & $9.490\cdot10^{-4}$ & $0.16$ & $1.024\cdot10^{-3}$ & $0.15$ & $8.792\cdot10^{-4}$ & $0.13$ & $9.585\cdot10^{-4}$ & $0.11$ \\
\hline
Wendland\strut & $16$
    & $4.288\cdot10^{-4}$ &  n.a.  & $4.576\cdot10^{-4}$ &  n.a. & $3.916\cdot10^{-4}$ &  n.a.  & $3.929\cdot10^{-4}$ &  n.a.  \\
($102\% \vec u$) & $32$
    & $2.283\cdot10^{-4}$ & $0.91$ & $2.338\cdot10^{-4}$ & $0.97$ & $1.260\cdot10^{-4}$ & $1.64$ & $1.294\cdot10^{-4}$ & $1.60$ \\
& $64$
    & $1.158\cdot10^{-4}$ & $0.98$ & $1.204\cdot10^{-3}$ & $0.96$ & $4.573\cdot10^{-5}$ & $1.46$ & $5.435\cdot10^{-5}$ & $1.25$ \\
\hline
\end{tabular}%
}
\caption{$l_1$ and $l_2$ norms and convergence rate of the
plane Poiseuille flow for a Newtonian fluid. Results are for the
implicit formulation, with dynamic and dummy boundary conditions, with
both the Gaussian and Wendland kernel. The Wendland kernel is also
compared against the solution it appears to converge to.}
\label{tab:newton-error}
\end{table}

In practical applications, a $2\%$ may be considered acceptable, and as
we shall see in the next section, the actual performance of the Wendland
kernel is also problem-dependent. Still, one might be led to ask about
the reason for this discrepancy.

The difference between the Gaussian and Wendland kernel is reminiscent
of the difference between splines of order~3 and~5 reported by
Morris~\cite{morris1997} (with the latter providing better results),
but our results shed additional light on the root cause: since
our Wendland kernel is an order~5 spline (like the best one used by
Morris), but with radius~2, it would seem that the discrepancy is not to
be sought (only) in the behavior of the second derivative of the kernel,
but also in the size of the kernel support itself.

In fact, a recent result by Violeau \& Fonty~\cite{violeaufonty2019} shows 
that the smoothing error is actually proportional to the square
of the kernel's standard deviation $\sigma$, rather than the smoothing
length $h$. Since the $\sigma/h$ ratio is kernel-dependent, the choice
of smoothing factor should be kernel-dependent to attain similar
results, whereas in our case we are using a fixed one for both
kernel choices.

Considering that $(\sigma/h)^2 = 4/15$ for the Wendland
kernel~\cite{violeaufonty2019}, and $(\sigma/h)^2 \approx 1/2$ for the
(truncated) Gaussian kernel, we can expect that using a smoothing factor
about $15/8$ times larger than the one used for the Gaussian kernel
would lead to comparable convergence ratios when using the Wendland
kernel. This is confirmed by running the Wendland case with a smoothing
factor of $2.5$ rather than the usual $1.3$
(table~\ref{tab:newton-error-wendland-25}).

\begin{table}
\centerline{%
\renewcommand{\arraystretch}{1.2}%
\setlength{\tabcolsep}{1ex}%
\begin{tabular}{r | l l | l l | l l | l l }
& \multicolumn{4}{c|}{Dynamic} & \multicolumn{4}{c}{Dummy} \\[1ex]
$1/\Delta p$ \strut&
\multicolumn{1}{c}{$l_1$} &
\multicolumn{1}{c}{rate $l_1$} &
\multicolumn{1}{c}{$l_2$} &
\multicolumn{1}{c|}{rate $l_2$} &
\multicolumn{1}{c}{$l_1$} &
\multicolumn{1}{c|}{rate $l_1$} &
\multicolumn{1}{c}{$l_2$} &
\multicolumn{1}{c}{rate $l_2$} \\
\hline
$16$ & $5.436\cdot10^{-3}$ &  n.a.  & $5.625\cdot10^{-3}$ &  n.a.  & $6.648\cdot10^{-4}$ &  n.a.  & $7.797\cdot10^{-4}$ &  n.a.  \\
$32$ & $2.588\cdot10^{-3}$ & $1.07$ & $2.673\cdot10^{-3}$ & $1.07$ & $5.095\cdot10^{-5}$ & $3.71$ & $1.502\cdot10^{-4}$ & $2.38$ \\
$64$ & $1.300\cdot10^{-3}$ & $0.99$ & $1.322\cdot10^{-3}$ & $1.02$ & $5.941\cdot10^{-5}$ & $-0.22$ & $6.696\cdot10^{-5}$ & $1.17$ \\
\hline
\end{tabular}%
}
\caption{$l_1$ and $l_2$ norms and ratios for growing resolutions of the
plane Poiseuille flow for a Newtonian fluid with dynamic and dummy
boundary conditions, using the Wendland kernel with smoothing factor $2.5$.}
\label{tab:newton-error-wendland-25}
\end{table}

Indeed, in this case the Wendland kernel exhibits convergence properties
similar to those seen for the Gaussian kernel with smoothing factor
$1.3$, at least up to numerical limits (seen in
table~\ref{tab:newton-error-wendland-25} for the dummy boundary case in
$l_1$ norm at $\Delta p = 1/64$). We remark also that the average number of
neighbors for Wendland\slash$2.5$ is also about twice as large as the
number of neighbors in the Gaussian\slash$1.3$ case, which is what we
expect given the different values of $\sigma/h$ for the two kernels.

\paragraph{Computational performance}

Due to the simplicity of the geometry in this test case, the results are
essentially the same for both the implicit and explicit
formulation. Moreover, the implicit formulation is actually
\emph{slower} than the explicit formulation, since the viscosity of the
fluid is not high enough, and the dominant time-step limitation comes
from the sound speed at most resolutions (table~\ref{tab:newton-times}).

\begin{table}
\centerline{%
\renewcommand{\arraystretch}{1.2}%
\setlength{\tabcolsep}{1ex}%
\begin{tabular}{r r | l l | l l | l l }
&&&& \multicolumn{4}{c}{Runtime}\\
Kernel & $\dfrac{1}{\Delta p}$ &&& \multicolumn{2}{c|}{Dynamic} & \multicolumn{2}{c}{Dummy}\\
&&
\multicolumn{1}{c}{$\Delta t_c$} &
\multicolumn{1}{c|}{$\Delta t_\nu$} & Explicit & Implicit & Explicit & Implicit \\
\hline
Gaussian & $16$
    & \cellcolor{lightgray}$3.85\cdot10^{-3}$ & $8.25\cdot10^{-3}$ & $7.7\cdot10^0$ & $4.6\cdot10^1$ & $7.8\cdot10^0$ & $6.8\cdot10^1$ \\
& $32$
    & \cellcolor{lightgray}$1.93\cdot10^{-3}$ & $2.06\cdot10^{-3}$ & $4.5\cdot10^1$ & $4.3\cdot10^2$ & $4.9\cdot10^1$ & $7.4\cdot10^2$ \\
& $64$
    & $9.63\cdot10^{-4}$ & \cellcolor{lightgray}$5.16\cdot10^{-4}$ & $1.3\cdot10^3$ & $9.7\cdot10^3$ & $1.4\cdot10^3$ & $1.3\cdot10^4$ \\
\hline
Wendland & $16$
    & \cellcolor{lightgray}$3.85\cdot10^{-3}$ & $8.25\cdot10^{-3}$ & $3.3\cdot10^0$ & $2.4\cdot10^1$ & $3.6\cdot10^0$ & $3.3\cdot10^1$ \\
& $32$
    & \cellcolor{lightgray}$1.93\cdot10^{-3}$ & $2.06\cdot10^{-3}$ & $2.0\cdot10^1$ & $2.5\cdot10^2$ & $2.1\cdot10^1$ & $3.7\cdot10^2$ \\
& $64$
    & $9.63\cdot10^{-4}$ & \cellcolor{lightgray}$5.16\cdot10^{-4}$ & $4.8\cdot10^2$ & $4.3\cdot10^3$ & $5.0\cdot10^2$ & $5.6\cdot10^3$ \\
\hline
%Wendland/2.5 & $16$
%    & \cellcolor{lightgray}$7.41\cdot10^{-3}$ & $3.05\cdot10^{-2}$ & $7.3\cdot10^0$ & $4.7\cdot10^1$ & $7.6\cdot10^0$ & $7.0\cdot10^1$ \\
%& $32$
%    & \cellcolor{lightgray}$3.71\cdot10^{-3}$ & $7.63\cdot10^{-3}$ & $4.4\cdot10^1$ & $5.3\cdot10^2$ & $4.8\cdot10^1$ & $9.5\cdot10^2$ \\
%& $64$
%    & \cellcolor{lightgray}$1.85\cdot10^{-3}$ & $1.91\cdot10^{-3}$ & $8.9\cdot10^2$ & $1.3\cdot10^4$ & $9.2\cdot10^2$ & $1.8\cdot10^4$ \\
%\hline
\end{tabular}%
}
\caption{Time-step limitation from speed of sound ($\Delta t_c$) and
viscosity ($\Delta t_\nu$), and effective simulation runtimes in seconds
for both boundary models with the explicit and semi-implicit formulations
in the Newtonian case. The kernel choice affects the runtime, but not
the time-step.}
\label{tab:newton-times}
\end{table}

Even at the highest resolution, where the viscous time-step limitation
becomes dominant, the benefit of being limited only by the sound speed in
the implicit case is not sufficient to offset the higher computational
cost of the semi-implicit formulation. However, it is interesting to
remark that, while the runtimes with the Gaussian kernel are
consistently more than twice longer than with the Wendland kernel in the 
explicit case, for the semi-implicit formulation the ratio is lower.

Both of these phenomena are explained by the difference between the
number of iterations required to solve the implicit system, and the
time-step choice.

Indeed, a full semi-implicit step at the highest resolution, requires on
average 3 (dynamic boundary) to 5 (dummy boundary) BiCGSTAB iterations
with the Wendland kernel, but only 1 (dynamic boundary) to 3 (dummy
boundary) iterations with the Gaussian kernels. Because of this, the
performance loss of the Gaussian kernel computation is partially
compensated by the faster convergence of the solver.

However, a full semi-implicit step is between $10\times$ (dynamic,
Gaussian) and $15.1\times$ (dummy, Gaussian) longer than the time needed
for a full explicit step. By contrast, adopting the semi-implicit
formulation leads to a reduction in the number of time-steps (thanks to
the removal of the viscous time-step constraint) which is much more
modest ($1.88\times$).

The performance difference between the explicit and semi-implicit scheme
is in line with the observations made in~\cite{zago_jcp_2018}, and
suggests that the turning point for the advantage of the semi-implicit
formulation would be a time-step constraint ratio of $20\times$ or more.
This could be achieved with an even higher resolution, but is better
illustrated by the non-Newtonian test-case presented in the
next section.

We remark that the implicit solver itself converges pretty quickly, but
this is only when implemented as described in
section~\ref{sec:new-bicgstab}. The standard BiCGSTAB implementation, in
single precision, is affected by frequent stalls, leading to much
noisier results, or even failing to converge altogether even at the
moderate resolution.

Note that for the dynamic boundary case we can also use the CG to solve
the linear system: tested with the Wendland kernel, runtimes are around
$10\%$ faster than the BiCGSTAB solver, despite the larger number of
iterations needed for convergence (4 instead of 1 at the lowest
resolution, 7 instead of 3 at the highest). The performance benefit of
the CG is thus not sufficient to make the semi-implicit formulation
convenient over the explicit scheme at these resolutions.

\subsection{Papanastasiou case}
\label{sec:results-papa}

\begin{figure}
\includegraphics{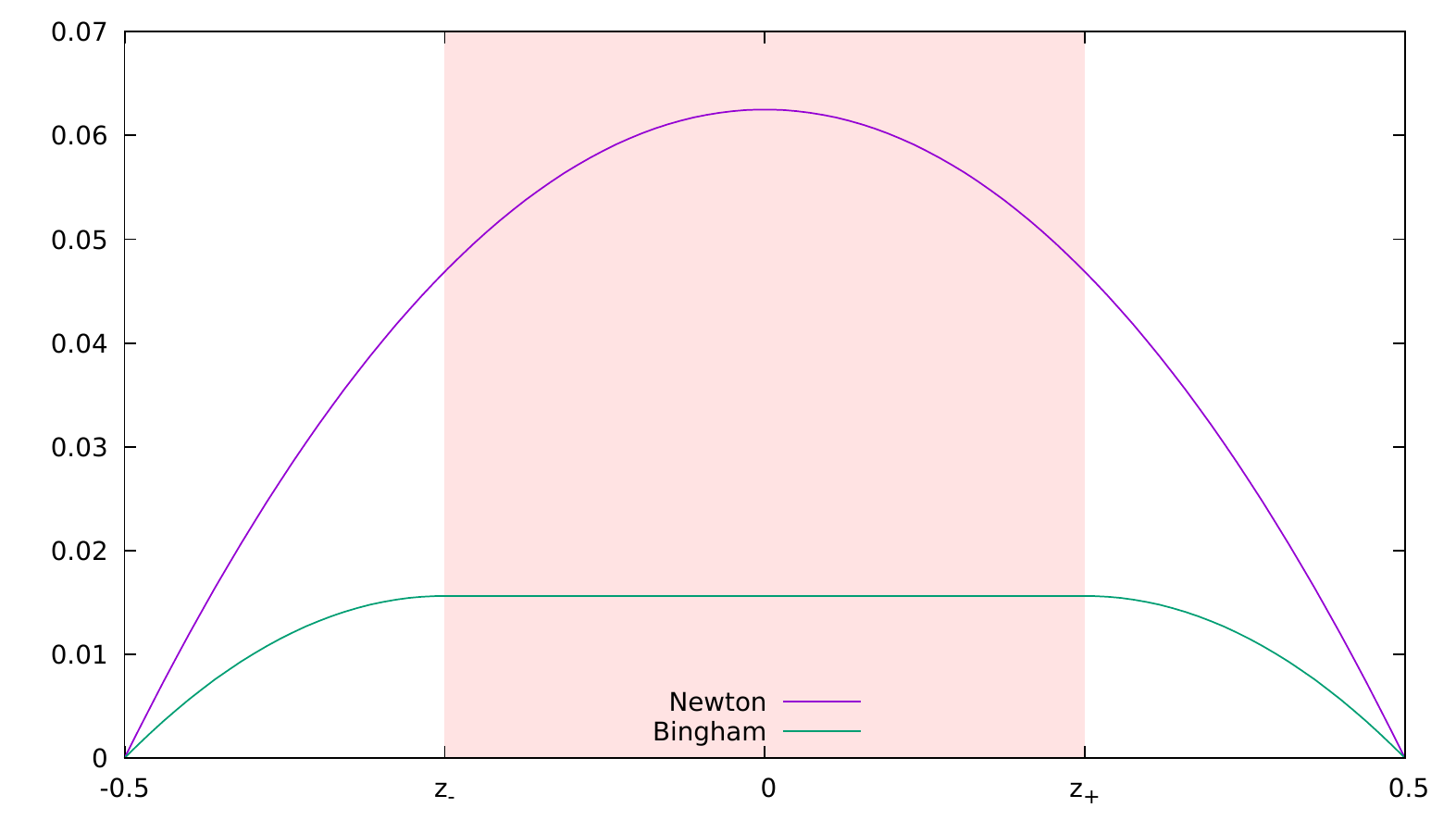}
\caption{Velocity profile for the plane Poiseuille flow of a Bingham
fluid, highlighting the presence of a plug (unyielded region) and its
shallowing effect in comparison to a Newtonian fluid.}
\label{fig:poiseuille-plug}
\end{figure}

For the Papanastasiou rheology, we compare the numerical results
against the analytical
solution for the velocity profile of a non-regularized Bingham fluid.
The flow is divided into a yielded $\invariant{\tau} \ge \tau_0$ and
unyielded $\invariant{\tau} < \tau_0$ regions
(figure~\ref{fig:poiseuille-plug}).
The unyielded region (\emph{plug}) spans the area $z_{-} < z < z_{+}$,
with $z_{+} = \tau_0/(\rho_0 g)$ and $z_{-} = -z_{+}$.

In the yielded region, the velocity profile is parabolic, with:
\[
u(z) = \frac12 \frac{\rho_0 g}{\mu_0} \left(\frac{L^2}{4} - z^2\right)
 - \frac{\tau_0}{\mu_0} \left(\frac{L}{2} - \abs{z}\right),
\]
whereas the plug moves at the \emph{plug velocity}
\[
u_p = u(z_{-}) = u(z_{+}) =
\left(\frac{L}{2} - \frac{\tau_0}{\rho_0 g}\right)
\frac{\rho_0 g L/2 - \tau_0}{2 \mu_0}.
\]

\paragraph{Convergence}

\begin{figure}
\centering{%
\includegraphics[width=.49\textwidth]{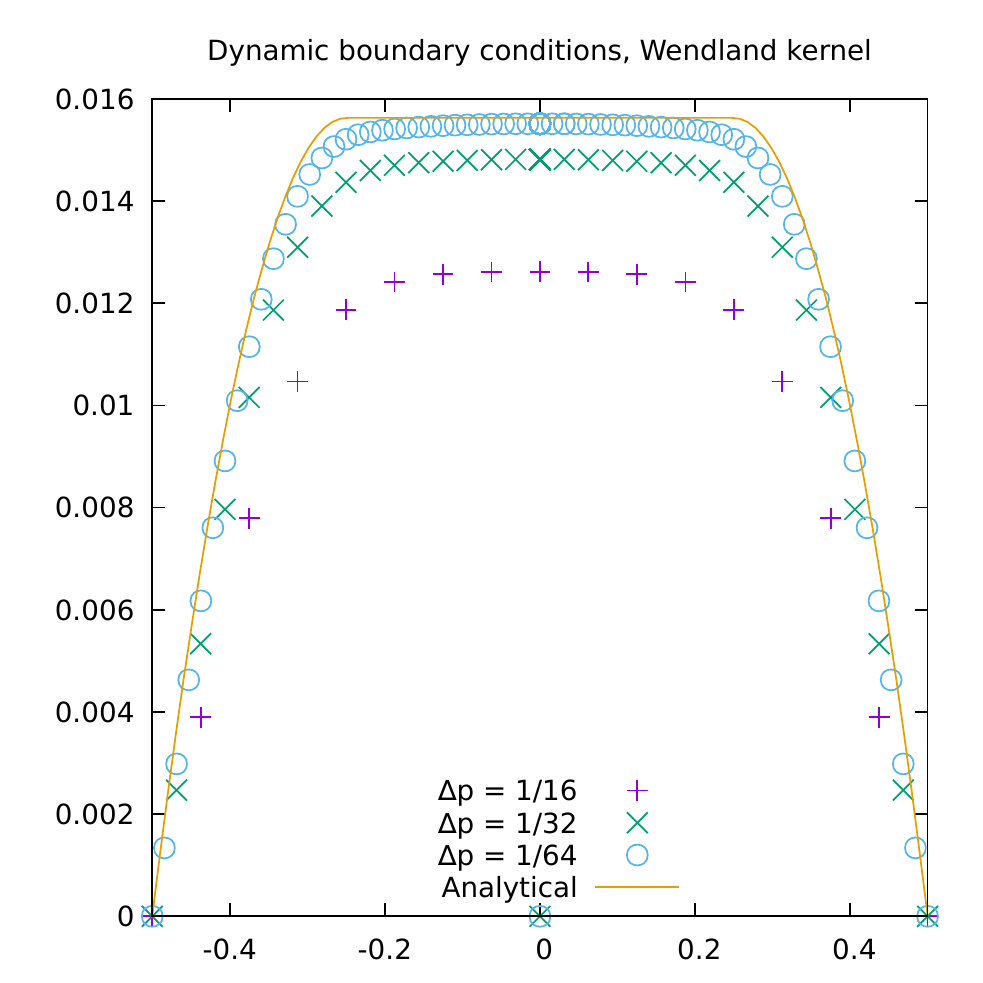}%
\hfill
\includegraphics[width=.49\textwidth]{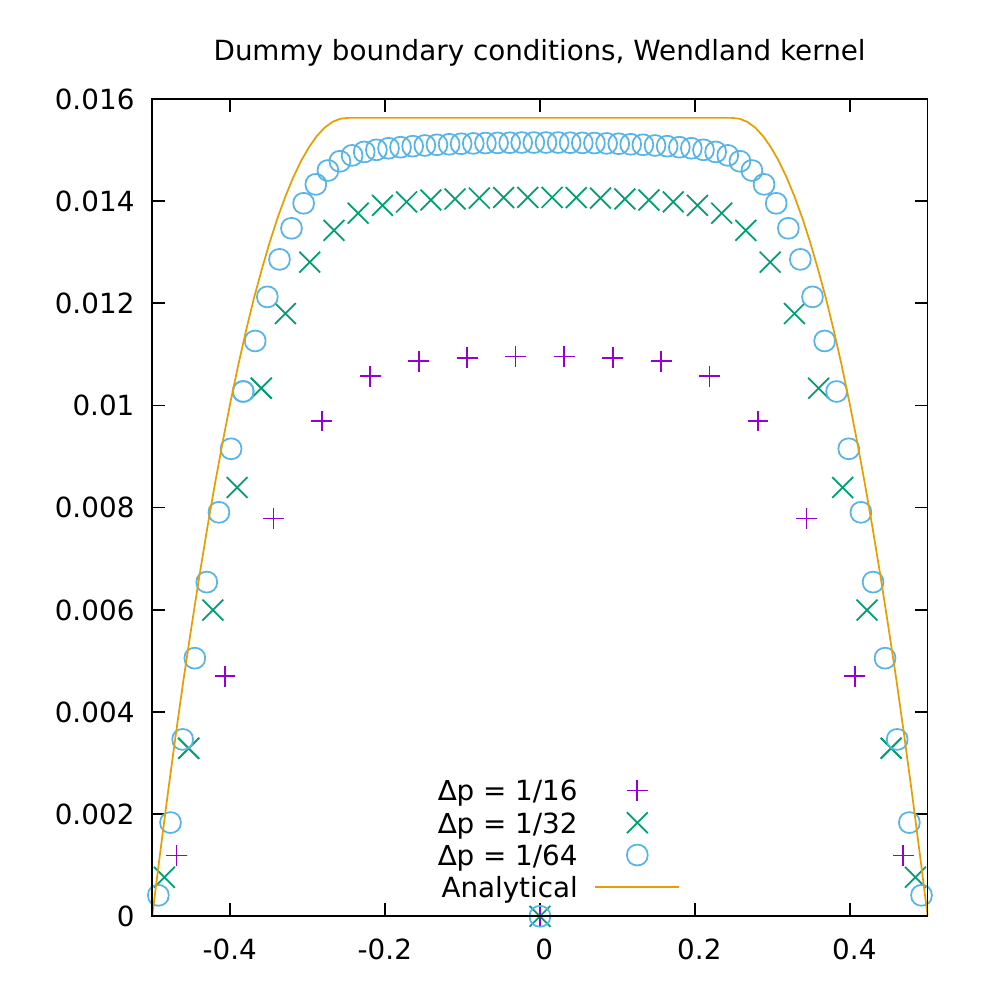}%
}
\caption{Analytical and numerical velocity profiles of the plane
Poiseuille flow for a Papanastasiou fluid, with dynamic (left) and dummy
(right) boundary conditions, using the Wendland smoothing kernel.}
\label{fig:papa-wendland}
\end{figure}

\begin{figure}
\centering{%
\includegraphics[width=.49\textwidth]{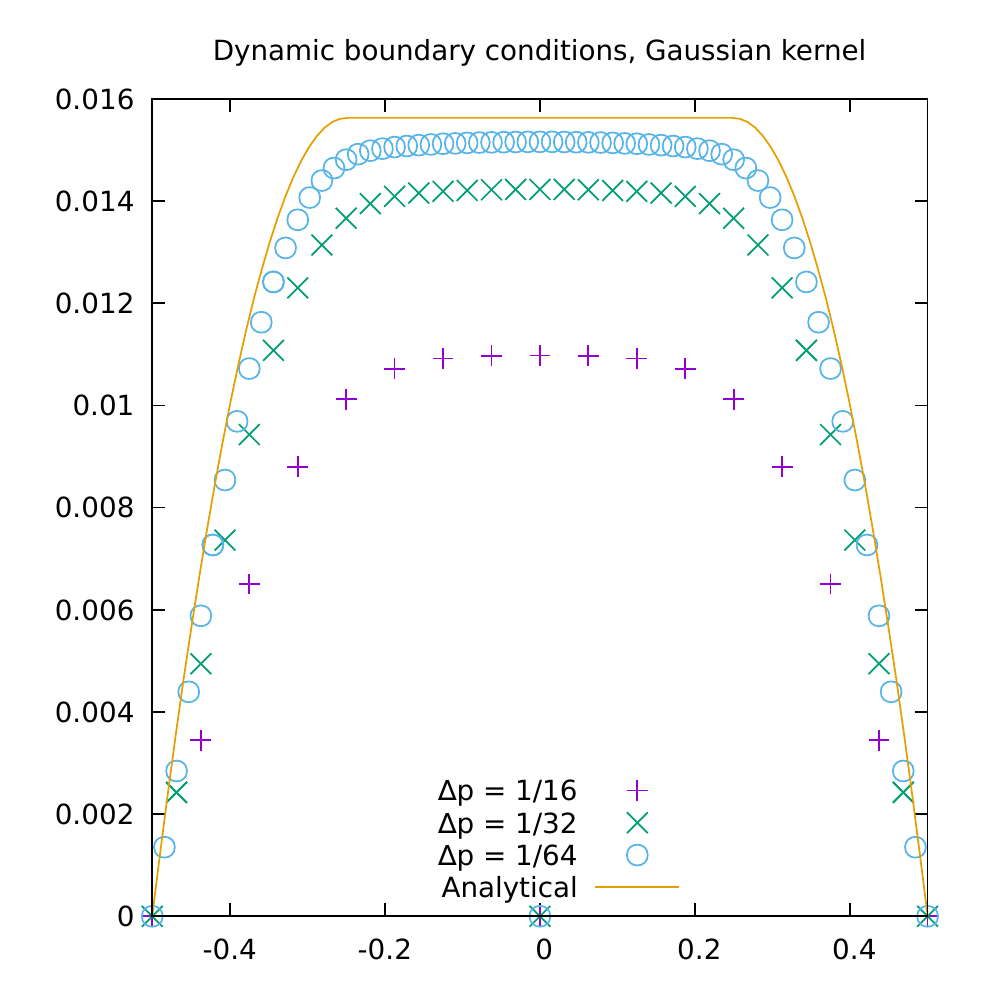}%
\hfill
\includegraphics[width=.49\textwidth]{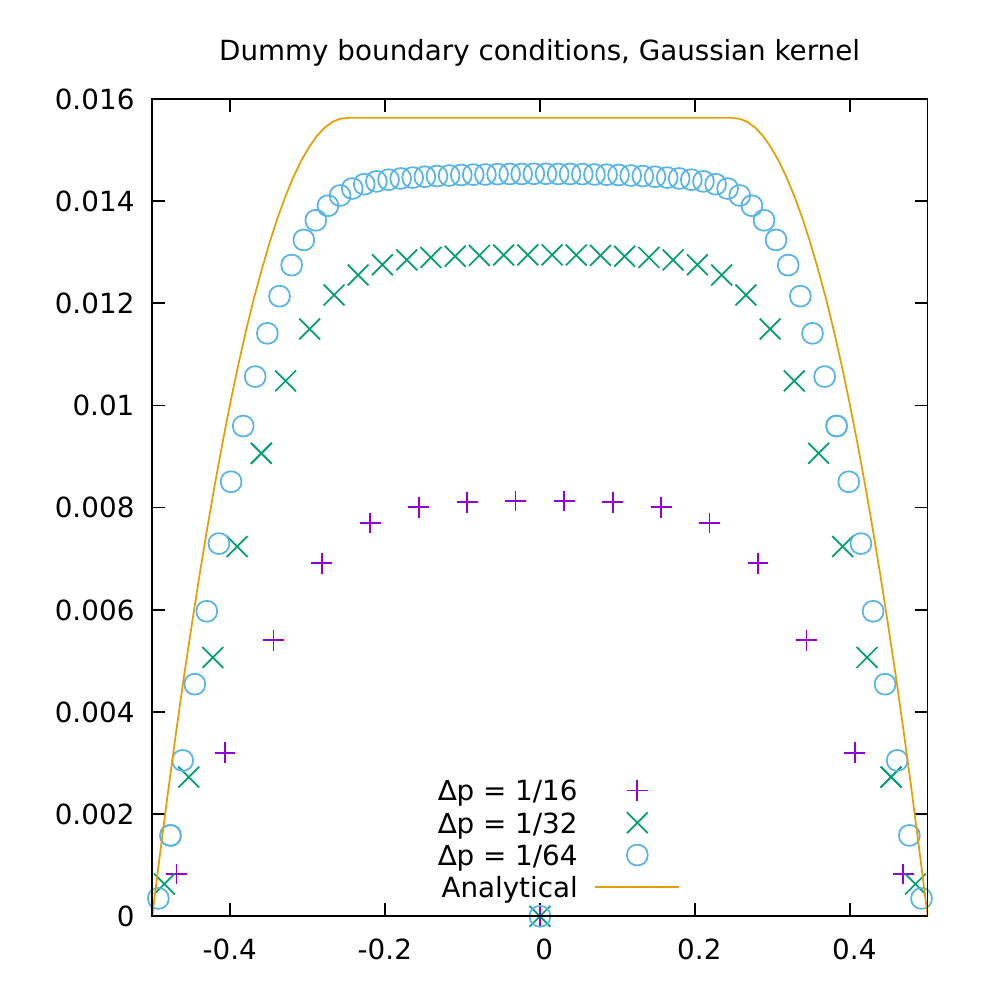}%
}
\caption{Analytical and numerical velocity profiles of the plane
Poiseuille flow for a Papanastasiou fluid, with dynamic (left) and dummy
(right) boundary conditions, using the Gaussian smoothing kernel.}
\label{fig:papa-gauss}
\end{figure}

In this test-case, the Wendland kernel with the standard smoothing
factor of $1.3$ shows good convergence to the
analytical velocity profile (figure~\ref{fig:papa-wendland}), even
better than the Gaussian kernel (figure~\ref{fig:papa-gauss}).

Again, as in the Newtonian case, the results are very clean (no
discernible discrepancy in the velocity for particles at the same $z$
coordinate). In contrast to the Newtonian case, however, the dummy
boundary model performs slightly worse than the dynamic boundaries.
While this can be at least partially ascribed to the lower effective
particle spacing used in the setup with the dummy boundary compared to
the dynamic boundary case, the results suggest that further research is
necessary to improve the behavior of the dummy boundary formulation in the
case of shear-dependent effective viscosity.

Details about the discrete error norms and convergence rates are shown
in table~\ref{tab:papa-error}. An important point to remark is that
these results include both the discretization error from the numerical
method and the discrepancy between the analytical solution for a Bingham
rheology versus the solution for its Papanastasiou regularization.
Despite the influence of the latter, however, the results obtained are
still very good.

\begin{table}
\centerline{%
\renewcommand{\arraystretch}{1.2}%
\setlength{\tabcolsep}{1ex}%
\begin{tabular}{r r | l l | l l | l l | l l }
Kernel & $\dfrac{1}{\Delta p}$ & \multicolumn{4}{c|}{Dynamic} & \multicolumn{4}{c}{Dummy} \\[1ex]
&&
\multicolumn{1}{c}{$l_1$} &
\multicolumn{1}{c}{rate $l_1$} &
\multicolumn{1}{c}{$l_2$} &
\multicolumn{1}{c|}{rate $l_2$} &
\multicolumn{1}{c}{$l_1$} &
\multicolumn{1}{c|}{rate $l_1$} &
\multicolumn{1}{c}{$l_2$} &
\multicolumn{1}{c}{rate $l_2$} \\
\hline
Gaussian & $16$
    & $4.569\cdot10^{-3}$ &  n.a.  & $4.769\cdot10^{-3}$ &  n.a.  & $7.027\cdot10^{-3}$ &  n.a.  & $7.227\cdot10^{-3}$ &  n.a.  \\
& $32$
    & $1.720\cdot10^{-3}$ & $1.41$ & $1.790\cdot10^{-3}$ & $1.41$ & $2.928\cdot10^{-3}$ & $1.26$ & $2.982\cdot10^{-3}$ & $1.27$ \\
& $64$
    & $7.245\cdot10^{-4}$ & $1.25$ & $7.635\cdot10^{-4}$ & $1.23$ & $1.336\cdot10^{-3}$ & $1.13$ & $1.360\cdot10^{-3}$ & $1.13$ \\
\hline
Wendland & $16$
    & $3.009\cdot10^{-3}$ &  n.a.  & $3.232\cdot10^{-3}$ &  n.a. & $4.724\cdot10^{-3}$ &  n.a.  & $4.815\cdot10^{-3}$ &  n.a.  \\
& $32$
    & $1.138\cdot10^{-3}$ & $1.58$ & $1.196\cdot10^{-3}$ & $1.43$ & $1.861\cdot10^{-3}$ & $1.34$ & $1.894\cdot10^{-3}$ & $1.34$ \\
& $64$
    & $3.715\cdot10^{-4}$ & $1.61$ & $6.833\cdot10^{-4}$ & $1.47$ & $7.339\cdot10^{-4}$ & $1.34$ & $7.631\cdot10^{-4}$ & $1.31$ \\
\hline
\end{tabular}%
}
\caption{$l_1$ and $l_2$ norms and ratios for growing resolutions of the
plane Poiseuille flow for a Papanastasiou fluid with dynamic and dummy
boundary conditions.}
\label{tab:papa-error}
\end{table}

\paragraph{Computational performance}

When simulating a non-Newtonian fluid, or more in general a fluid with
non-constant viscosity, the upper bound on the time-step controlled
by the viscosity should change according to the highest effective
viscosity encountered during the simulation. However, in this case the
presence of the plug (where the viscosity is always the maximum possible
value) effectively enforces the same time-step throughout the whole
simulation. Moreover, due to the (finite but) very high viscosity of the
plug, the viscous condition on the time-step is always dominant when
using the explicit integration scheme (table~\ref{tab:papa-times}).

\begin{table}
\centerline{%
\renewcommand{\arraystretch}{1.2}%
\setlength{\tabcolsep}{1ex}%
\begin{tabular}{r r | l l | l l | l l }
&&&& \multicolumn{4}{c}{Runtime}\\
Kernel & $\dfrac{1}{\Delta p}$ &&& \multicolumn{2}{c|}{Dynamic} & \multicolumn{2}{c}{Dummy}\\
&&
\multicolumn{1}{c}{$\Delta t_c$} &
\multicolumn{1}{c|}{$\Delta t_\nu$} & Explicit & Implicit & Explicit & Implicit \\
\hline
Gaussian & $16$
    & $3.85\cdot10^{-3}$ & \cellcolor{lightgray}$6.55\cdot10^{-5}$ & $7.1\cdot10^2$ & $1.3\cdot10^2$ & $7.3\cdot10^2$ & $1.8\cdot10^2$ \\
& $32$
    & $1.93\cdot10^{-3}$ & \cellcolor{lightgray}$1.64\cdot10^{-5}$ & $1.0\cdot10^4$ & $1.7\cdot10^3$ & $1.2\cdot10^4$ & $2.6\cdot10^3$ \\
& $64$
    & $9.63\cdot10^{-4}$ & \cellcolor{lightgray}$4.09\cdot10^{-6}$ & $3.6\cdot10^5$ & $4.3\cdot10^4$ & $3.9\cdot10^5$ & $6.1\cdot10^4$ \\
\hline
Wendland & $16$
    & $3.85\cdot10^{-3}$ & \cellcolor{lightgray}$6.55\cdot10^{-5}$ & $2.7\cdot10^2$ & $7.4\cdot10^1$ & $3.1\cdot10^2$ & $9.5\cdot10^1$ \\
& $32$
    & $1.93\cdot10^{-3}$ & \cellcolor{lightgray}$1.64\cdot10^{-5}$ & $4.3\cdot10^3$ & $1.0\cdot10^3$ & $4.9\cdot10^3$ & $1.4\cdot10^3$ \\
& $64$
    & $9.63\cdot10^{-4}$ & \cellcolor{lightgray}$4.09\cdot10^{-6}$ & $1.2\cdot10^5$ & $2.0\cdot10^4$ & $1.3\cdot10^5$ & $2.7\cdot10^4$ \\
\hline
\end{tabular}%
}
\caption{Time-step limitation from speed of sound ($\Delta t_c$) and
viscosity ($\Delta t_\nu$), and effective simulation runtimes in seconds
for both boundary models with the explicit and semi-implicit formulations
in the Papanastasiou case.}
\label{tab:papa-times}
\end{table}

The number of BiCGSTAB iterations needed to converge is also higher in
this case; as remarked also in~\cite{zago_jcp_2018}, this is most likely
due to the diagonal dominance of the matrix (for the rows that
are SDD) diminishing as the viscosity grows higher. With the Wendland
kernel, the implicit solver converges in 12 iterations on
average at the lowest resolution with the dynamic boundary model, (14
for the dummy boundary model), in 18 iterations on average at the
moderate resolution (21 for the dummy boundary model), and in 25
iterations at the highest resolution (32 for the dummy boundary model).

At the highest resolution, with both boundary formulations, the BiCGSTAB
solver is sometimes affected by early returns due to stalls. However,
the only component that is affected is the transverse direction $y$,
where the velocity component should be zero analytically. Numerically,
the order of the $y$ component of the velocities is $10^{-8}$, which is
negligible, but still manages to affect the solver. This is the only
case where even our restructured BiCGSTAB
(section~\ref{sec:new-bicgstab}) is affected by the low numerical
precision.

Overall, the semi-implicit formulation in this case is 3 to 5 times
faster than the explicit scheme, when using the Wendland kernel. As
in the Newtonian test case, the break-even point can be computed
comparing the ratio between the viscous and sound-speed time-step
limitation with the runtime for a single time-step, which in the
semi-implicit case is proportional to the number of iterations needed by
the implicit solver.

While in the Newtonian case the break-even was computed to be around
$20\times$ with an average of 5 implicit solver iterations, with the
slower convergence in the Papanastasiou case the break-even is expected
to be 2 to 6 times higher (thus $40\times$ to $120\times$) depending on
resolution, while we have $\Delta t_c/\Delta t_\nu$ ranging from $170$
(low resolution) to $425$ (high resolution). The ratio of ratios matches
up with the effective speed-up achieved.

We can also observe again that the computationally more expensive
Gaussian kernel benefits more from the semi-implicit scheme, thanks to
the faster convergence (6\slash 11\slash 16 iterations on average  at
the low\slash middle\slash high resolution with the dynamic boundary
conditions, and 9\slash 14\slash 21 with the dummy boundary conditions.)
This leads to simulations that are over~8 times faster with the
semi-implicit scheme than the explicit formulation, at the highest
resolution tested.

\subsection{Cross-flow velocities}

Even though the Poiseuille flow is essentially uni-directional,
we see particles developing a non-zero velocity in the cross-flow directions
$y$ (periodic) and $z$ (wall-to-wall) directions, reaching at most in the order of
$10^{-8}\,\text{m}\,\text{s}^{-1}$ and
$10^{-5}\,\text{m}\,\text{s}^{-1}$ respectively by the end of the runs.
These are independent of the integration scheme, and are related instead to the
nature of the adopted formulation for the viscous term, that does not
strictly conserve angular momentum \citep{huAdams-angular}.

As shown in the results, the development of these spurious velocities does not
affect the numerical results in the considered timeframe
(although they should be addressed for longer simulations). However,
they do influence the convergence speed of the linear solver when adopting the
semi-implicit scheme, due to the resulting vanishing but non-zero tolerance
value in the linear system for the corresponding coordinates.

The effect is particularly strong in the periodic direction $y$, where the
velocities (and the corresponding thresholds for convergence) are significantly
smaller, often bordering the minimum representable single-precision floating-point
value \fltmin.

Consider as an example the Papanastasiou case with dummy
boundary conditions and $\Delta p = 1/64$: during the corrector step at iteration 390
the thresholds for $x$, $y$ and $z$ are respectively $2.31\cdot 10^{-16}$, $9.37\cdot 10^{-38}$
and $2.48\cdot10^{-24}$. After 42 solver steps, the algorithm bails out
due to a stall in the convergence of the $y$ component of the velocity,
with residual norms $2.89\cdot10^{-22}$, $3.70\cdot10^{-37}$ (failed to converge)
and $2.15\cdot 10^{-27}$ respectively. The user is notified of the failure,
and the simulation continues. In this case, this is an acceptable loss of
accuracy, and we decide to keep the results.

While these extreme situations are an excellent test-bed for the robustness
of our improved CG and BiCGSTAB solvers, in practical applications they
may lead to undesired increases in runtimes due to the slower convergence.
Heuristics (either automatic or user-guided) to determine negligible
components for the stopping criterion are currently being explored as a
possible solution.

\section{Test case: flow in a periodic lattice of cylinders}
\label{sec:lattice}

Since the plane Poiseuille flow is effectively a simple unidirectional shearing flow,
it doesn't fully capture the effect of particle disorder on the solver robustness
within the timeframe needed to reach a steady-state flow:
perturbations are small and the effect is mostly seen in the loss of convergence speed
in the cross-flow directions as pointed out at the end of the previous section.

To illustrate the semi-implicit solver performance in a more complex test case,
including the effect of a distorted (non-uniform) particle distributions,
we show here the results in the case of a flow in a periodic lattice of cylinders.
Following~\cite{morris1997}, two variants of the test case are presented,
one with $Re=1$ (which we refer to as Low Reynolds Number or LRN case),
and one with $Re=0.03$ (Very Low Reynolds Number or VLRN case).

\begin{figure}
\centerline{%
\includegraphics[width=.5\textwidth]{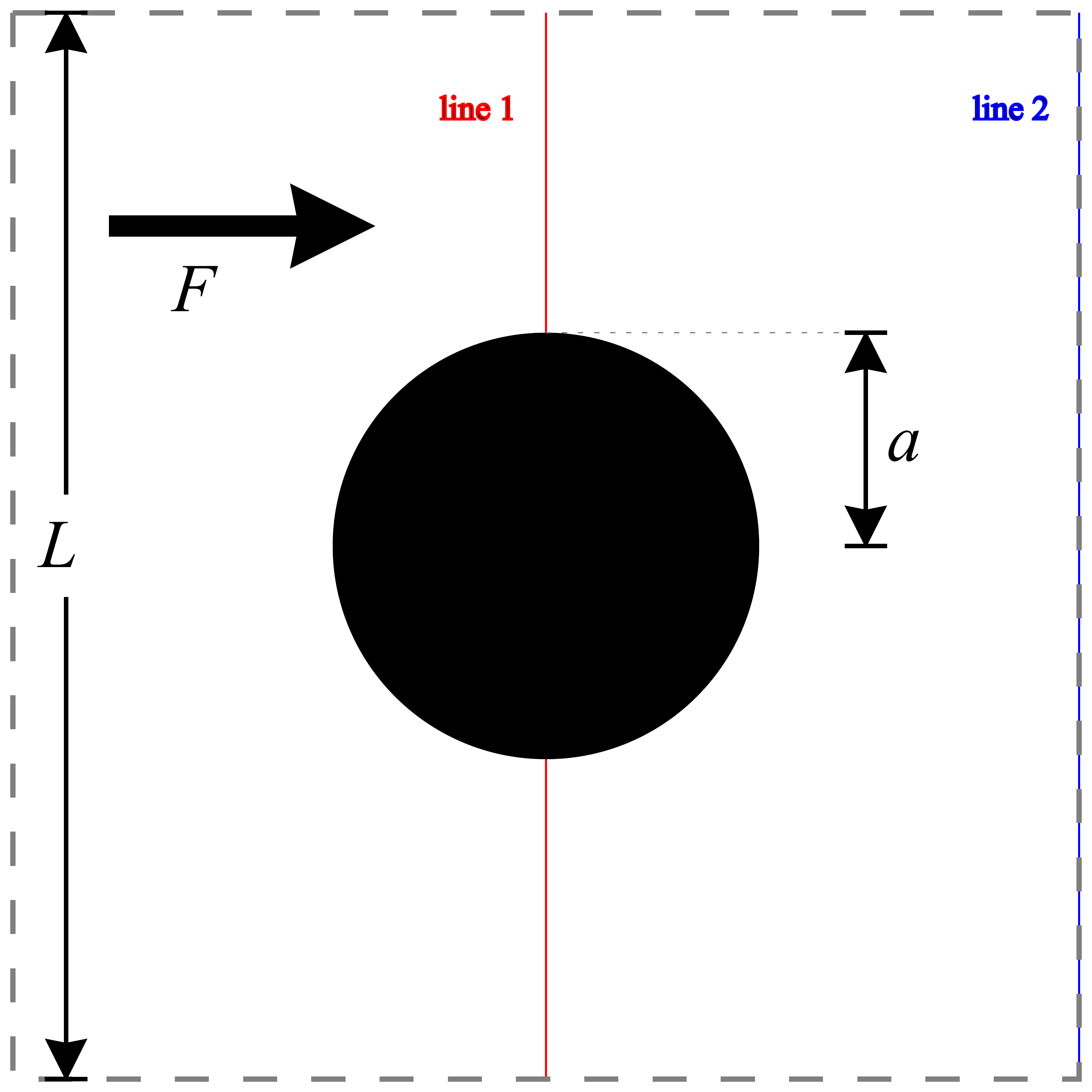}%
}
\caption{Geometry for the flow in a periodic lattice of cylinders test cases.
The marked lines 1 and 2 are the ones through which the velocity profiles are taken
to validate our results.}
\label{fig:lattice-geom}
\end{figure}

\begin{table}
\centerline{%
\renewcommand{\arraystretch}{1.2}%
\begin{tabular}{r l | l | l }
& & LRN & VLRN \\[1ex]
\hline
$F$ & ($\text{ms}^{-2}$) & $1.5\cdot 10^{-7}$ & $5\cdot 10^{-5}$ \\
$\nu$ & ($\text{m}^2\text{s}^{-1}$) & $10^{-6}$ & $10^{-4}$ \\
$c_0$ & ($\text{ms}^{-1}$) & $6\cdot10^{-4}$ & $10^{-2}$ \\
$t_{\text{max}}$ & (s) & $1.6\cdot10^{3}$ / $8\cdot10^{3}$ & $7\cdot10^{2}$ / $1.4\cdot10^{5}$
\end{tabular}
}
\caption{Parameters for the Low Reynolds Number (LRN) and Very Low Reynolds Number (VLRN)
flows in a periodic lattice of cylinders. The two end times $t_{\text{max}}$ refer to the
``short'' and ``long'' simulations as discussed in the text.}
\label{tab:lattice-params}
\end{table}

In both cases the domain is a 2D periodic square cell with side $L = 0.1\,\text{m}$,
with a cylinder of radius $a = 0.02\,\text{m}$ centered in it (Figure~\ref{fig:lattice-geom}).
The fluid is initially at-rest, subject to a driving force in the $x$ direction.
The magnitude $F$ of the driving force per unit mass, kinematic viscosity $\nu$ and
at-rest sound-speed $c_0$ depend on the test case and are illustrated in
Table~\ref{tab:lattice-params}. We use $\rho_0 = 1\,\text{kg}\,\text{m}^{-3}$ in both test cases.

Especially in the VLRN, the effect of particle disorder cannot be fully appreciated
when the steady state is reached,
so we run both LRN and VLRN for a ``short'' time (to the steady state)
as well as for a ``long'' time. The simulated runtimes are also indicated in
Table~\ref{tab:lattice-params}, and correspond approximately to the number of timesteps
indicated by~\cite{morris1997}.

\subsection{Domain discretization}

\begin{figure}
\centerline{%
\includegraphics[width=0.49\textwidth]{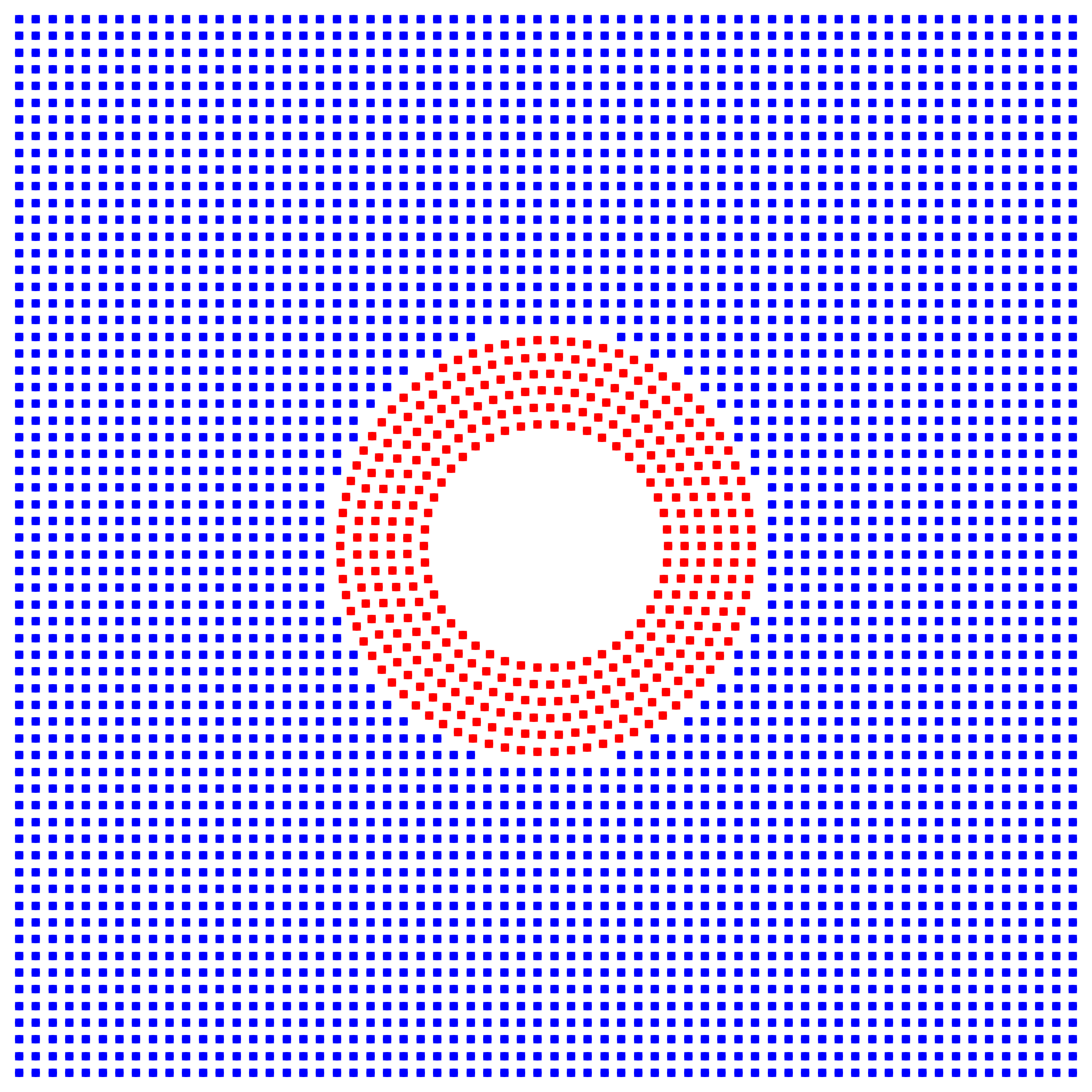}%
\hfill
\includegraphics[width=0.49\textwidth]{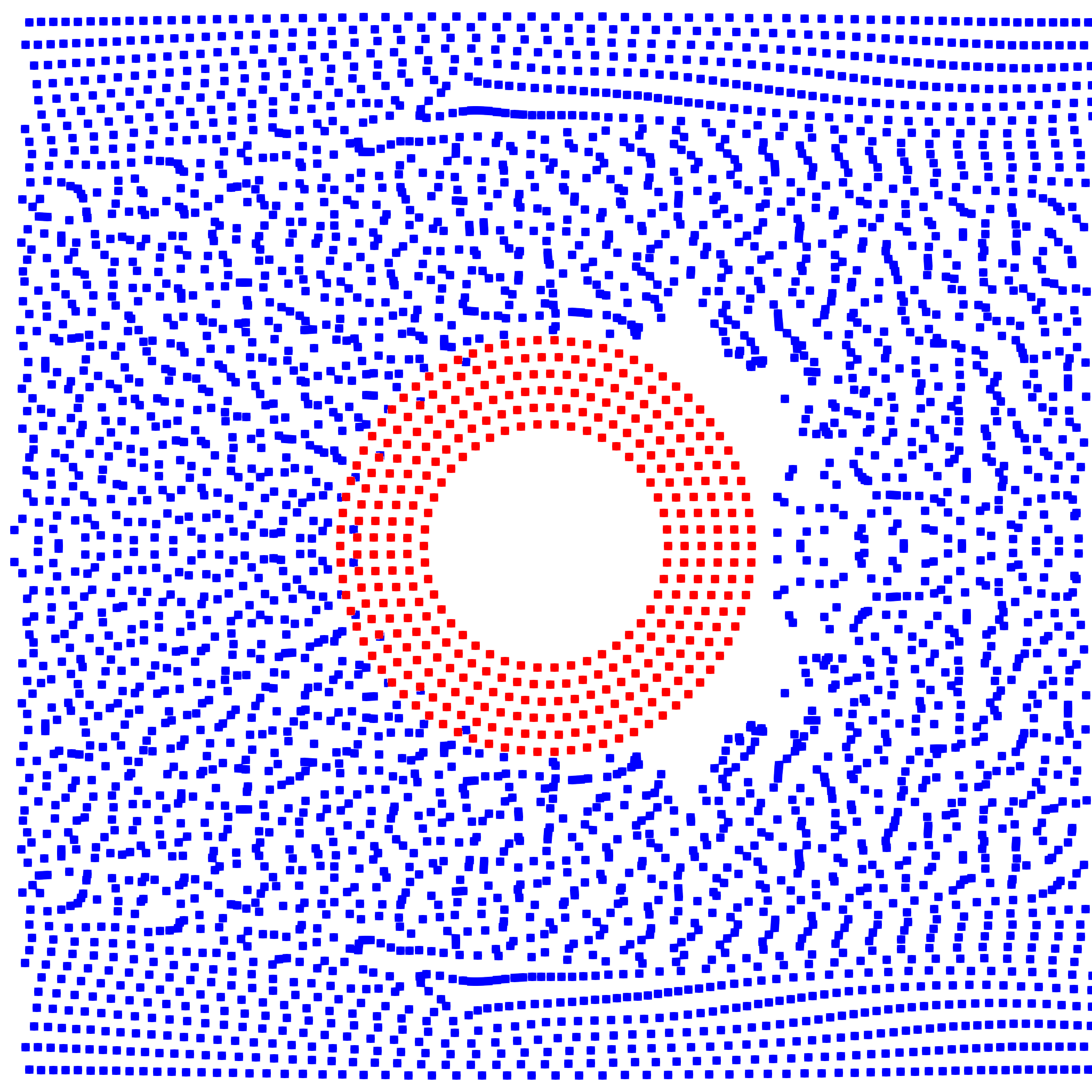}%
}%
\centerline{%
\includegraphics[width=0.49\textwidth]{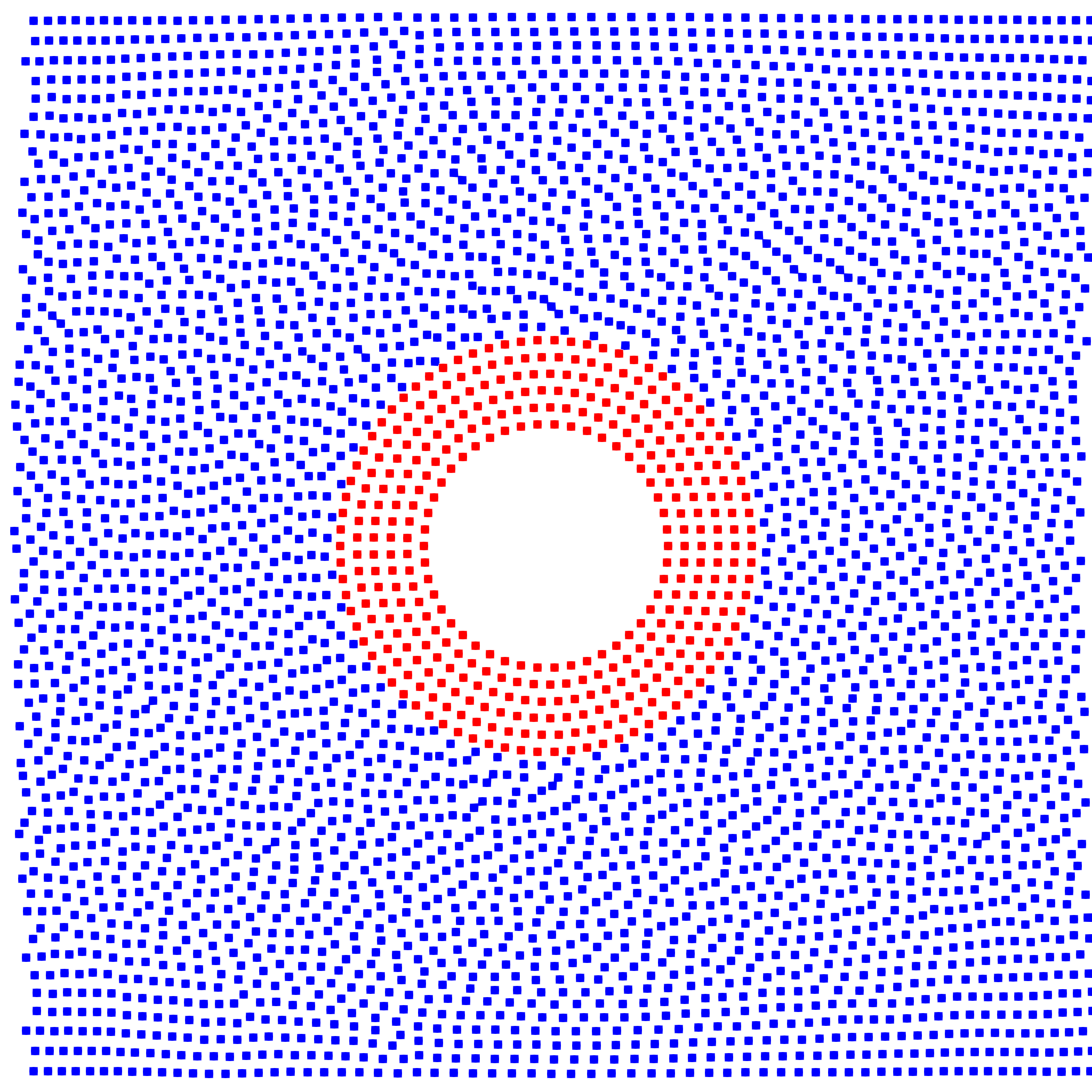}%
\hfill
\includegraphics[width=0.49\textwidth]{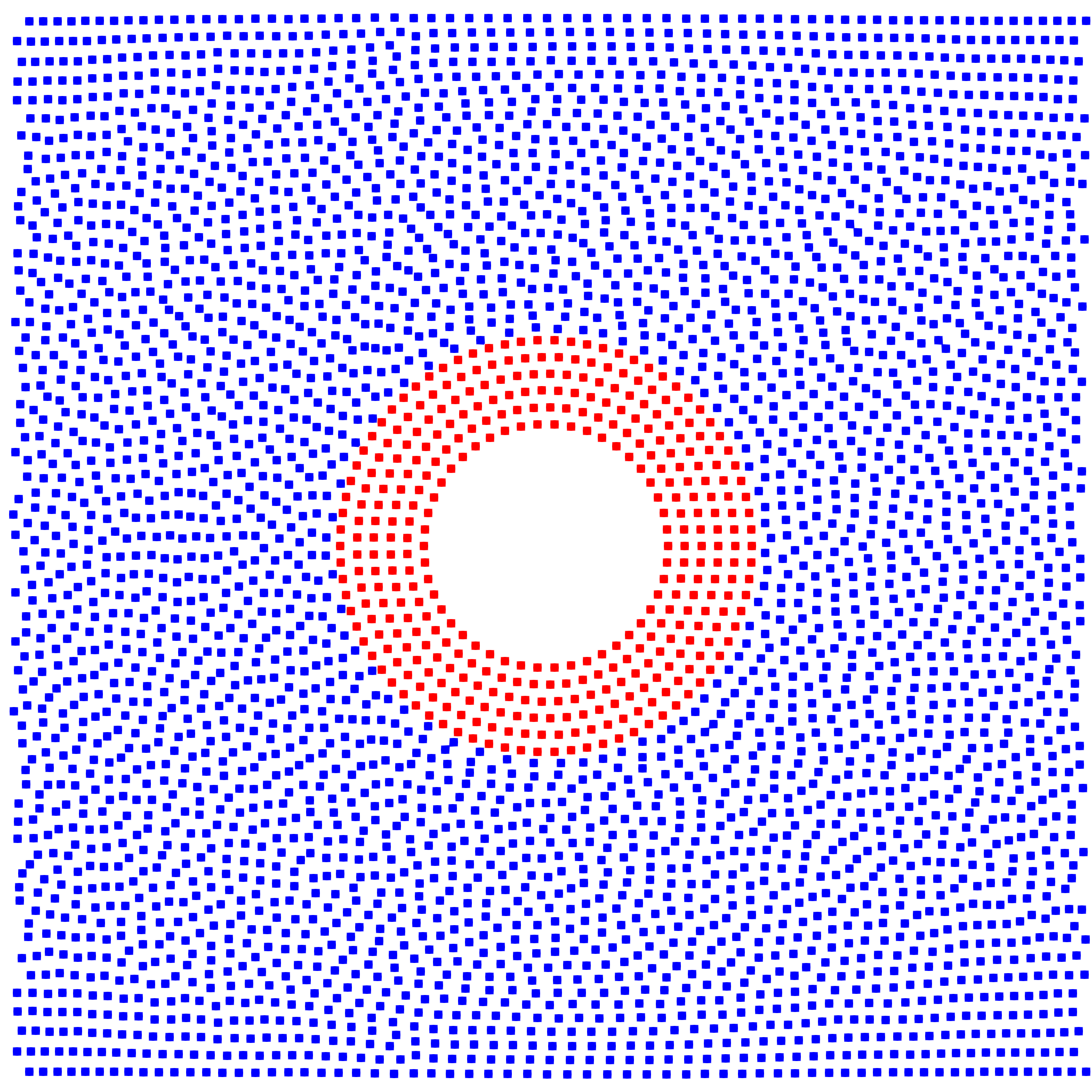}%
}%
\caption{Initial particle configuration (top left),
effect of tensile instability with Cole's equation of state (top right),
and layout at $t = 4000\,\text{s}$ with the explicit (bottom left)
and semi-implicit (bottom right) integrators when using the simplified
equation of state.}
\label{fig:lattice-distribution}
\end{figure}

Our initial setup (shown in the top left of Figure~\ref{fig:lattice-distribution})
has a regularly spaced grid of particles for the fluid stopping at a distance of $\Delta p/2$
from the cylinder surface, whereas the cylinder itself is discretized
with dummy boundary particles distributed along concentric circles starting $\Delta p/2$ inside.

For a complete assessment of the computational performance,
we run simulations with a number of resolutions, defined in terms of the number of particles
spanning the length $L$ (p.p.$L$, particles per $L$), starting from 64 p.p.$L$
(close to the resolution used in~\cite{morris1997}) and doubling up to 2048 p.p. $L$.
The ''long'' simulations in the VLRN case were only run up to a 512 p.p.$L$ resolution due to
the prohibitive computational times for the explicit integration scheme.
Some of the considerations about the choices of SPH formulation will show examples from
the lowest resolution, and the ''long'' run validation tests will show the results at the highest resolution.

The well-known tensile instability of SPH~\cite{swegle1995}
leads to numerical problems such as particle clumping and the creation of voids
in the low-pressure region past the cylinder when using Cole's equation of state~\eqref{eq:cole-pressure},
as show in Figure~\ref{fig:lattice-distribution} (top right).
Since the test case does not feature a free surface, the issue can be solved
by adopting the simpler equation of state
\[
P(\rho)=c_0^2 \rho
\]
proposed by~\cite{morris1997}.
We remark that the choice of integrator has no influence on the tensile instability,
although it does result in different particle distributions in the long term
(as shown in the bottom to subfigures of Figure~\ref{fig:lattice-distribution}).

To reduce the numerical noise in the pressure field, we adopt the density diffusion
mechanism described in~\cite{molteni_colagrossi_2009}, adding a diffusive term
to the mass continuity equation~\eqref{eq:part_continuity}
\[
\frac{D\rho_\beta}{Dt} = \sum_{\alpha} m_\alpha \vec u_{\alpha\beta}
\cdot \vec r_{\alpha\beta} F_{\alpha\beta} +
\xi h c_0 \sum_{\alpha} m_\alpha \Psi_{\alpha\beta} F_{\alpha\beta}
\]
where $\xi \in [0, 1]$ is the diffusive coefficient, for which we use $\xi = 0.1$, and
\[
\Psi_{\alpha\beta} = \begin{cases}
2\left( \dfrac{P_\beta}{P_\alpha} - 1 \right) &
\text{if }\dfrac{\abs{P_{\alpha\beta}}}{\rho_\beta \abs{\vec g\cdot \vec r_{\alpha\beta}}} > 1 \\
0 & \text{otherwise}
\end{cases}
\]
is the diffusive contribution.
We prefer this to the more recent and more sophisticated forms proposed for the diffusive term~\cite{antuono_2010,marrone_2011},
because these come at the cost of higher computational loads and are aimed at improving the solution near the free surface,
which is not present in our test cases.

\begin{figure}
\centerline{%
\includegraphics[width=0.49\textwidth]{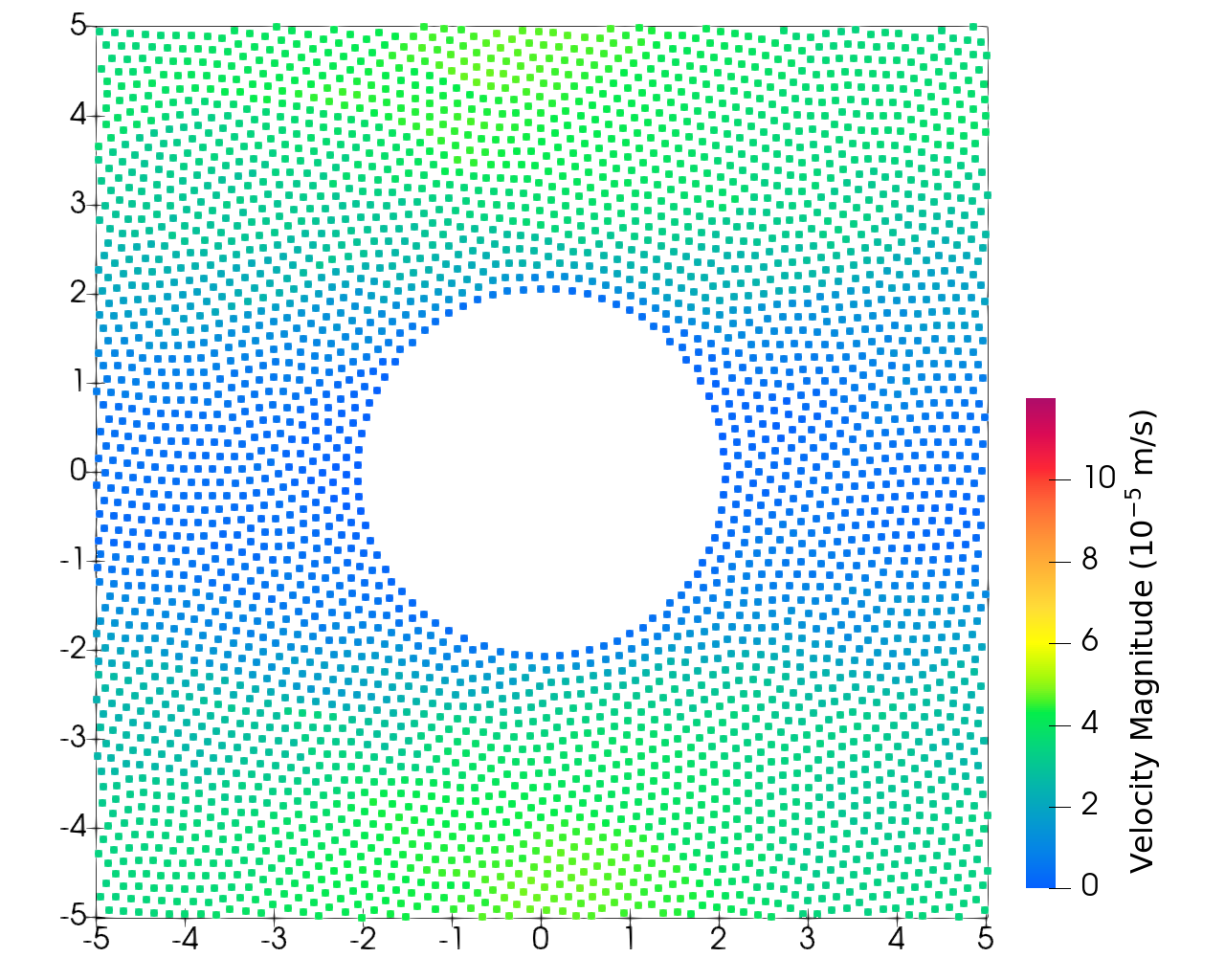}%
\hfill
\includegraphics[width=0.49\textwidth]{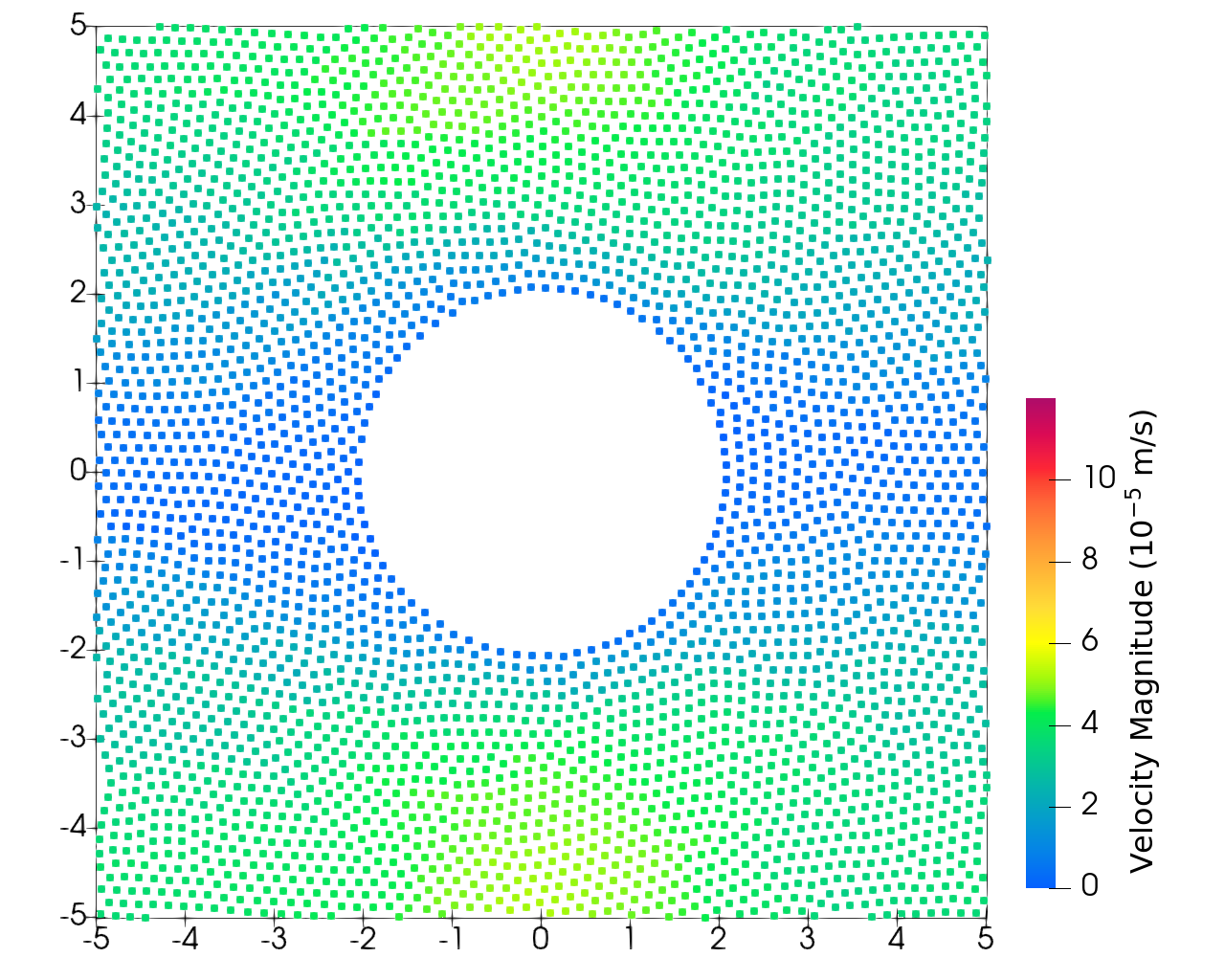}%
}%
\centerline{%
\includegraphics[width=0.49\textwidth]{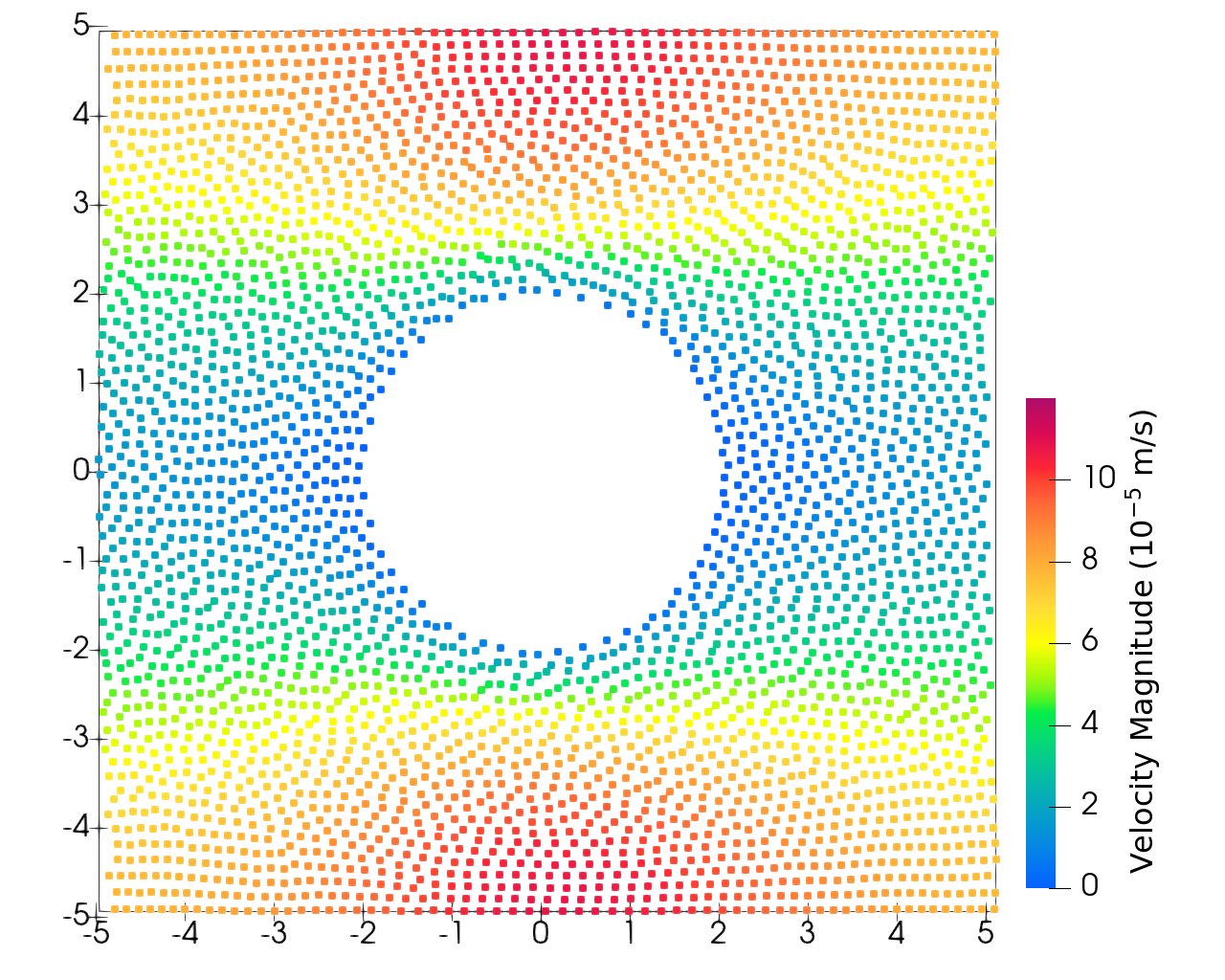}%
\hfill
\includegraphics[width=0.49\textwidth]{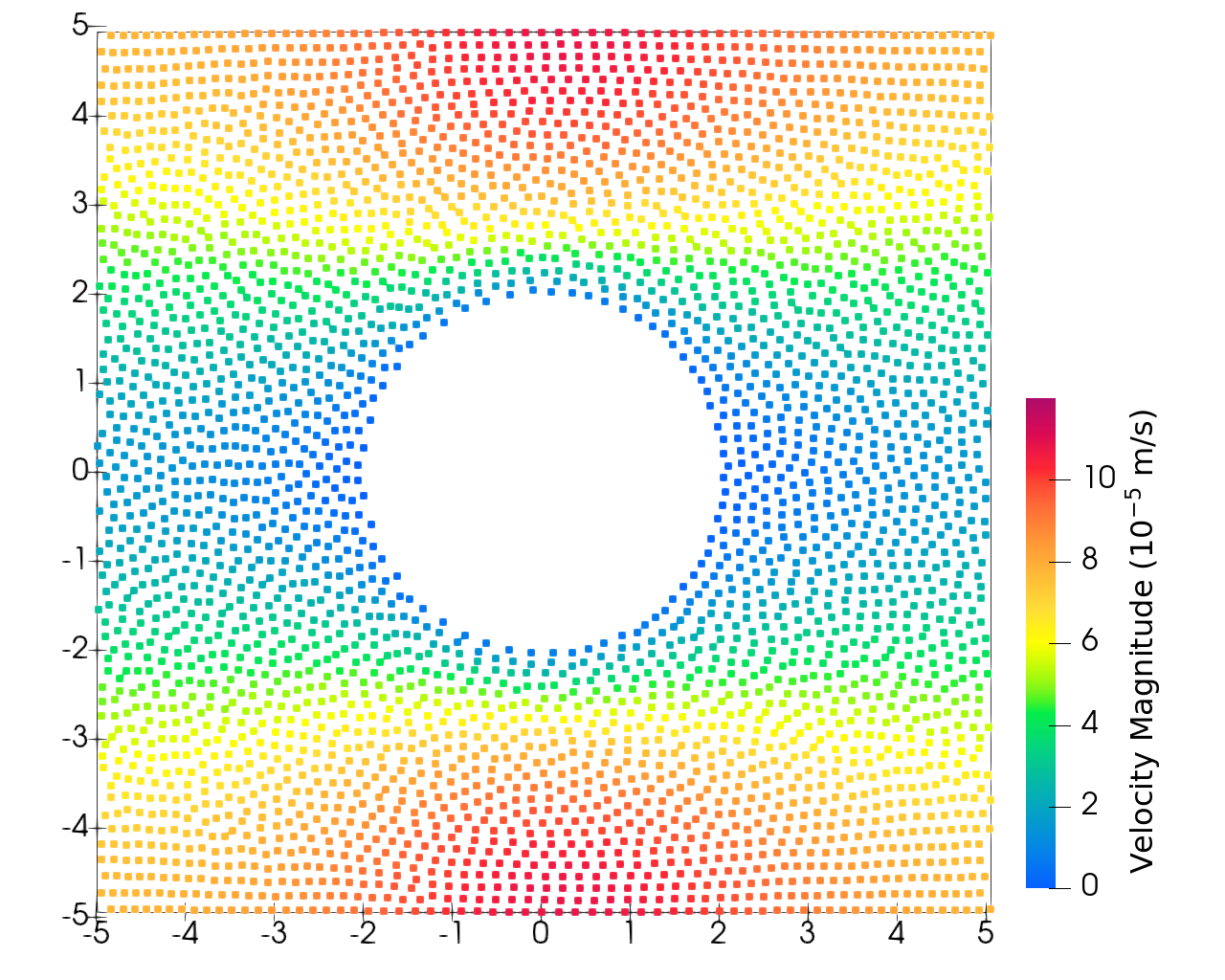}%
}%
\caption{Distribution of the velocity magnitude at a steady state
for the explicit (left) and semi-implicit (right) integrator
when using smoothing factor $1.3$ (top) versus $2.25$ (bottom) in the LRN case
at the lowest resolution employed in our tests.}
\label{fig:lattice-vel}
\end{figure}

We use the Wendland kernel described in section~\ref{sec:kernel-choice},
with a smoothing factor for $2.25$.
The choice of the smoothing factor is dictated by the excessive numerical dissipation observed
at smaller values (Figure~\ref{fig:lattice-vel}).
Indeed, in this test case the default value of $1.3$ for the smoothing radius
results in an effective speed that is about 50\% of the expected value at a steady state.
Such excessive viscous dissipation is independent from the chosen integration scheme,
and has been thoroughly studied in the context of SPH application to gravity waves
(see e.g. \cite{Colagrossi_2013,antuono_2011,Antuono_2013,oger2007}),
with the most common solution being the adoption of larger smoothing factors~\cite{Colagrossi_2013},
sometimes coupled with appropriate kernel choices~\cite{Chang_2017,antuono_2011}.

The larger smoothing factor is, in our case, a simple and effective solution,
consistent with the already-discussed results seen for the plane Poiseuille test case (section~\ref{sec:convergence}).
Additionally, our test-case does not have a free surface, and is thus immune to the ``rarifying'' free surface
side-effect of larger smoothing factors when studying gravity waves discussed e.g. in~\cite{zago_ccsph}.
Alternative strategies such as the use of kernel corrections~\cite{guilcher_2007,xiao_2020,zago_ccsph},
that may require specific adaptation to the semi-implicit numerical scheme are being considered as future work.

%\begin{figure}
%\centerline{%
%\includegraphics[width=0.33\textwidth]{img/sph-implicit-lrn-2048-short-vel}%
%\hfill
%\includegraphics[width=0.33\textwidth]{img/sph-explicit-lrn-2048-short-vel}%
%\hfill
%\includegraphics[width=0.33\textwidth]{img/fem-re1-hires-vel}%
%}%
%\centerline{%
%\includegraphics[width=0.33\textwidth]{img/sph-implicit-vlr-2048-short-vel}
%\hfill
%\includegraphics[width=0.33\textwidth]{img/sph-explicit-vlr-2048-short-vel}
%\hfill
%\includegraphics[width=0.33\textwidth]{img/fem-re03-hires-vel}%
%}%
%\caption{Distribution of the velocity magnitude for the LRN (top) and VLRN (bottom) cases
%at a steady state
%for our semi-implicit integrator at 2048 p.p.$L$ (left),
%explicit integrator at 2048 p.p.$L$ (middle)
%versus the FVE model used for validation (right).
%Axes are in units of $10^{-2}\,\text{m}$.}
%\label{fig:lattice-valid}
%\end{figure}

\begin{figure}
\centerline{%
\includegraphics[width=0.33\textwidth]{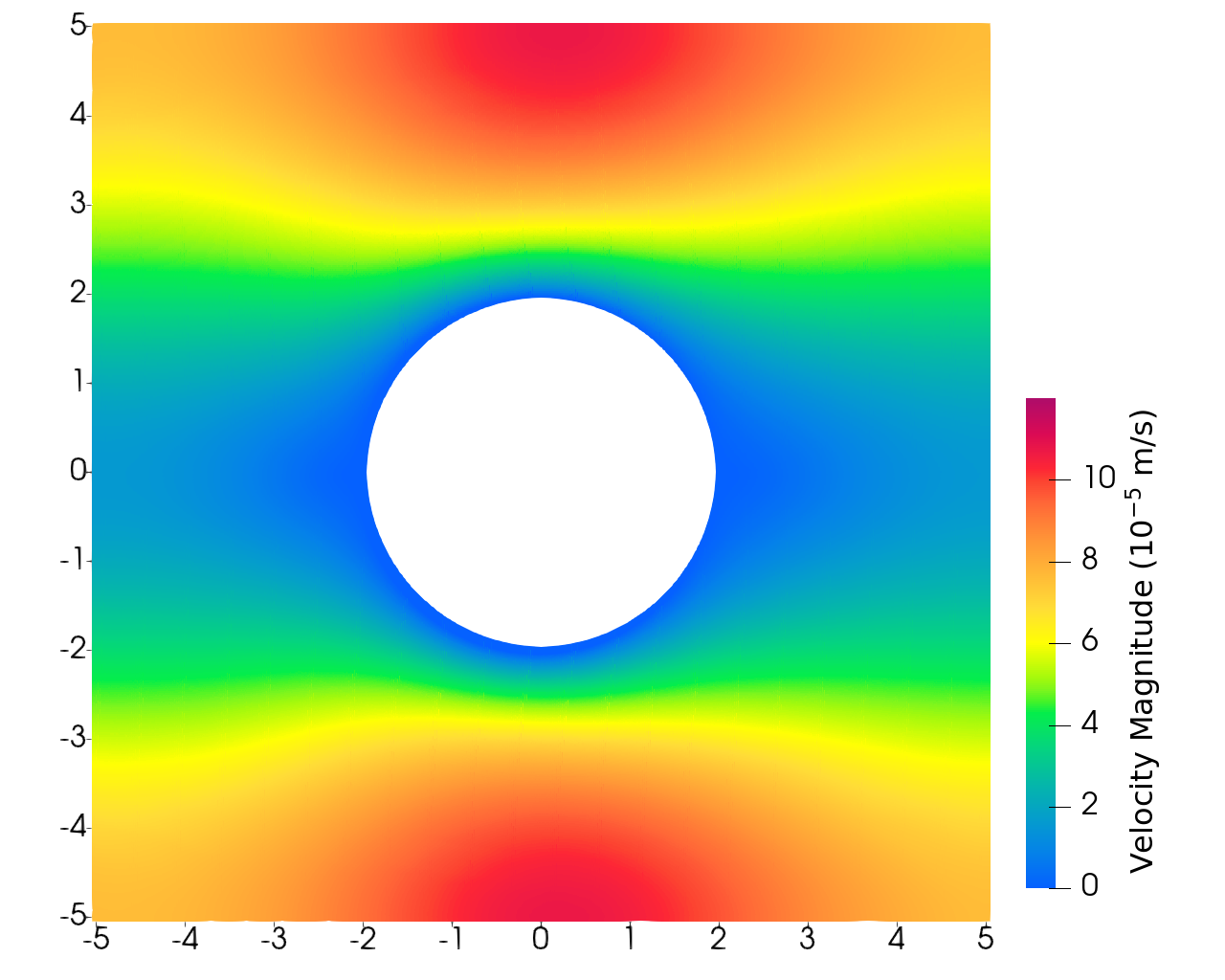}%
\hfill
\includegraphics[width=0.33\textwidth]{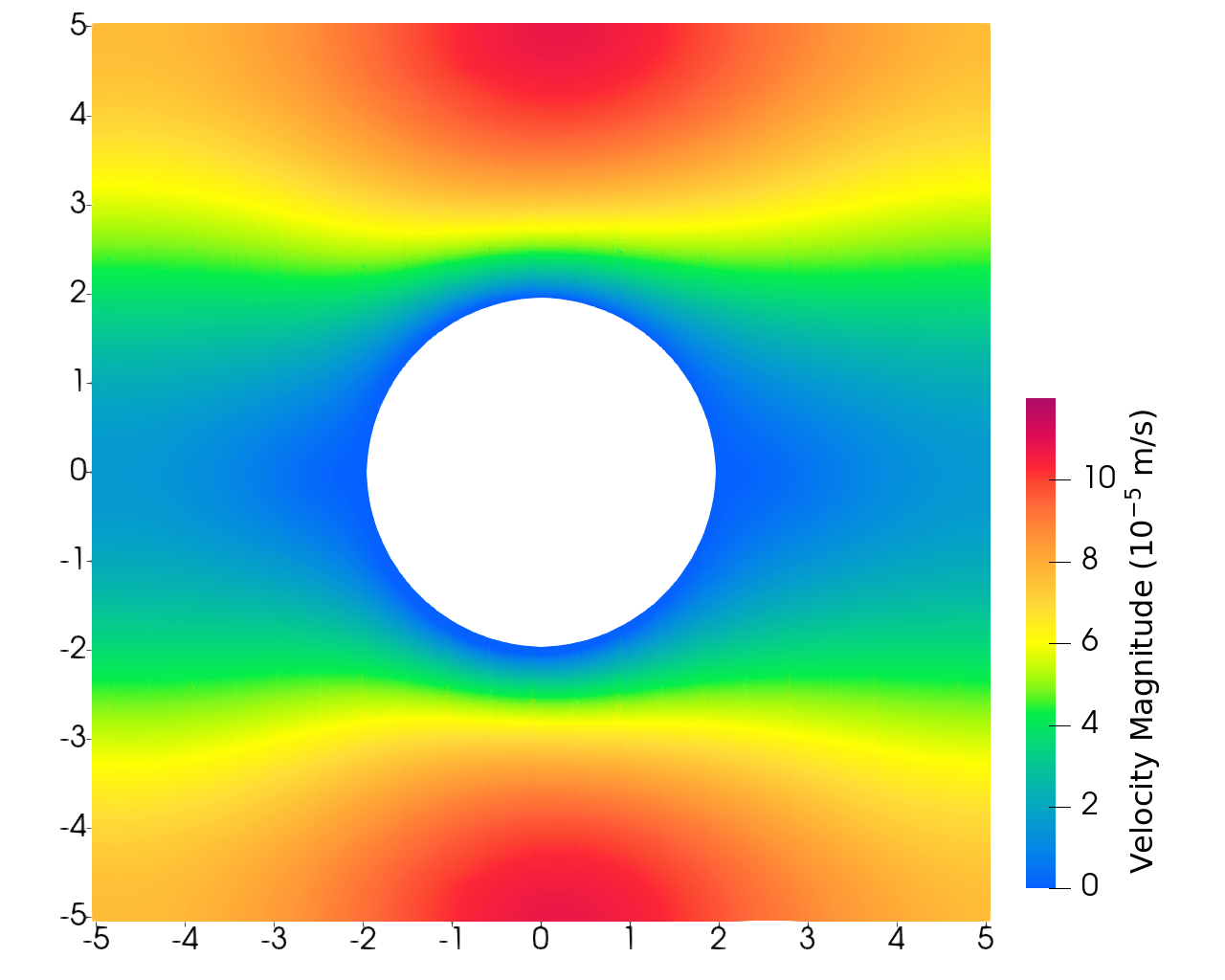}%
\hfill
\includegraphics[width=0.33\textwidth]{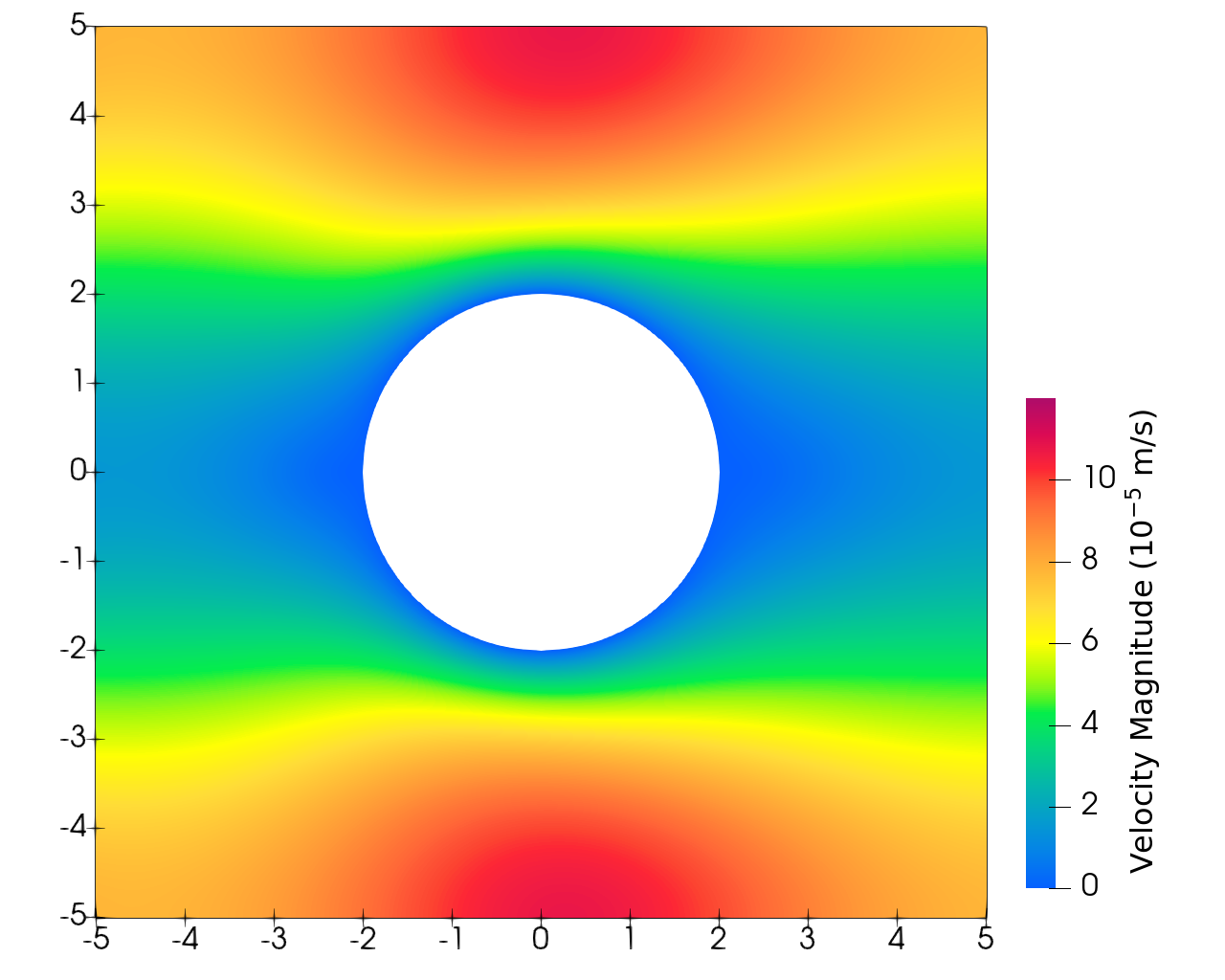}%
}%
\centerline{%
\includegraphics[width=0.33\textwidth]{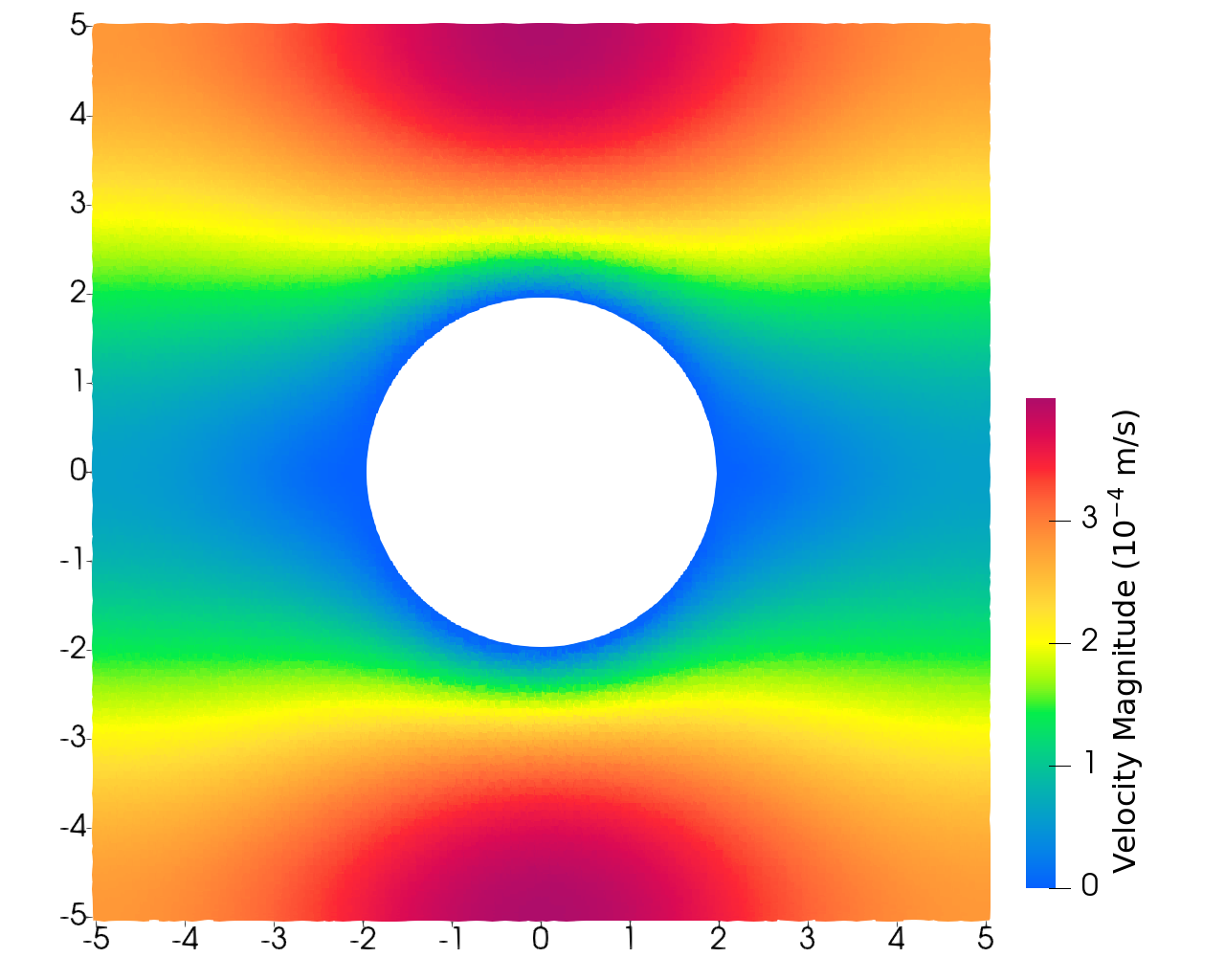}
\hfill
\includegraphics[width=0.33\textwidth]{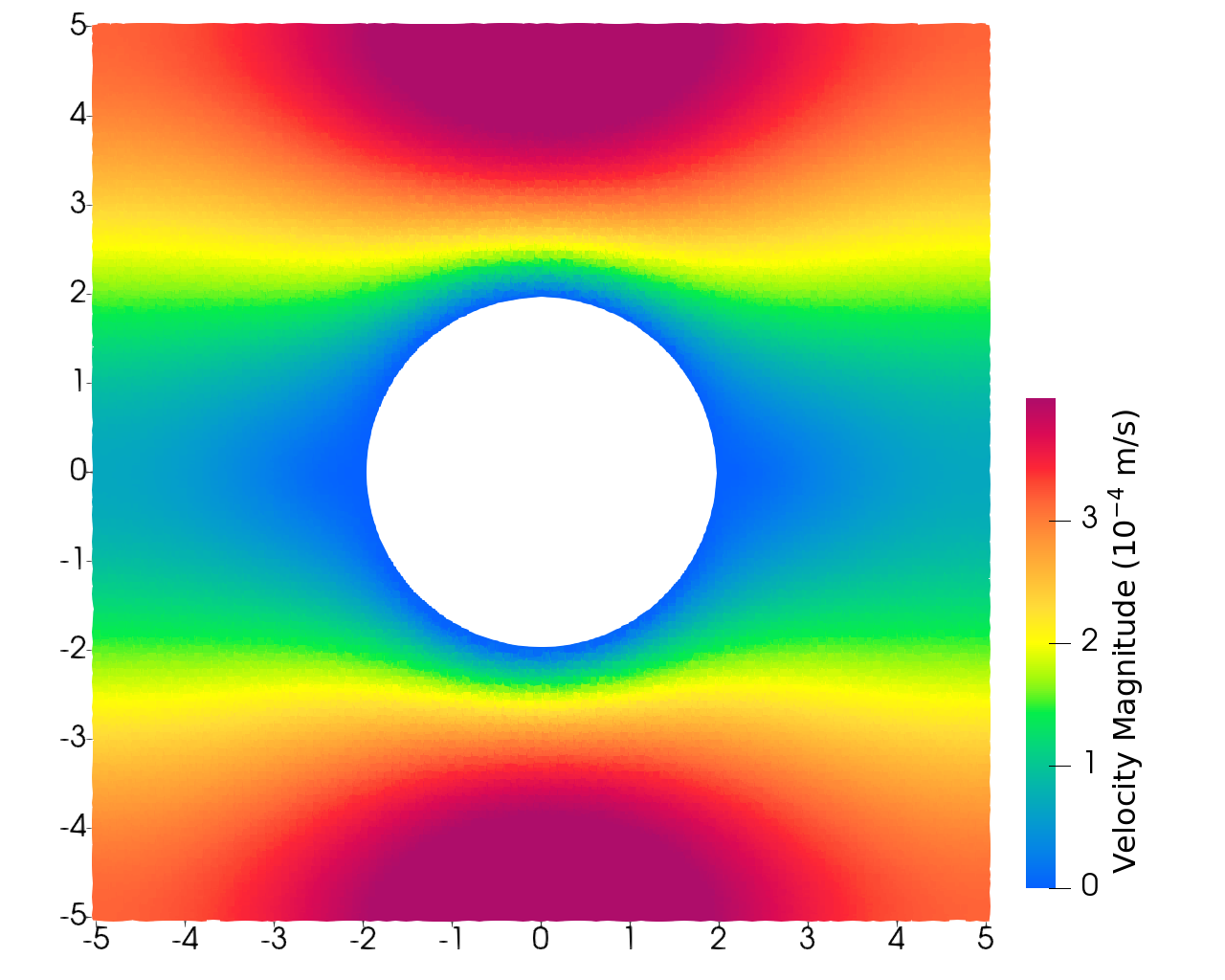}
\hfill
\includegraphics[width=0.33\textwidth]{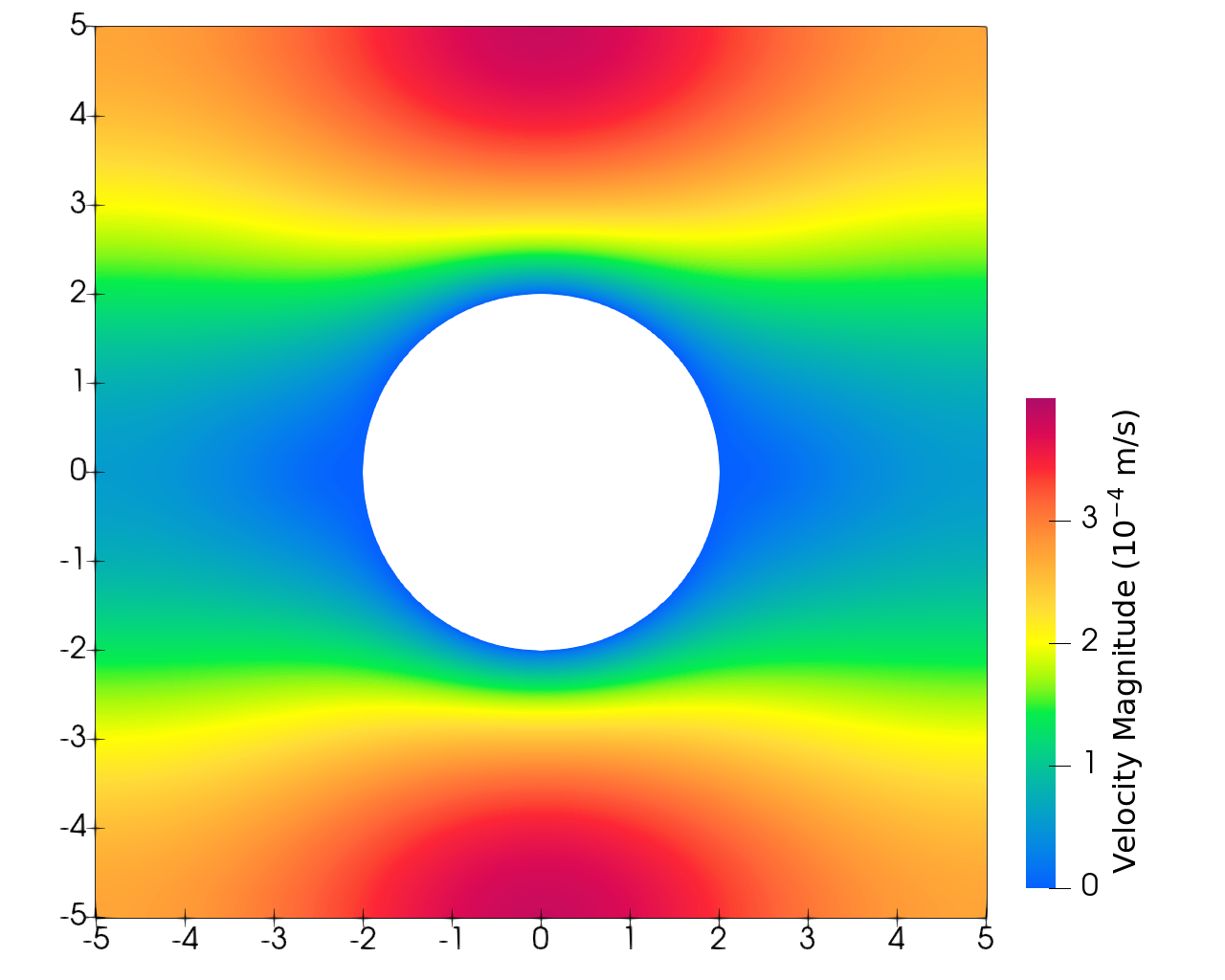}%
}%
\caption{Distribution of the velocity magnitude for the LRN (top) and VLRN (bottom) cases
in the ''long'' run
for our semi-implicit integrator at high p.p.$L$ (left),
explicit integrator at high p.p.$L$ (middle)
versus the FVE model used for validation (right).
The high p.p.$L$ is 2048 for the LRN case and 512 for the VLRN case.
Axes are in units of $10^{-2}\,\text{m}$.}
\label{fig:lattice-valid}
\end{figure}

%\begin{figure}
%\centerline{%
%\includegraphics[width=0.49\textwidth]{img/lattice-re1-vel-plot}%
%\hfill
%\includegraphics[width=0.49\textwidth]{img/lattice-re03-vel-plot}%
%}%
%\caption{Velocity along the middle and boundary line for the LRN (left)
%and VLRN (right) case at a steady state
%for our semi-implicit integrator at 64 and 2048 p.p.$L$ (SPH) versus the FVE model used for validation.}
%\label{fig:lattice-valid-lines}
%\end{figure}

\begin{figure}
\centerline{%
\includegraphics[width=0.49\textwidth]{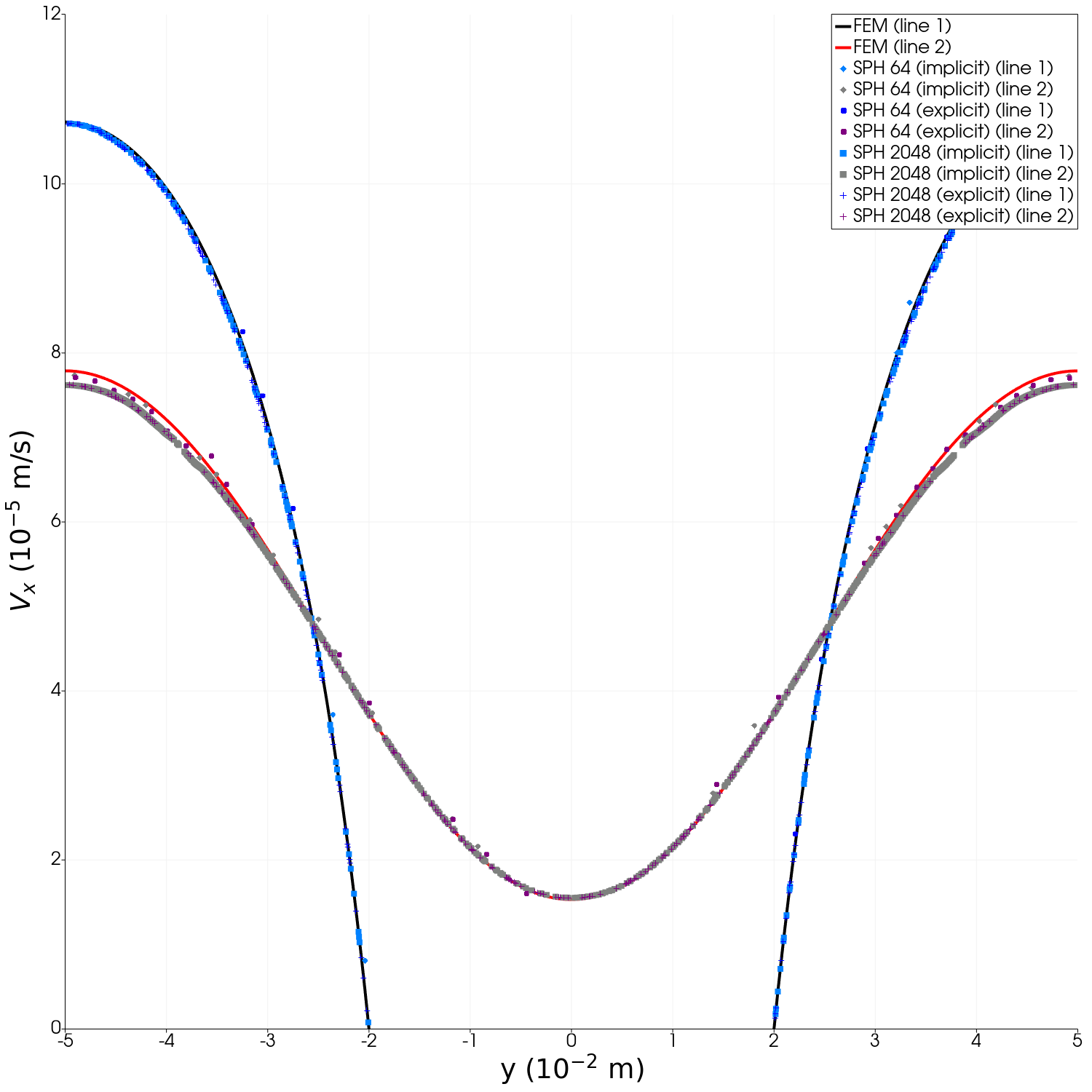}%
\hfill
\includegraphics[width=0.49\textwidth]{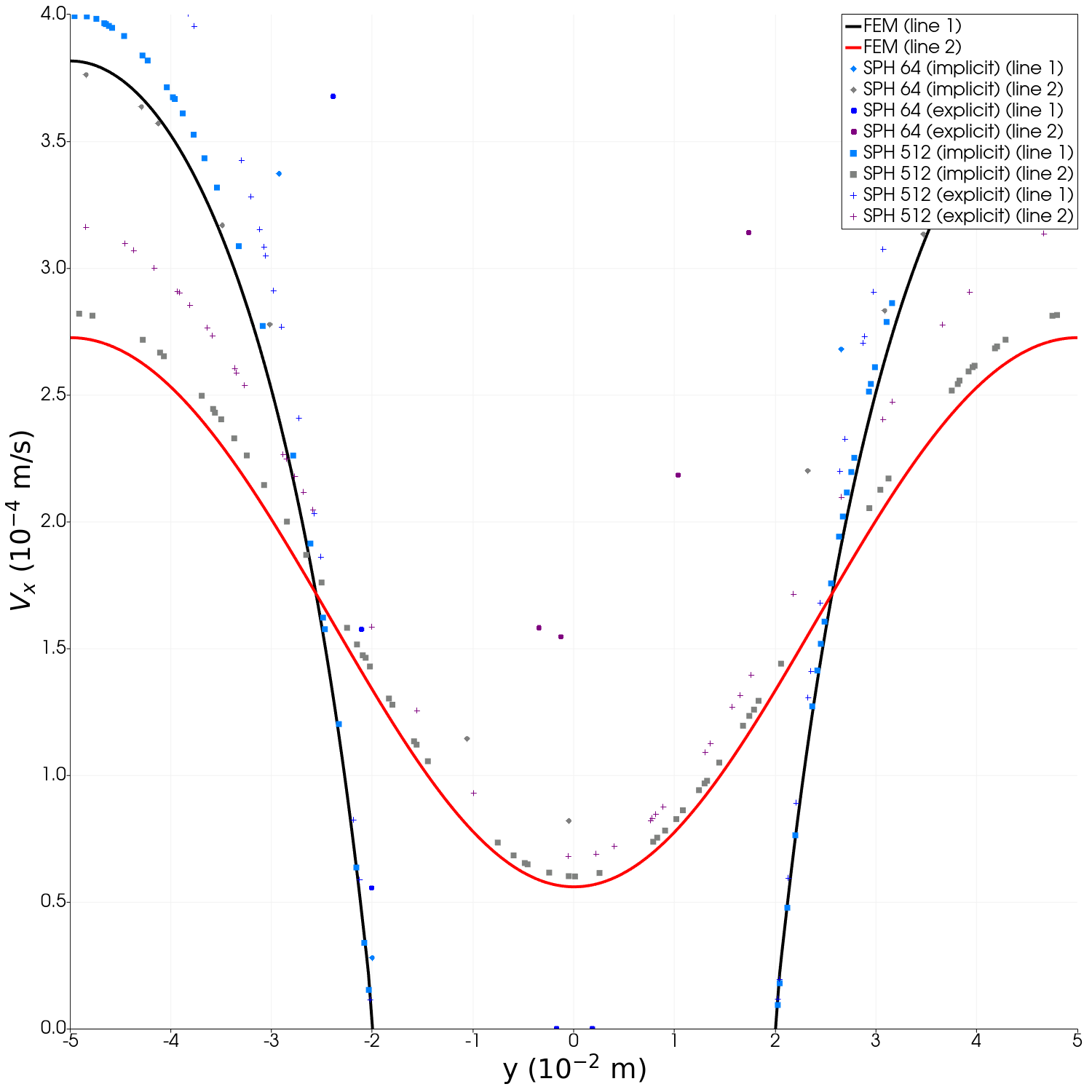}%
}%
\caption{Velocity along the middle and boundary line for the LRN (left)
and VLRN (right) case for the ``long'' runs,
for our semi-implicit integrator at 64 and and a higher p.p.$L$ (SPH) versus the FVE model used for validation.
The higher p.p.$L$. is 2048 for the LRN case and 512 for the VLRN.}
\label{fig:lattice-valid-lines-long}
\end{figure}

To validate the model, we compare our results against those obtained using the
TRUST code~\cite{trust}, an open source platform developed at the Nuclear Energy Division of
the French Atomic Energy Commission (CEA).
A Finite Volume Element (FVE) scheme~\cite{zhiqiang1990,fortin2017mixed}
is employed on triangular cells to integrate the Navier--Stokes equations, in
conservative form, over the control volumes belonging to the calculation
domain~\cite{Emonot1992MthodesDV,Laucoin2008DveloppementDP,fortin2006methode}.
As in the classical Crouzeix--Raviart element, both vector and scalar
quantities are located at the centers of the faces. The pressure, however, is
located at the vertices and at the center of gravity of a tetrahedral element.
This discretization leads to very good pressure/velocity coupling and has a
very dense divergence free basis. Along this staggered mesh arrangement, the
unknowns, i.e. the vector and scalar values, are expressed using non-conforming
linear shape-functions (P1-nonconforming). The shape function for the pressure
is constant for the center of the element (P0) and linear for the vertices
(P1). The spatial discretization scheme is of second order for both convection
and diffusion terms. The time integration scheme is implicit. All linear
systems are solved by a direct Cholesky solver. Further information concerning
the TRUST code and the numerical methods can be found in~\cite{saikali2018numerical,saikali2019highly,saikali2021hydro}.
The obtained results are consistent with the ones shown in~\cite{morris1997}.

Since for this test case we are particularly interested in the effect of particle disorder,
we look at the velocity profiles for the ''long'' SPH runs, comparing them with the FVE results.
We observe that in the LRN case, the semi-implicit and explicit SPH integrators give essentially the same results,
and are in good accordance with the FVE results, as shown
in both Figure~\ref{fig:lattice-valid} (top row) and~\ref{fig:lattice-valid-lines-long} (left).
The situation is quite different in the VLRN case:
the very small timestep in the explicit integrator leads to an accumulation of numerical error
that manifests as increasing turbulence and a higher velocity magnitude (Figure~\ref{fig:lattice-valid}, bottom row, center column),
which reflects in mismatched velocity profiles (Figure~\ref{fig:lattice-valid-lines-long}, right).
By contrast, the semi-implicit integrator manages to preserve a good match with the FVE velocity
distribution (Figure~\ref{fig:lattice-valid}, bottom row, left vs right column),
and profiles (Figure~\ref{fig:lattice-valid-lines-long}, right) even in the ''long'' run.

\subsection{Computational performance}

\begin{figure}
\centerline{%
\includegraphics[width=0.49\textwidth]{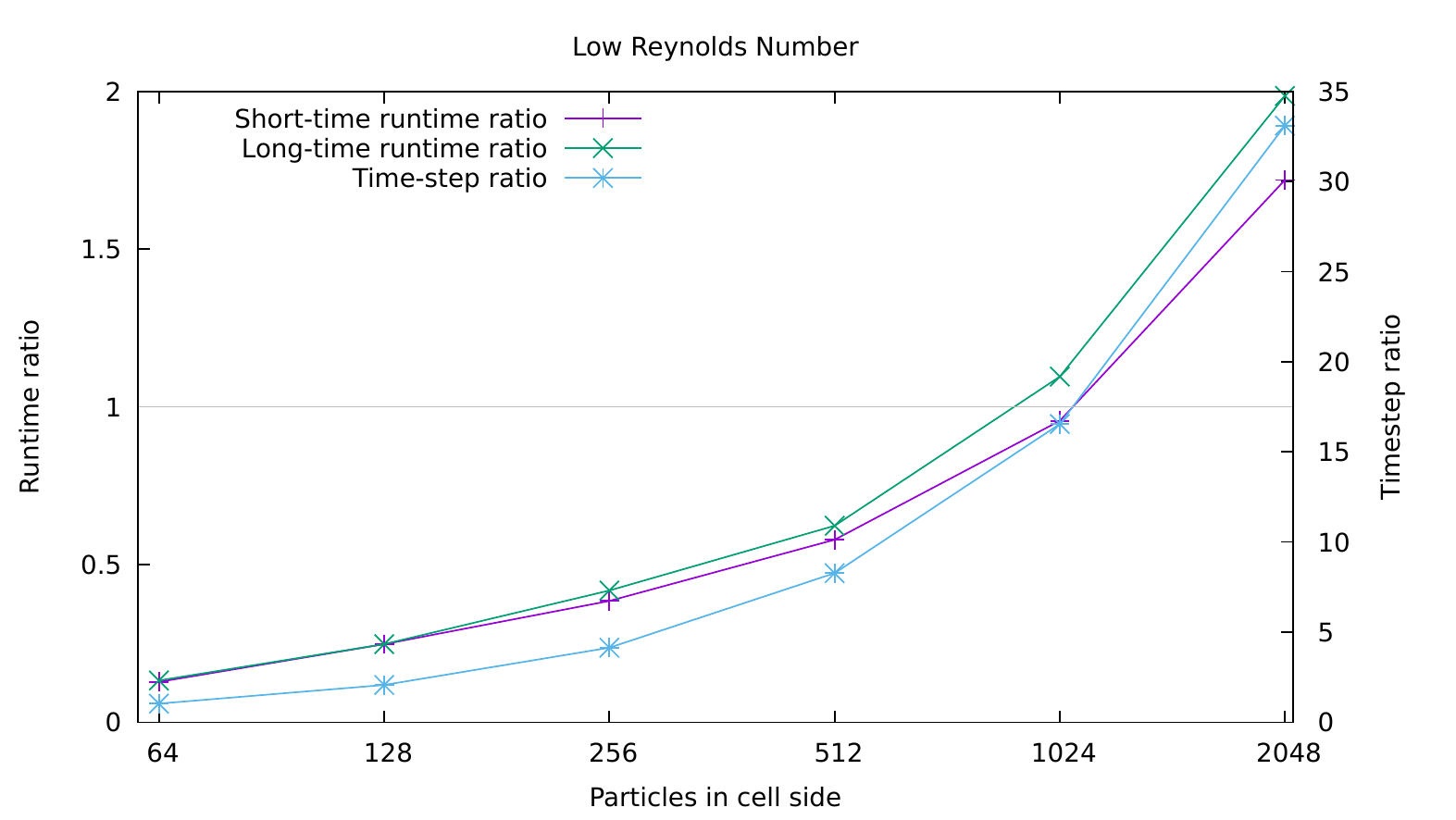}%
\hfill
\includegraphics[width=0.49\textwidth]{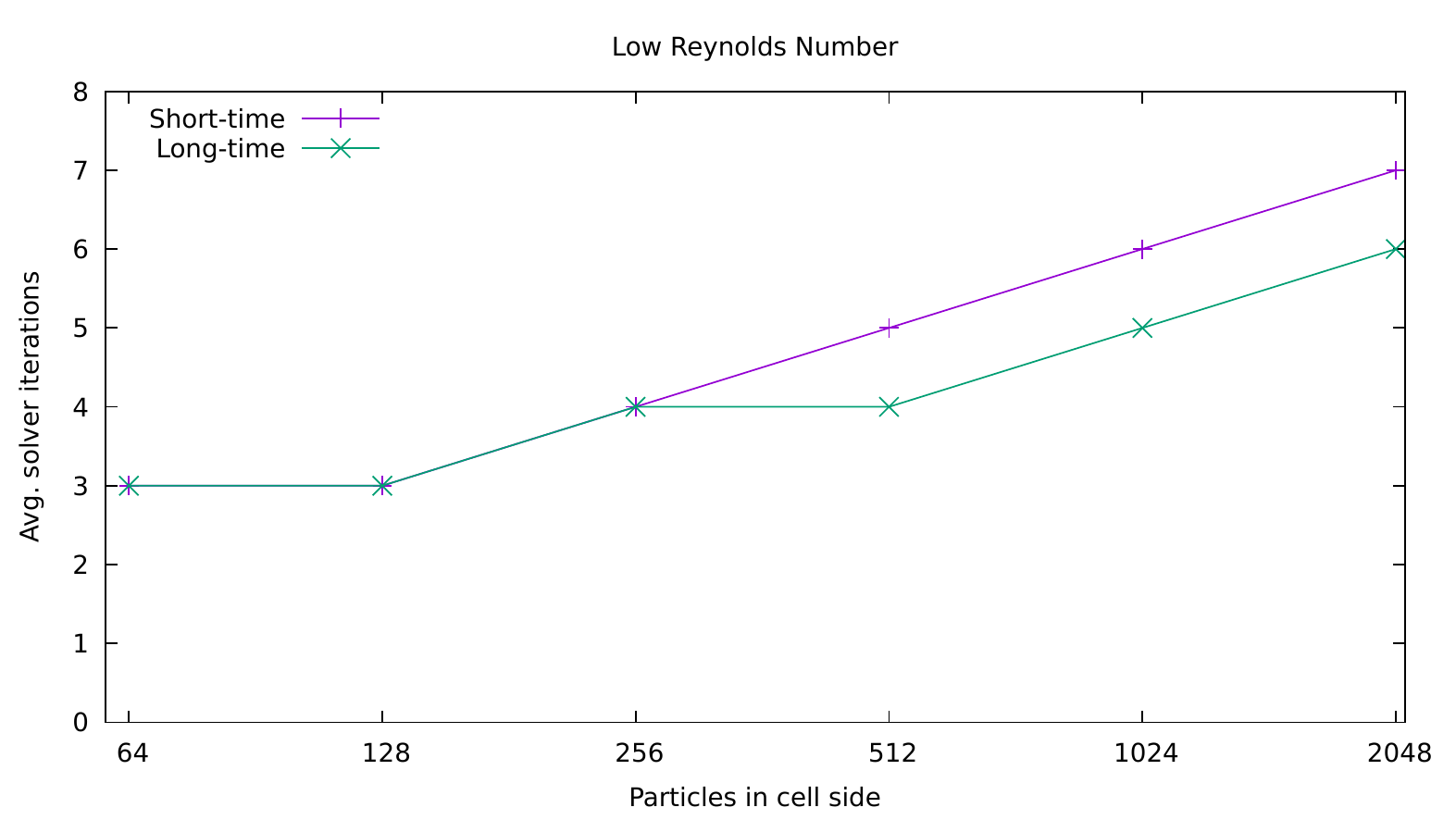}%
}%
\centerline{%
\includegraphics[width=0.49\textwidth]{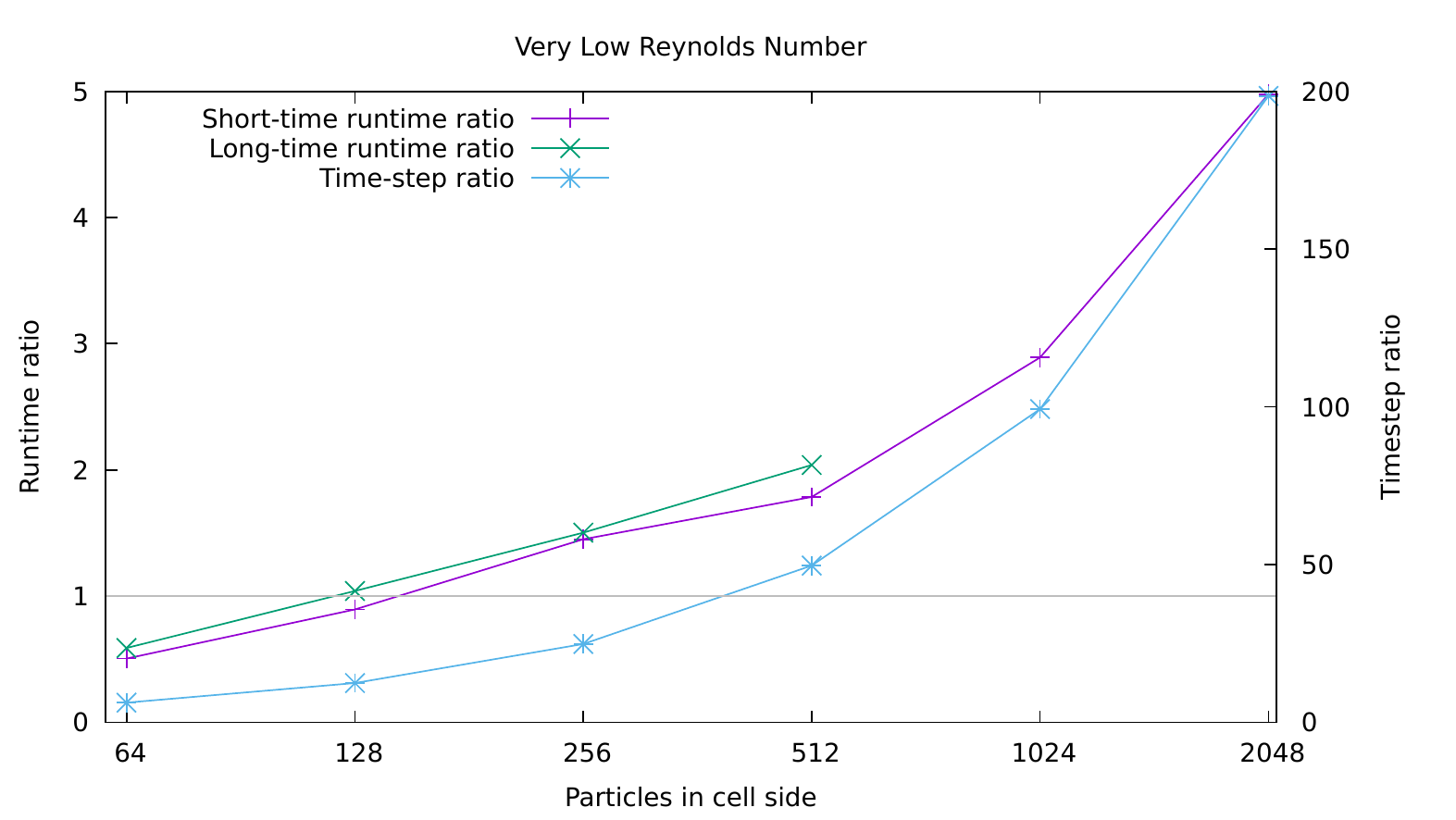}%
\hfill
\includegraphics[width=0.49\textwidth]{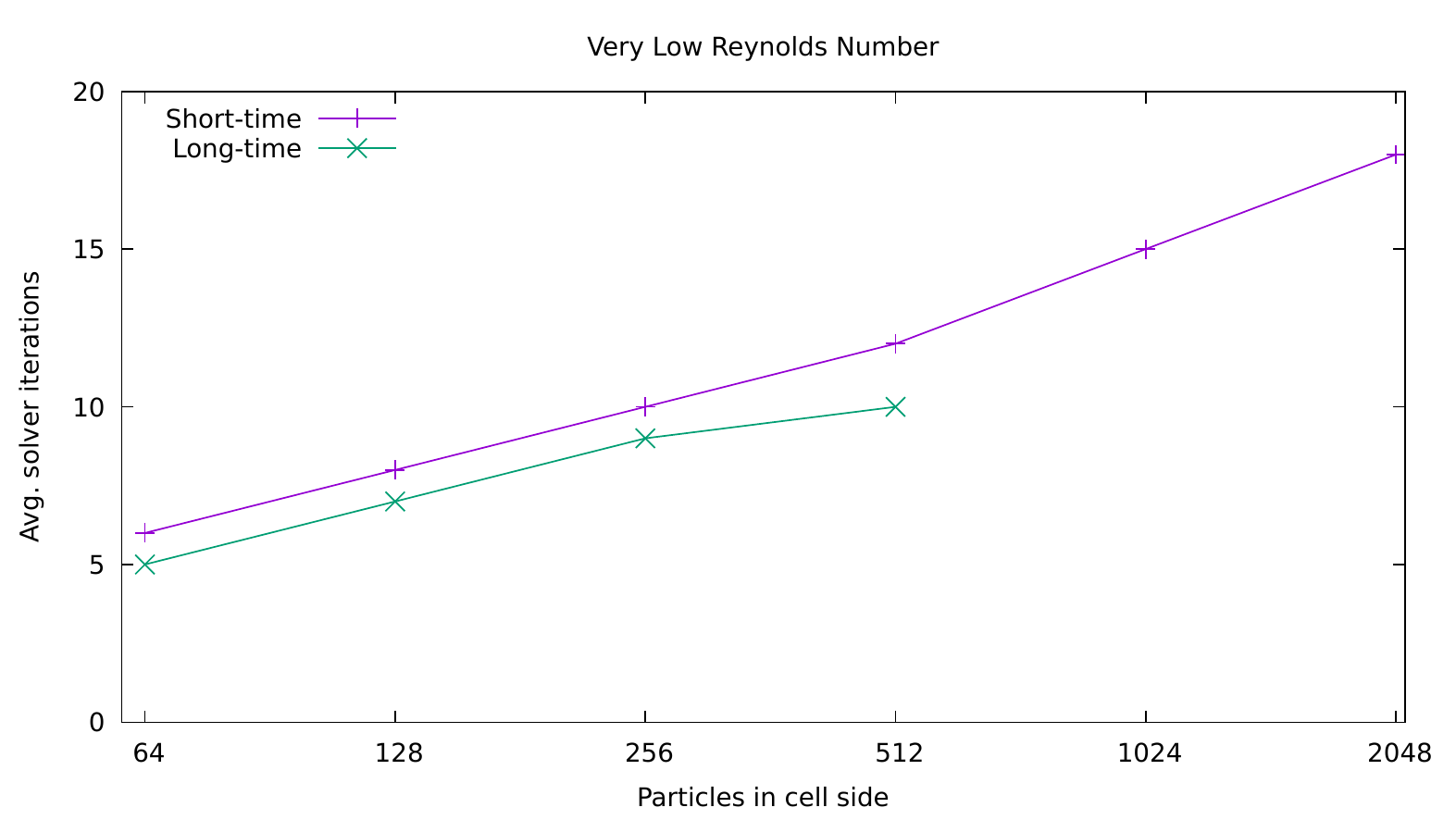}%
}%
\caption{Left: runtime and timestep ratio between the explicit and semi-implicit integrator.
Right: average number of implicit solver iterations (right).
Results are for the Low Reynolds Number (top) and Very Low Reynolds Number (bottom) test cases
as a function of resolution, for both short and long simulation times.}
\label{fig:lattice-perf}
\end{figure}

Consistently with the results presented in~\cite{zago_jcp_2018},
the validation of our semi-implicit integration scheme shows a numerical improvement in test-cases
where the viscous time-step constraint is dominant and may lead to higher numerical error accumulation
in the explicit scheme. Even when the results between the explicit and semi-implicit integration scheme
match, however, the semi-implicit scheme can still make a difference in computational performance.

We are particularly interested in seeing if (and how) particle disorder affects the
resolution at which it becomes convenient to switch from the explicit to the semi-implicit integrators
(``switch-over''). The results are illustrated in Figure~\ref{fig:lattice-perf},
where the runtimes for each resolution are taken as the median over three runs
(the maximum deviation between the runs is less than $0.5\%$).
We show the ratio of the semi-implicit runtime to the explicit runtime,
the ratio between the time-step used with the two integrators, and
the average number of iterations needed for the solver to converge in the semi-implicit case.
``Long'' runs for the VLRN case at the higher resolutions are excluded due to the extremely long runtime
that would be needed with the explicit integrator.

We remark that particle disorder does not seem to affect the solver convergence negatively:
in fact, the average number of solver iterations \emph{decreases} in the ``long'' run
compared to the early phase (``short'' runs).
The effect is more visible in the VLRN case than in the LRN case,
due to a combination of factors including the difference in physical parameters,
the lower number of iterations needed in the LRN even for the ``short'' run,
and the fact that we compute the average number of iterations only up to the nearest integer.
The benefit can be seen more clearly in the small improvement
in the explicit\slash semi-implicit runtime ratio for the long run over the short run.

This would not be particularly surprising in an Eulerian framework,
since the (Eulerian) velocity remains constant once the steady state is reached.
However, in a Lagrangian framework such as SPH, the velocity vector
of the particles is \emph{not} constant even after reaching a steady state,
since particles flowing around the cylinder do not move on a straight line:
the predictive power of the velocity at the previous time-step as
initial guess for the next time-step is thus reduced, although
we can see by the reduction in solver iterations that it remains a good choice,
and one that is not adversely affected by the increasing particle disorder.

Since the number of solver iterations is a key element in determining the convenience of the
semi-implicit over the explicit integrator, in this case the net effect is that in the ``long''
run the switch-over resolution is coarser than in the ``short'' run
(down to $1024$ p.p.$L$ from $2048$ p.p.$L$ in the LRN case,
down to $128$ p.p.$L$ from $256$ p.p.$L$ in the VLRN case).

\section{Multi-GPU}

A significant advantage of the BiCGSTAB and CG implementations presented here
compared to the CG solver used in~\cite{zago_jcp_2018} is the support
to parallelize the code across multiple GPUs, either on the same machine
or across multiple machines, using the existing GPUSPH infrastructure
for this~\cite{rustico_advances_2014,rustico_multi-gpu_2014}.

The key change when distributing the computations across multiple GPUs
is the transition from shared-memory parallelism (single GPU) to
distributed memory parallelism, and the subsequent need to choose which
data to exchange between devices, and when to exchange it. The objective
is generally to minimize the number of transfers, minimize the amount of
data transferred, and if possible overlap data transfers with
computations to (at least partially) cover the data transfer latency.
In the implementation we present here, we have aimed to minimize data
transfer, but no effort has been made (yet) to overlap computations and
data transfers, so we can expect some overhead to be present in the
multi-GPU case.

The performance of a multi-GPU implementation can be assessed by looking
at strong and weak scaling metrics. Ideal (linear) strong scaling is
achieved when running the same problem on $n$ devices takes $1/n$ the
time of running on a single device, although in practice this is limited
by how much of the program can be parallelized, following Amdahl's
law~\cite{amdahl}. Ideal (linear) weak scaling is achieved when
running a workload $n$ times larger on $n$ devices takes the same time
as running the original workload on a single device~\cite{gustafson}.

Weak scaling is not always easy to measure, because most workloads are
not perfectly linear in the number of particles, so that calibrating the
problem size for the tests is frequently non-trivial.
We run the multi-GPU tests on the Poiseuille test cases, and for the weak scaling
tests we increase the domain size in the periodic cross-flow
direction: due to the problem setup, making the domain $n$ times larger
gives us exactly $n$ times more particles which —under the assumption
that the number of iterations for the implicit solver does not change
significantly despite the larger size of the implicit matrix— leads to a
linear scaling of the work-load.

The domain size in the cross-flow periodic direction is defined by an
\emph{aspect} parameter $a$, such that the width in the given direction
is $aH$ (giving us the original setup for $a=1$). In practice, we
observed that the work-load only scales approximately linearly with
growing aspect; weak scaling measurements should thus be considered only
approximate.

\subsection{Cross-device data transfers}

In a multi-GPU configuration, a geometric section of the domain is
assigned to each GPU; the particles present in the corresponding domain
are considered \emph{internal} to the GPU. Additionally, to allow the
device to compute e.g. the forces acting on these particles, each GPU
also stores a copy of the particles that are \emph{neighbors} to the
internal particles, but belong to geometrically adjacent devices: these
are known as \emph{outer edge} particles. Finally, we define \emph{inner
edge} particles as particles which are internal to the given GPU, but
whose data must be copied to other GPUs (i.e. particles which are
``outer edge'' for an adjacent GPU).

We can differentiate two classes of computational kernels. If
information about neighboring particles is needed (e.g. forces
computation), then in a multi-GPU context, each device can only process
internal particles, and the data about the outer edge particles must be
exchanged; these will be termed \emph{internal-only} computational
kernels. However, if processing of a particle does not require
neighbors information (e.g. integration in the explicit scheme), each
device can update both its internal particles and the copies of the
outer edge particle, avoiding a data exchange; these will be termed
\emph{in-place} computational kernel.

As mentioned in~\ref{sec:impl-notes}, we use a Jacobi
preconditioner for our solvers. This requires us to compute the diagonal
of the (original) matrix during the initialization phase, and since this
requires the contributions from the neighbors, each device computes the
diagonal entries corresponding to its internal particles, and the
inner\slash outer edge particle data is then exchanged with the adjacent
devices.

During the initialization phase, we will also compute and exchange $\vec
r_0$ and, for BiCGSTAB, $\hat{\vec r}_0$ which is stored separately
(even though it holds the same values as the residual).

During the solver loop, the only vectors that need to be updated across
devices are $\vec A\cdot\vec p_k$ and, in the BiCGSTAB case, $\vec A \cdot \vec
r\star$, since all other vector updates can be done in-place on each
device with the available information.

Steps that require parallel reductions (such as the computation of the
residual norm or the updates of the $\alpha, \beta, \gamma, \delta,
\omega$ coefficients) will produce partial results on each device, which
will then be unified on host to obtain the final result (e.g. by adding
up the partial results) and, in the multi-node case, across nodes using
MPI primitives.

Due to the very small number of data transfers required during the
semi-implicit solver execution, we have provisionally decided to not
leverage concurrent transfer and computation, leaving this as a
potential optimization opportunity in the future.

\subsection{Performance results}

We present the result for the Newtonian and Papanastasiou rheology
in the plane Poiseuille flow test case, with
dummy and dynamic boundary conditions at resolutions of 32 and 64 particles
per unit length, and domain aspect from 1 to 4. For each aspect $a$, we
test $1$ to $a$ GPUs, resulting in 80 configurations total. To
determine the accuracy of the time measurements, we have run each
configuration 3 times, resulting in 240 total runs.

We ran the tests on an older machine equipped with 4 NVIDIA GeForce GTX
TITAN X (Maxwell architecture, compute capability 5.2): each GPU has
12GB of VRAM an 24 Compute Units (CUs) with 128 ``CUDA cores'' per CU.
The GPUs have peer access (i.e. the possibility to directly access each
other's memory without passing through the host) only in pairs (i.e. device 0 with
device 1, and device 2 with device 3), which limits the performance of
data exchanges when non-peering devices are involved (and thus in 3- and
4-GPU simulations).

The relative difference between the minimum and maximum runtimes for the
80 configurations is at most $1\%$, with a median of less than $0.1\%$
and a 90th percentile of less than $0.6\%$. Given the stability of the
result, we can rely on the median runtime for each configuration in our
analysis.

\begin{figure}
\centering{%
\includegraphics[width=.49\textwidth]{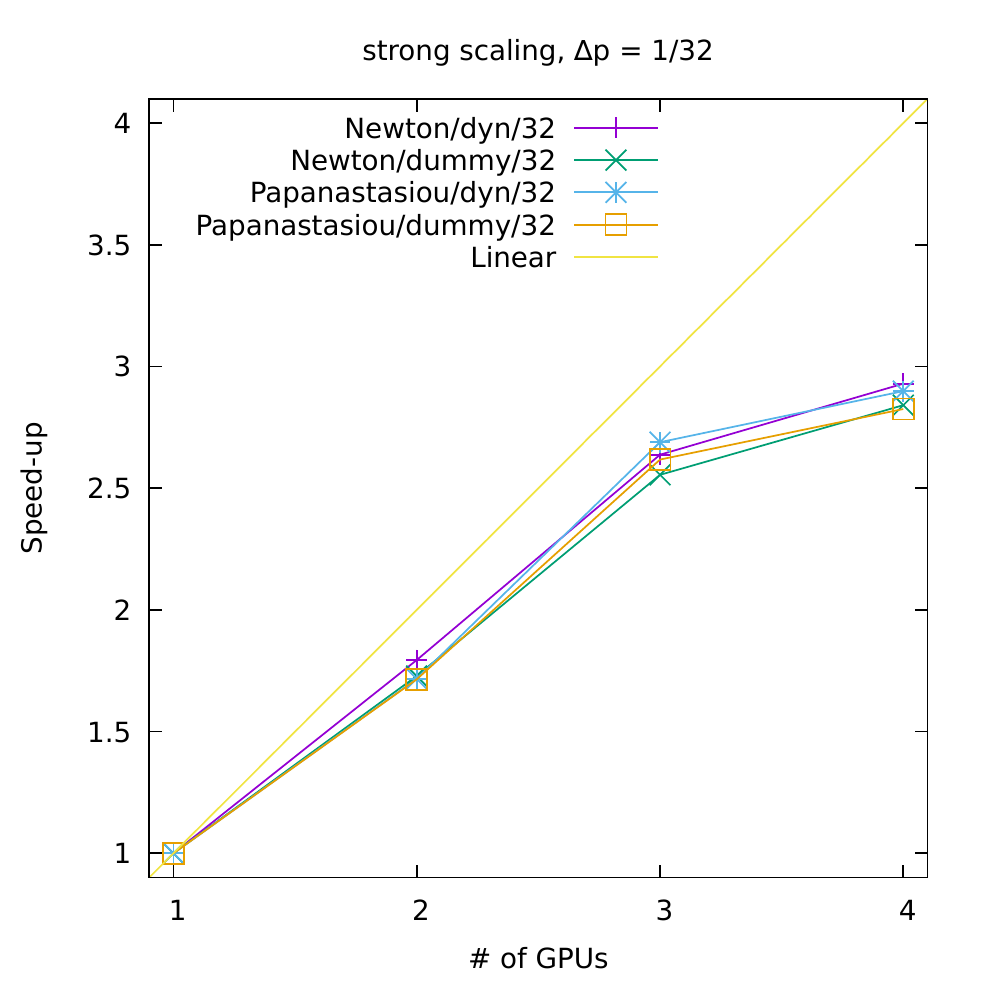}%
\hfill
\includegraphics[width=.49\textwidth]{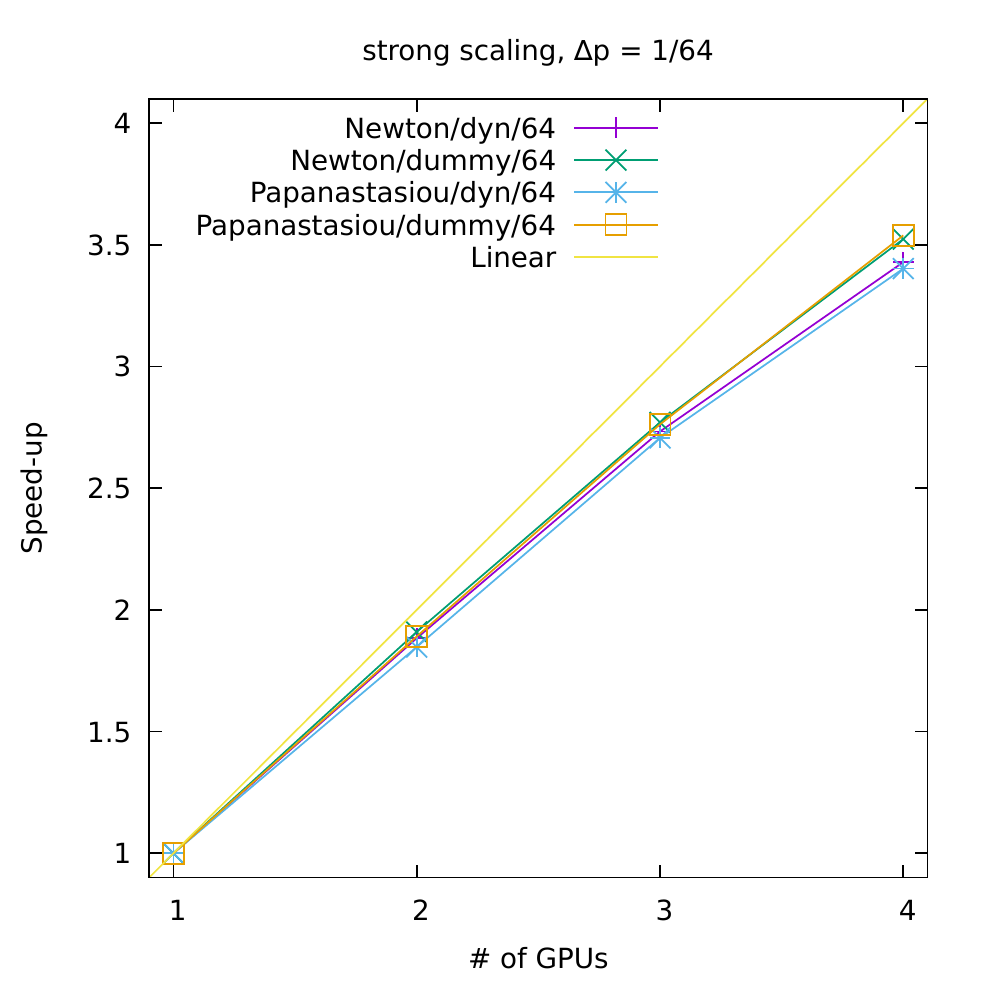}%
}
\caption{Strong scaling performance for our multi-GPU implementation,
with $\Delta p = 1/32$ (left) and $\Delta p = 1/64$ (right) with
different combinations of rheology and boundary conditions, using the
BiCGSTAB solver and the Wendland kernel.}
\label{fig:strong-scaling}
\end{figure}

For the strong scaling, we compare the runtime performance of the
problems with aspect $a=4$ (figure~\ref{fig:strong-scaling}). At the
coarser resolution this amounts to about 160,000 particles, while at the
finer resolution we have slightly less than 1.2 million particles.
The results show close to linear scaling (efficiency between $85\%$ and
$96\%$), provided there is enough load: the only case that drops below
$85\%$ efficiency is the 4-GPU case with 32 particles per unit length,
indicating that in this case there aren't enough particles (less than
40,000) to fully
saturate all devices, despite the large aspect.

\begin{figure}
\centering{%
\includegraphics[width=.49\textwidth]{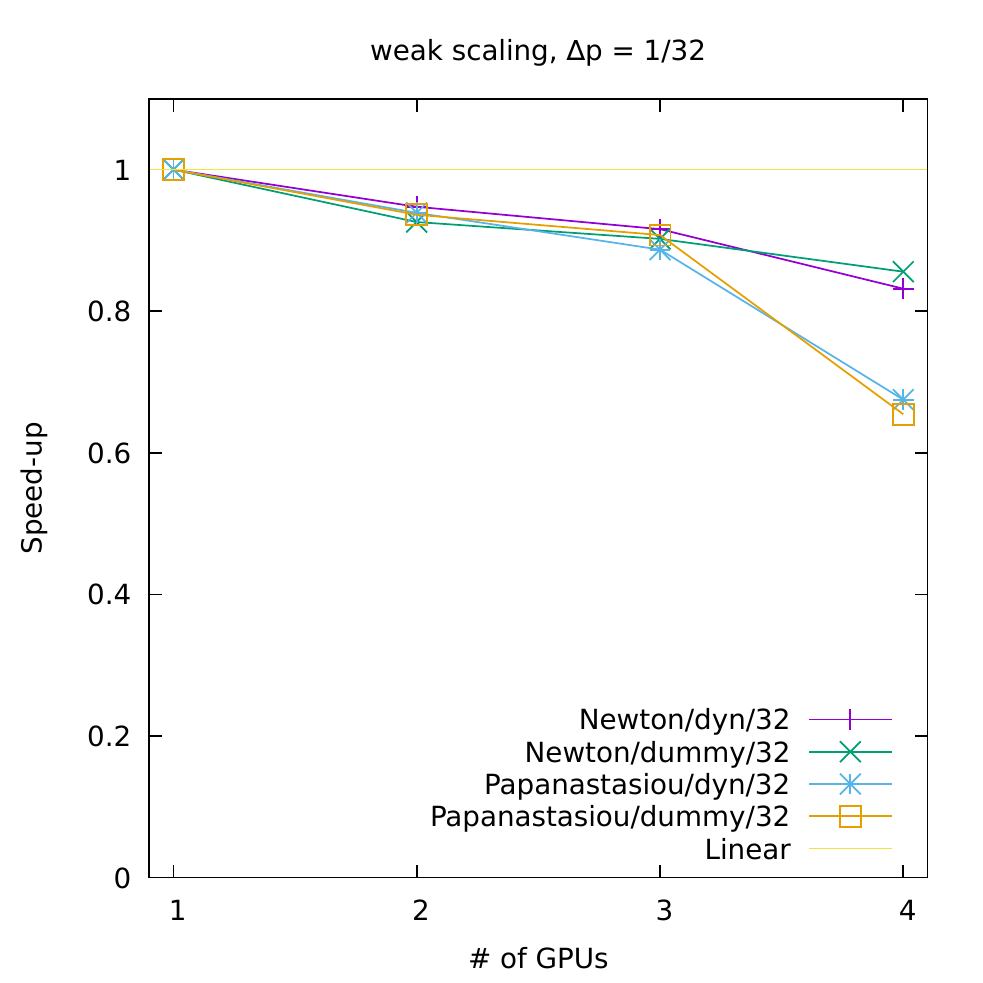}%
\hfill
\includegraphics[width=.49\textwidth]{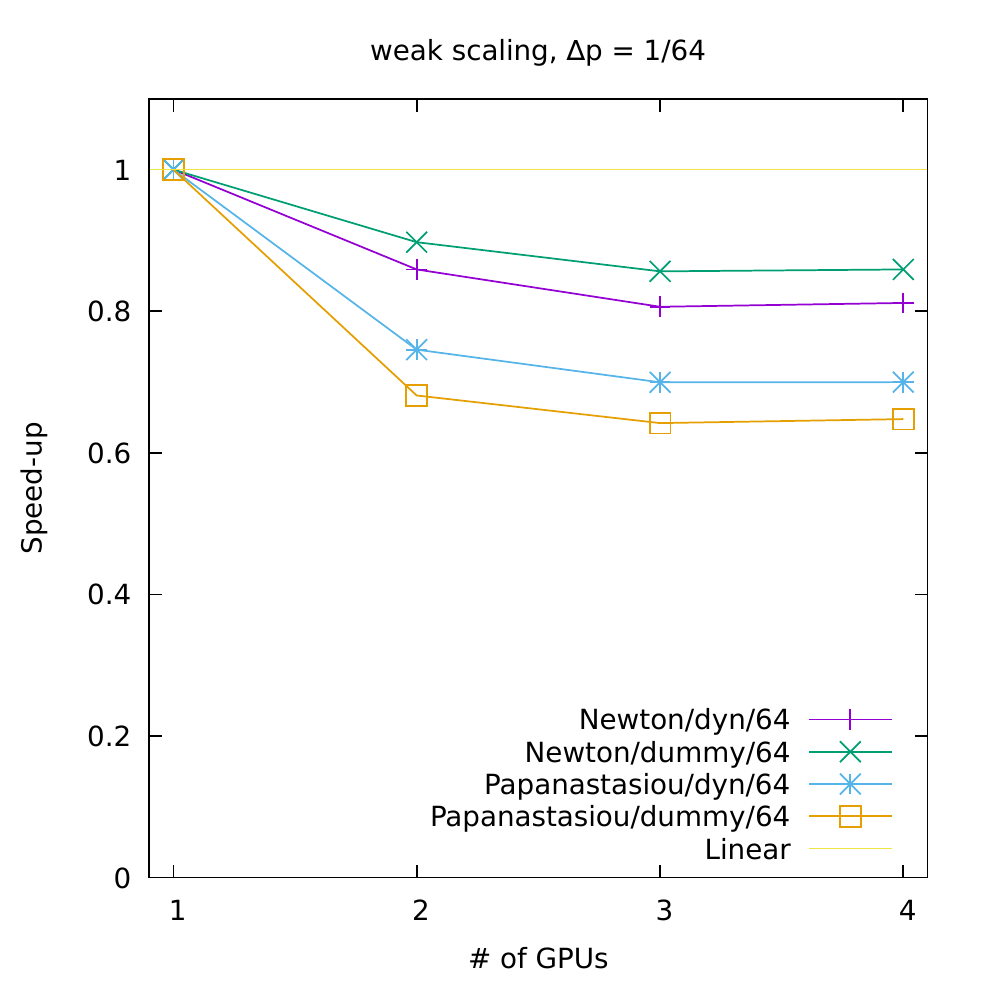}%
}
\caption{Weak scaling performance for our multi-GPU implementation,
with $\Delta p = 1/32$ (left) and $\Delta p = 1/64$ (right) with
different combinations of rheology and boundary conditions, using the
BiCGSTAB solver and the Wendland kernel.}
\label{fig:weak-scaling}
\end{figure}

For the weak scaling, we compare the runtime performance of the
single-GPU case with aspect $a=1$ with the multi-GPU case where the
number of GPUs is equal to the aspect (figure~\ref{fig:weak-scaling}).
As mentioned before, these results should be considered approximate,
since the work-load scaling itself is only being approximately linear as
a function of the domain aspect. Still, we get an efficiency of over
$80\%$ when using the Newtonian rheology, although the efficiency drops
by about $15\%$ in the Papanastasiou case.

The main difference between the two rheologies is the much higher number
of implicit solver iterations needed to converge at each time-step
(Papanastasiou requiring between 5 and 8 times more iterations than
Newton). The lower weak scaling efficiency in this case may thus be
indicative of the overhead cost of the data transfers during the solver
iterations. Indeed, as mentioned before, we do not overlap computations
and data transfers in this case, due to the small number of data
transfers needed during each iteration, but as the number of iteration
grows, this overhead becomes apparently large enough to measurably
affect the runtime.

\section{Conclusions and future work}

The semi-implicit integration scheme for the viscous term in SPH,
introduced in~\cite{zago_jcp_2018}, has been extended to support the
dummy boundary model, that requires the
boundary neighbors viscous velocity to be computed together with the
fluid particles. Due to the difference in the expressions for the
boundary\slash fluid contribution versus the fluid\slash boundary
contribution, the resulting coefficient matrix for the implicit system
is not strictly diagonally dominant, but the system is still solvable
due to it being weakly-chained diagonally dominant.

We adopt a modified version of the biconjugate gradient stabilized
method, that is both more robust numerically (particularly important in
single precision) and more apt for parallelization in distributed memory
configurations (e.g. multi-GPU). Our current multi-GPU implementation
achieves very efficient strong scaling (over $85\%$) given a
sufficiently large number of particles, although the data transfer
overhead has a measurable impact on the weak scaling performance when
the implicit solver requires a large number of iterations (typically,
for very high values of the viscosity). We expect this to be
significantly reduced by introducing striping (overlapping computation
and data transfers) in the solver routines.

While the improvements to BiCGSTAB were designed and implemented here
in the context of SPH, the numerical and computational benefits
of our approach are independent of the numerical method that generates the matrix,
and we expect similar benefits in terms of stability and parallelism
to be gained in any application adopting BiCGSTAB as a solver.

The computational benefit of the semi-implicit solver over the explicit
method in SPH can only be observed at fine resolutions, or when the viscosity
is extremely high. Due to the higher cost of the implicit solver, the
break-even point for a gain in performance can be estimated at around a
factor of $20$ between the time-step restriction for the sound-speed
versus the time-step restriction for the viscosity. The actual
performance gains are then dependent on the number of iterations needed
by the implicit solver. An interesting result that was found is that
smoothing kernels with a larger influence radius (and thus a larger
average number of neighbors per particle) produce systems that require
less solver iterations to converge, partially compensating for the
higher computational cost of the kernel evaluation and lowering the
break-even point for the convenience of the semi-implicit scheme over
the explicit one.

With the availability of a solver that does not require symmetry,
further extensions of this approach are possible to other boundary
models not explored in this paper (e.g. the semi-analytical boundary
models~\cite{ferrand2013}).

A prospective improvement to the approach presented so far would be the
adoption of an appropriate preconditioner. This should be fast to
compute, and ideally preserve the neighborhood structure of the matrix.
While the Jacobi preconditioner we have adopted fits these criteria, its
benefits in term of convergence speed are limited.

A better choice for the initial guess may also provide meaningful
speed-ups, but the computational cost for more sophisticated
alternatives must be taken into consideration, especially in the case of
the dummy boundary model, that requires the computation of the initial
guess to be split in two separate phases (computation of an initial
guess for the fluid particles, followed by the interpolation to produce
the initial guess for the boundary particles).

Finally, the good results achieved in the non-Newtonian case indicate
that using the viscosity computed from the velocity at the previous
time-step is accurate enough to provide superlinear convergence to the
theoretical solution, although further improvements can be expected by
bringing the computation of the new viscosity into the implicit system.
On the other hand, we can expect the resulting non-linearity of the
system to have non-trivially higher computational requirements,
suggesting the need for a study of the performance\slash accuracy
benefits. Our tests additionally reveal that for shear-dependent
viscosity, further tuning to the dummy boundary formulation is necessary
to improve the results, and that the excessive numerical dissipation
at small smoothing factors and\slash or influence radius deserves more attention,
and may require the need to introduce kernel correction terms in our semi-implicit formulation.

To this end, we are running a more complete set of validation tests,
covering all of the several non-Newtonian rheological models currently
implemented in GPUSPH and a wider variety of test cases.
This is made possible by leveraging the computational benefits of the
integrator scheme described here and its particular advantage
in the case of non-Newtonian rheologies.



\section*{Acknowledgements}

This work was developed within the framework of the ATHOS Research
Programme organized by the Laboratory of Technologies for Volcanology
(TechnoLab) at the Istituto Nazionale di Geofisica e Vulcanologia
(Catania, Italy) and the Department of Civil and Environmental
Engineering at the Northwestern University (Evanston, USA),
and was partially supported by the EDF-INGV partnership contract
8610-5920039040. More references and acknowledgments about GPUSPH can be
found in the website http://gpusph.org



\bibliographystyle{elsarticle-num}
\bibliography{gpusph,nongpusph}

\appendix

\section{Numerical pitfalls in the viscous matrix symmetry}
\label{sec:numsym}

Recall (equations~\eqref{eq:kab} and~\eqref{eq:offdiag}) that, for fluid
particles, the non-zero off-diagonal coefficients of the viscous matrix
are
\[
a_{\beta\alpha} = \Delta t K_{\alpha\beta}
\]
where
\[
K_{\alpha\beta} = -\frac{2\bar\mu_{\alpha\beta}}{\rho_\alpha \rho_\beta}
F_{\alpha\beta} m_\alpha;
\]
sufficient condition for matrix symmetry is that $m_\alpha = m_\beta$
(true if all particles have the same mass, numerically), $F_{\alpha\beta} =
F_{\beta\alpha}$ and $\bar\mu_{\alpha\beta} = \bar\mu_{\beta\alpha}$.

The two latter equalities are true analytically. Numerical differences
however may arise depending on how the expressions are computed. If the
matrix coefficients are not stored, but are computed on-the-fly as in
our case, it is also impossible to enforce symmetrization (e.g. by
storing the arithmetic mean of $a_{\beta\alpha}$ and $a_{\alpha\beta}$).

This has implications in contexts where the matrix should be symmetric,
allowing the use of the conjugate gradient as solver, but ends up not
being such numerically, potentially enough to prevent the CG from
converging. Our tests have exposed in particular two instances, that we
here present, discussing also ways in which they can be prevented.

\paragraph{Fused Multiply-Add} (FMA) is an operation that has been
introduced in the 2008 revision of the IEEE-754 standard for
floating-point~\cite{ieee754}. It takes three arguments $a, b, c$ and
computes $a\cdot b + c$ with a single rounding (i.e. as if the product
was computed with infinite precision). Modern hardware has good support
for this operation, that can be completed in a single cycle, allowing
for fast and robust computation of e.g. polynomial in Horner's
form~\cite{horner1819}. As pointed out by Kahan~\cite{kahan1996},
FMA is a ``mixed blessing'', since improper usage (by either the user or
the compiler) can produce issues.

This is exactly the case for the computation of the average viscosity,
in some circumstances. Assume for example that the code tracks the
kinematic viscosity $\nu$ of the particles, so that the dynamic
viscosity is computed as-needed with $\mu = \nu \rho$; assume also that
the arithmetic mean is being used as average. Then:
\[
2\bar\mu_{\alpha\beta} = \nu_\alpha\rho_\alpha + \nu_\beta\rho_\beta,
\]
and with FMA this can be computed using two instructions (a
multiplication and an FMA) rather than three (two multiplications and an
addition).

However, as Kahan remarks, without further indications there is no way
to know if FMA would be used as
$\fma(\nu_\alpha, \rho_\alpha, \nu_\beta\rho_\beta)$ or the converse
(i.e. which multiplication will be fused, and which not). For us, either
choice will lead to a non-symmetric matrix, since e.g.
\[
2\bar\mu_{\alpha\beta} = \fma(\nu_\alpha, \rho_\alpha, \nu_\beta\rho_\beta)
\ne \fma(\nu_\beta, \rho_\beta, \nu_\alpha\rho_\alpha) =
2\bar\mu_{\beta\alpha}.
\]
(and likewise for the other choice of FMA).

In our tests, the compiler may apply this optimization even if
$\mu_\alpha, \mu_\beta$ are precomputed. In such circumstances the
off-diagonal matrix terms end up with differences in the order of
$10^{-8}$, which are sufficient to affect the convergence of the CG
(even if not to prevent it altogether). Interestingly, we have only seen
this happen for the arithmetic mean; in particular, this does not happen
when using the harmonic mean, which is the more physically correct
choice of averaging operator with non-uniform viscosity.

A way to avoid this situation is to prevent the compiler from
``optimizing'' the computation beyond what is allowed by the correct
application of the IEEE-754 standard. How to achieve this depends on the
specifics of the language and compiler used. In C-based languages, for
example, contraction of numerical expressions can be disabled by adding
the \verb+STDC FP_CONTRACT OFF+ pragma. Tracking the dynamic viscosity
as particle property also avoids this issue altogether, although it may
also eliminate other optimization opportunities.

\paragraph{Relative position computation}

The computation of $F_{\alpha\beta}$ is symmetric if the computation of
$r_{\alpha\beta}$ is symmetric, which is the case if $\vec
r_{\alpha\beta} = \vec r_\alpha - \vec r_\beta$ is computed directly.

This however is not the case when the strategy discussed
in~\cite{herault_accuracy_2014,saikali_2020} is adopted to ensure uniform accuracy
throughout the computational domain, which is essential when simulating
large domains with fine resolutions, and especially when using
single-precision. With the uniform accuracy strategy, the global
position $\vec r_\alpha$ of the particle is not stored in the code.
Instead, particle positions are tracked as positions $\vec
r'_\alpha$ relative to the centre of the cells $\vec c_\alpha$ used to
speed-up the neighbors search and to split the computation across
multiple devices.

Let $\vec l$ be the vector of the cell side lengths. Then, the distance
between two particles can be computed as:
\[
\vec r_{\alpha\beta} = \vec r'_\alpha - \vec r'_\beta + \vec l \cdot
(\vec c_\alpha - \vec c_\beta).
\]
When computing the contributions from all neighbors to a specific
particle, it is convenient to pre-compute the offset term $\vec
o_{\alpha\beta} = \vec r'_\alpha + \vec l \cdot
(\vec c_\alpha - \vec c_\beta)$, which can be used for all neighbors in
sharing a cell. However, due to rounding differences, we then have:
\[
\vec r_{\alpha\beta} = \vec o_{\alpha\beta} - \vec r'_\beta \ne
\vec r'_\alpha - \vec o_{\beta\alpha} = -\vec r_{\beta\alpha}.
\]

This rounding difference propagates to $F_{\alpha\beta} \ne
F_{\beta\alpha}$, ultimately leading to a non-symmetric matrix. In this
case, the discrepancies can add up to the point of preventing the CG
from converging when it should.

The solution in this case is to only precompute $\vec o'_{\alpha\beta} =
\vec l \cdot (\vec c_\alpha - \vec c_\beta)$, and computing the distance
as
\[
\vec r_{\alpha\beta} = \vec o'_{\alpha\beta} + (\vec r'_\alpha - \vec r'_\beta).
\]
Grouping the computation of $\vec r'_{\alpha\beta}$ ensures that
roundings are done in the same sequence independently from the particle
order, resulting in a numerically symmetric matrix, as expected.

\section{Implementation notes}
\label{sec:impl-notes}

We have implemented our own set of Basic Linear Algebra Subprograms
(BLAS) routines, covering the essential
operations used in the CG and BiCGSTAB algorithms:

\begin{description}
 \item[axpby] computes $\alpha \vec x + \beta \vec y$ for scalars
  $\alpha, \beta$ and vectors $\vec x, \vec y$; as an optimization,
  if a coefficient is zero the corresponding vector is not read from
  memory; since the condition is uniform (in the parallel
  sense) and memory access is the primary bottleneck in this operation,
  this leads to small but measurable performance improvements;
  the most important benefit of this approach however is that a vector
  may have undefined content (i.e. not be initialized) if the corresponding
  coefficient is null, which simplifies the initialization of the CG;
 \item[dot product] computes $\vec x \cdot \vec y$ for vectors $\vec x,
  \vec y$; on GPU this is implemented using the \emph{parallel
  reduction} paradigm, with an efficient two-step implementation;
  additional cross-GPU and cross-node reductions are necessary if more
  than one GPU or node is being used;
 \item[squared norm] i.e. the dot product specialized for the case
  $\vec y = \vec x$, used to compute the norm of $\vec b$ during
  initialization, and the norm of $\vec r_{i+1}$ during the CG
  iterations;
 \item[gemv] computes $\alpha \vec A \vec x + \beta \vec y$ for scalars
  $\alpha, \beta$, vectors $\vec x, \vec y$ and matrix $\vec A$; in our case
  the matrix is not stored, but expressed as a functional, and our
  implementation leverages the fact that the structure of $\vec A$
  depends on the neighborhood structure of the underlying SPH method;
  null coefficients are handled efficiently and robustly also for
  this operator.
\end{description}

The main reason behind our choice to reimplement the routines rather
than relying on existing implementations (such as \texttt{cuBLAS}) is
that the additional control on the implementation allows us to tune it
for our specific use-cases and existing data structures. This has been
particularly important for the \texttt{axpby} and \texttt{gemv}
routines, as mentioned.

An additional important aspect of our implementation is that all
summations with multiple terms (i.e. the ones in the dot product,
squared norms and matrix\slash vector products) use compensated
summation algorithms to improve numerical accuracy, essential due to our
use of single precision. In particular, we use the Kahan
summation algorithm~\cite{kahan1965}, pending the evaluation of the accuracy\slash
computational cost assessment of the Kahan--Babu\v{s}ka--Neumaier
compensated summation~\cite{neumaier1974, klein2006}.

As a bonus, not being tied to existing libraries will reduce migration
costs for the adoption of different (i.e. not CUDA) computational
technologies.

Storage requirements for CG are of one vector each for $\vec p, \vec r, \vec x$
and $\vec A \vec p$, plus one scalar each for $\gamma_i = \vec r_i \cdot \vec
r_i, \gamma_{i+1}, \alpha_i$ and $\beta_i$. Separate storage of the
scalars and vectors takes into account shared\slash reused
sub-expressions as well as the need to hold temporary values for the
cross-GPU and cross-node reductions of the dot products and squared
norms, but taking into account the possibility to reuse storage (e.g. we
can reuse the storage for $\beta_i$ to hold $\vec p_i\cdot \vec A\vec p_i$
after the $\vec p_i$ update and up to the $\alpha_i$ computation).

For BiCGSTAB, we need to store two additional vectors compared to CG,
$\hat{\vec r}_0$ and $\vec A \vec r_\star$, and one additional scalar,
$\omega$; $\alpha'$ and $\alpha$ can share storage, since they are never
needed concurrently.

In our experiments, we have further observed an improvement in
convergence by adopting a Jacobi preconditioner. This requires two
additional computational kernels (one to compute the diagonal terms
$\vec D$
of the matrix, one to divide the right-hand side vector by them), which
are executed before the first iteration, and an additional vector to
store $\vec D$ itself, which is needed for all iterations.

For both the CG and BiCGSTAB solver, the initial guess is currently
taken to be the solution at the previous time-step. More sophisticated
choices are being explored, in an effort to reduce the number of solver
iterations necessary to reach convergence.

\end{document}